\documentclass[usenatbib]{mn2e}

\usepackage[dvips]{graphicx}
\def\insitu{{in situ~}}
\begin{document}

\title[Chemical properties of stellar haloes]
  {Stellar haloes in Milky-Way mass galaxies: From the inner to
    the outer haloes}

\author[Tissera et al. ]{Patricia B. Tissera $^{1,2}$,Timothy
  C. Beers$^{3,4}$, Daniela Carollo$^{5}$, Cecilia Scannapieco$^{6}$\\
$^1$  Consejo Nacional de Investigaciones Cient\'{\i}ficas y T\'ecnicas, CONICET, Argentina.\\
$^2$ Instituto de Astronom\'{\i}a y F\'{\i}sica del Espacio, Casilla de Correos 67, Suc. 28, 1428, Buenos Aires, Argentina.\\
$^3$ National Optical Astronomy Observatory, Tucson, AZ, 85719 USA.\\
$^4$ JINA:  Joint Institute for Nuclear Astrophysics. \\
$^5$ Department of Physics and Astronomy, Macquarie University,
Sydney, 2109 NSW, Australia.\\
$^6$ Leibniz-Institut für Astrophysik Potsdam (AIP), An der Sternwarte 16, D-14482 Potsdam, Germany.\\
}

\maketitle

\begin{abstract}
We present a comprehensive study of the chemical properties of
the stellar haloes of Milky-Way mass galaxies, analysing the transition between the inner to the outer haloes. We find
the transition radius between the relative dominance of the inner-halo
and outer-halo stellar populations to be $\sim15-20 $ kpc for most
of our haloes, similar to that inferred for the Milky Way from recent
observations. While the
number density of stars in the simulated inner-halo populations
decreases rapidly with distance, the outer-halo populations contribute
about $20-40 $ per cent in the fiducial solar neighborhood, in
particular at the lowest metallicities. We have determined [Fe/H]
profiles for our simulated haloes; they exhibit flat or mild gradients,
in the range [--0.002, --0.01 ] dex kpc$^{-1}$. The metallicity
distribution functions exhibit different features, reflecting the
different assembly history of the individual stellar haloes. We find
that stellar haloes formed with larger contributions from massive
subgalactic systems have steeper metallicity gradients. Very
metal-poor  stars are mainly contributed to the halo
systems by lower-mass satellites. There is a clear trend among the
predicted metallicity distribution functions that a higher fraction of
low-metallicity stars are found with increasing radius.  These properties are consistent with the
 range of behaviours observed for  stellar haloes of nearby galaxies. 
\end{abstract}

\begin{keywords} Galaxy: structure, galaxies:formation, galaxies:evolution, cosmology: theory
\end{keywords}

\section{Introduction}

The formation of diffuse, extended stellar haloes is a natural result of
hierarchical clustering scenarios, where most of the stars are acquired
by the accretion of satellites that are completely or partially
disrupted \citep[e.g.][]{bullock2005,johnston2008,cooper2010, font2011b,
tissera2012}. Numerical simulations have also  indicated
the existence of an \insitu stellar halo component formed 
dissipatively in the inner regions and/or by disc-heated stars
\citep[e.g.][]{zolo2009, mcc2012}. The contributions of each type of
stellar populations reflect the histories of assembly of the stellar
haloes, as well as the physics included in the simulations. 
Disentangling their individual effects is a challenging task. In this
 paper, we study a suite of six Milky-Way mass galaxies, resolved
with  intermediate numerical resolution, with the aim of
constructing a more complete picture of the assembly of galaxies and
their stellar haloes.

Most of the baryonic physics in current galaxy formation models are
described through the use of subgrid physics; these models have improved
significantly in the last few decades. As a consequence, it is now
possible to describe, on  a more physical basis, the formation
of galaxy discs, mass-loaded gas outflows and the chemical evolution of
gas and stars\citep[e.g.][]{arge2009,gove2009,scan09,guedes2011}. Other
trends, such as the dependence of star-formation efficiency on total
mass, remain to be fully explained \citep[e.g.][]{aumer2013,derossi2013}. 
Nevertheless, all such models carry with them theoretical or
numerical caveats that can be constrained through confrontation with
observations. 

 It is now well-established that the observed chemodynamical
patterns of galaxies provide powerful tools for the interpretation of
observations and for efforts to improve galaxy formation models.
In particular, regarding the Milky Way (MW), the plethora of such
information that will flow in the near future from the Gaia satellite,
as well as from other large-scale ground-based efforts
\citep[for a recent review, see][]{ivesic2012} will dramatically expand
the scope of the observational constraints. There  is already
evidence for the existence of two stellar  halo populations in the
MW, with different properties  \citep[e.g.][]{caro2007,caro2010,
beers2012,an2013}. Recent reports on nearby galaxies, such as M31
and M81,  also show evidence for the existence of stellar haloes
with complex  structure  \citep{gilbert2013,ibata2013,
monachesi2013}. Observations are also starting to provide information
on the stellar haloes at high redshift, opening the possibility of
constraining the assembly of the stellar haloes directly
\citep{trujillo2012}. 

 The properties of the baryons and the dark matter in the
 re-simulated version of the Aquarius haloes by \citet{scan09} have been
analysed in a series of papers. From these works, we know that none of
these haloes hosts a disk-like galaxy resembling the MW, but instead,
galaxies with a variety of morphology have been formed. 
\citet{scan09} and \citet{scan11} find that the discs are
generally young, with stars spanning a wide range in stellar age. The
younger stars are on circular orbits, defining thin discs, while the
oldest stars determine thicker discs with 2-3 times larger velocity
dispersion than  the thin components. Most of the stars in the discs formed
from  gas accreted onto the main progenitor  to first build up gaseous discs. The bulge components formed
early, and in shorter starbursts. The stellar haloes are reported to have
a complex structure, with most of their mass coming from accreted
satellites. The chemical patterns of these dynamical components have
been extensively discussed by \citet[hereafter, Paper~I]{tissera2012}
and \citet[hereafter, Paper~II]{tissera2013}. In Paper~I we analysed the
chemical properties of the discs, bulge,  and inner- and outer-halo
populations in relation to their assembly histories, finding that the
[Fe/H] and the $\alpha$-elements trends are in very good agreement with
observations. In particular, in Paper~II, we highlighted how one might
understand a range of the properties observed in the stellar haloes of
MW-mass galaxies (e.g., metallicity distribution functions, MDFs,
$\alpha$-element enrichment histories, density distributions, and
kinematics) in the context of their $\Lambda$CDM hierarchical halo
assembly histories. In our previous works, we analysed the  inner- and
outer-halo populations as separate components.

In this  paper, we continue our analysis of the nature of the
predicted stellar populations of a suite of MW-mass galaxies, selected
from the Aquarius simulation project, but focus on the stellar haloes as
 a whole, and their dependence on the accretion history of
satellites. We aim at studying in more detail the superposition of the
inner-halo populations and outer-halo populations, how they vary from
one simulated system to another, and how the transition from the
inner-halo regions to the outer-halo regions can be linked to their
assembly histories. We concentrate on the analysis of the accreted
satellites, since they are the main contributors to the formation
of the stellar haloes in  our adopted cosmological model, and how
the history of their assembly could be linked to the characteristics of the
MDF as a function of radius.  We discuss implications for the
interpretation of observational data in the MW and other large spirals,
and on the development of predictions that might be used to test and
refine future simulations. 

In Section 2 we describe the simulations and the main aspects of our
chemical-evolution model. In Section 3 we consider the transition
between the inner-halo populations and the outer-halo populations, as
well as the predicted combined MDFs and the expected abundance
profiles. In Section 4 we analyse the nature of the satellites that
contributed to the formation of the stellar haloes, and compare how they
differ with respect to the contribution of stars with different
metallicities to the formation of the stellar haloes. In Section 5 we
study very metal-poor stars in particular, and compare their frequency
as a function of distance with available observations. We summarize our
findings in Section 6, and provide a brief discussion.

\section{The simulated stellar haloes}

We continue our analysis of the suite of six high-resolution MW-mass
systems described in Papers~I and II. These six haloes are part of the
Aquarius simulation project presented by \citet{scan09}. The Aquarius
haloes were selected from a cosmological volume of a $100 \ {\rm Mpc \
h^{-1}}$ box, with only a mild isolation criterion imposed at z=0
\citep[see][for further details on the generation of the initial
conditions]{sprin2008}. The initial conditions used in this paper have
been modified to include baryons \citep{scan09}.

The simulations were performed with a $\Lambda$CDM cosmogony with
$\Omega_{\rm m}=0.25,\Omega_{\Lambda}=0.75, \sigma_{8}=0.9, n_{s}=1$,
and $H_0 = 100 \ h \ { \rm km s^{-1} Mpc^{-1}}$ with $h =0.73$. The
galactic systems are identified at the virial radius ($r_{200}$), and
have virial masses in the range $\approx 5 - 11 \times 10^{11} {\rm
M_\odot} h^{-1}$. They are numerically resolved using $\approx 1$
million total particles within the virial radius, so that dark matter
particles have masses of the order $\approx 10^{6}{\rm M_\odot }h^{-1}$,
while initially the gas particles have masses of $\approx 2 \times
10^{5}{\rm M_\odot} h^{-1}$. Table ~\ref{tab1} summarizes their
principal characteristics.

The simulations were run using a modified version of {\small GADGET-3}
\citep{scan05, scan06}, which explicitly describes the physics of
the baryons, including their chemical evolution from the contributions
of Type II and Type Ia supernovae (hereafter, SN). The SN feedback
allows for the triggering of galactic mass-loaded outflows,
self-regulated by the potential well of the systems. A detailed
description of the chemical model and  SN-feedback mechanisms can be
found in \citet{scan09}. This  SN-feedback scheme has proven
successful for the regulation of the star-formation activity during violent and
quiescent stages of evolution. One of its most important features is the
non-existence of mass-size dependent parameters, which makes this
implementation  well-suited to study  of the formation of
galaxies in a cosmological context. However, in the current set of 
Aquarius-simulated haloes, there is still an overproduction of stars at
high redshift. This is a complex, open problem for current galaxy
formation models, and an area of active research \citep[e.g.][]{aumer2013, hopkins2013,stinson2013}. Nevertheless, these haloes
allow us to study and compare their properties in relation to their
history of assembly, given that all of them have been run with the same
numerical model. 

In Papers~I and II, the effects of numerical resolution on the
physical properties of the structure have been extensively discussed. In
Paper~I we analysed three different levels of resolution, in order to
assess numerical effects on the derived chemical properties. We found
that, for the global analysis we performed, the current resolution was
sufficient.  We estimated the mass fraction of stars with
[Fe/H]$<-2.5$ within the virial radius, for the three level of
resolution, finding that the percentages converge from the very 
low resolution ($\sim 7\%$ in Aq-E-7) to the highest available ($\sim
5\%$ in Aq-E-6 and Aq-E-5).  However, some properties of the
simulated haloes might be more sensitive to numerical effects; below we
alert the reader where those might be a concern.

\subsection{The Aquarius stellar haloes}

In Paper~I, three main dynamical components were defined: the discs,
central spheroids, and stellar haloes, by adopting simple binding-energy
criteria. Briefly, this method uses the ratio between the angular
momentum along the major axis of rotation and the total angular momentum
of the systems, $\epsilon$, to define the disc component ($\epsilon >
0.65$). The remaining particles are assigned membership in the central
spheroids, the inner-halo populations (hereafter, IHPs) and the
outer-halo populations (hereafter, OHPs) by application of a
binding-energy criterion, as described in detail in Papers~I and II.
Simple energy criteria were used to define these haloes, with the
higher-energy (less-bound) stars belonging to the OHPs, and the
lower-energy (more-bound) stars belonging to the IHPs. The surviving
satellites are not considered in this analysis. There are, however,
certainly debris streams that have not been totally disrupted present in
the simulated stellar haloes, as can be seen from see Fig. 1 in Paper I.
Their quantification as separate substructures might be limited by
numerical resolution \citep{gomez2013}, but this does not directly
impact our results. We are interested in analysing them as part of the
OHPs or IHPs to which they are assigned on the basis of their dynamical
properties. 

The binding-energy criterion we employ does not establish a sharp
spatial boundary between the populations of these components; rather,
they are superposed on one another. Hence, it is convenient to
distinguish, as was done in Paper~II, between the inner-halo and
outer-halo regions (IHRs and OHRs, respectively), within which the
corresponding IHPs and OHPs dominate. We refer to the radial distance
where the transition from the IHRs to the OHRs occurs as the halo
transition radius, $r_{\rm HTR}$. 

According to the results of our previous work, the OHPs mainly comprise
low-metallicity, highly $\alpha$-element enriched stars (the so-called
debris subpopulation) assembled from accreted satellites. The level of
enrichment of the OHPs correlates with the fraction of these stars which
were formed in more-massive satellites, with dynamical masses larger
than $10^9M_{\odot}$. Some satellites will be able to survive farther
into the potential well of their parent galaxies, and although they are
disrupted by dynamical friction during their infall, they can contribute
to the formation of a weak metallicity gradient in the outer haloes.
These systems could be gas-rich, so that star formation might continue
within the virial haloes of their parent galaxies \citep[see][for a
detailed analysis of the evolution of the satellites in
Aq-C-5]{sawala2012}. These stars were called the endo-debris
subpopulation, and were found to be low metallicity, but less
$\alpha$-enriched, than the debris stars. In Paper~II it was shown that
the metallicity gradients in the OHPs increased as the contribution from
stars formed in more-massive (more-bounded) satellites increased, as
expected. Since no clear difference was found in the kinematics of the
stars arising from the debris and the endo-debris stars, they were
considered as a single population in the outer haloes. 

The inner haloes are more complex, and in our simulations their
assembly appears to have involved three different components: debris
stars, endo-debris stars and disc-heated stars. Each of the Aquarius
haloes exhibit the three subpopulations in different proportions,
depending on their formation history. It was shown in Paper~II that
accreted stars cover the entire range of binding energies of the inner
haloes, while the endo-debris and disc-heated stars are more
gravitationally  bound. Most of the IHP stars have ages larger than
$\sim 8$ Gyr, with an average age around $\sim 11$ Gyr. The endo-debris
and disc-heated stars exhibit lower levels of $\alpha$-element
enrichment. The younger disc-heated stars are supported more by
rotational motions, while the older of them exhibit cooler kinematics,
similar to those of the accreted stars. The endo-debris stars exhibit
slightly larger velocity dispersions.

Overall, our simulated haloes are assembled from stars formed in
satellites that were accreted by the main progenitor galaxy, or from the
gas retained by incoming satellites that was later transformed into
stars and/or tidally disrupted. A mean contribution of $\sim 25$ per
cent of stars in the IHPs comes from disc-heated stars\footnote{The
fraction of disc-heated stars in our simulations should be considered
upper limits, due to numerical limitations. Because of the gravitational
softening is $\sim1$ kpc$ h^{-1}$, the discs are already thick.}.
The fraction varies between $3-30$ per cent as a consequence of their
different assembly histories. The disc-heated stars in our simulations
are, in general, older than 8 Gyr; only the younger population is able
to recall its origin, because of  the remaining excess of rotational
velocity in comparison with other stellar-halo stars (see
fig. 1 of Paper~II). And, while having been formed  in a disc structure, they were
disturbed and dynamically  heated by violent events, as found in
previous works \citep[e.g.][]{abadi2003,purcell2009}. On average, these
stars have lower $\alpha$-enhancement and lower metallicity than those
formed in accreted satellites \citep{zolo2010}. Recent  observations have reported the
existence of stars  with dynamical and chemical properties in agreement
with  their being disc-heated stars in the stellar haloes of the
 MW \citep{nissen2010} and  M31 \citep{dorman2013}.
The frequency and properties of these stars can be reproduced by our simulations.

\begin{figure}
\hspace*{-0.02cm}\resizebox{7.0cm}{!}{\includegraphics{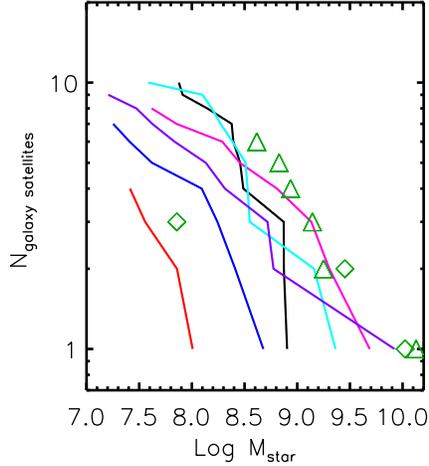}}
\caption{Number of surviving satellites in the simulated Aquarius
haloes as a function of their stellar masses (see Table 1 for colour
code).  The  observed distributions for  the MW (green
diamonds) and M31 (green triangles) taken from \citet{mcconnachie2012} and \citet{watkins2013}, respectively, have been included for comparison.
 }
\label{fmsat}
\end{figure}

Since, in this work, we focus our analysis on the  entire
stellar haloes of massive spiral galaxies, it is interesting to 
ask if the surviving satellite systems in the simulated haloes are
comparable to the current state-of-knowledge of the orbiting
satellites in the MW and M31. Therefore, we  estimated the number
of satellites as a function of the stellar masses, and compared them
with observations from \citet{mcconnachie2012} and \citet{watkins2013}. We only consider systems
resolved with at least 1000 particles, since otherwise their properties
could be strongly affected by  the resolution of the simulations.
These satellite systems have been also analysed in Paper~I, 
where it was
shown that they possess diverse star-formation histories and
metallicity distributions.
In fact, on average, our surviving satellites tend
to be $\alpha$-element poor with respect to those satellites
accreted to form the stellar haloes (which  also tend to be
$\alpha$-element poor). As can be seen from Fig.~\ref{fmsat}, the
analysed Aquarius haloes show a variety of  satellite
distributions, as expected. The observed relations corresponding to MW
and M31 are within the simulated values, at least for the mass range
analysed in this work.\footnote{We assume a mass-to-light ratio of 4.8
to covert observed luminosities in the V-band to stellar masses.}

\begin{table*}
  \begin{center}
  \caption{General characteristics of the Aquarius stellar haloes. The
left columns indicate the encoding name, the colours and symbols of the
simulations used in our figures, all simulated with numerical resolution
level-5. $N_{\rm dm}$ and $N_{\rm bar}$ are the total number of particles in each
mass component per halo. The parameters $r_{\rm 200}$ and $M_{\rm 200}$ are the virial
radius and mass, respectively. $M_{\rm s}$ is the stellar mass of the
main galaxies hosted by the halo measured within the galaxy's radius.
The last two columns show 
 $r_{\rm
HTR}$ , the transition radius between the IHPs and OHPs, and  [Fe/H]$_{\rm
slope}$, the gradient for the metallicity profiles measured from $r_{\rm
HTR}$ to $r_{\rm 200}$.}
\label{tab1}
\begin{tabular}{|l|c|c|c|c|c|c|c}\hline
 {Systems}   &$N_{\rm dm}$&$N_{\rm bar}$ & $r_{\rm 200}$& $M_{\rm 200}$&$M_{\rm s}$ & $r_{\rm HTR}$ & [Fe/H]$_{\rm slope}$\\ 
   & && kpc $h^{-1}$ &$10^{12}{\rm M_{\odot}h^{-1}} $ &$10^{10}{\rm M_{\odot}}h^{-1} $&kpc & dex kpc$^{-1}$ \\\hline
   Aq-A-5 (black crosses) &529110 &425737 &	169 &  1.10    & 5.92 & 18& $-$0.006\\
   Aq-B-5 (red squares)   &435330 &354976 & 132 &  0.52    & 2.53 &36& $-$0.007\\
Aq-C-5 (magenta asterisks) & 681143 &647325&  173  &  1.18 & 5.93 &15 & $-$0.002\\
Aq-D-5 (cyan open circles) &599438&460845 &	    171 &  1.09& 4.41&19& $-$0.008\\
Aq-G-5 (violet crosses)& 679177&778397&  143  & 0.68&   5.63&16 & $-$0.002\\
Aq-H-5 (light green filled circles)&515392&569595 &  133   & 0.53& 4.73&14& $-$0.007\\
\end{tabular} 
 \end{center}
\vspace{1mm}
\end{table*}

\section{The transition between the IHPs and OHPs}

As shown in Papers~I and II, the IHPs and OHPs of the different
simulated haloes exhibit different chemical and dynamical properties,
reflecting their different assembly histories. In these previous papers,
their physical properties were studied individually. However, observers
detect the superposed populations ``as is,'' requiring them to interpret
their findings in order to satisfactorily confront them with 
galaxy formation models. Motivated by this simple reality, we have
considered the nature of the superposed IHPs and OHPs together, and
attempt to glean information about their observed characteristics in
relation to the different halo-assembly histories.

\begin{figure}
\hspace*{-0.02cm}\resizebox{4.0cm}{!}{\includegraphics{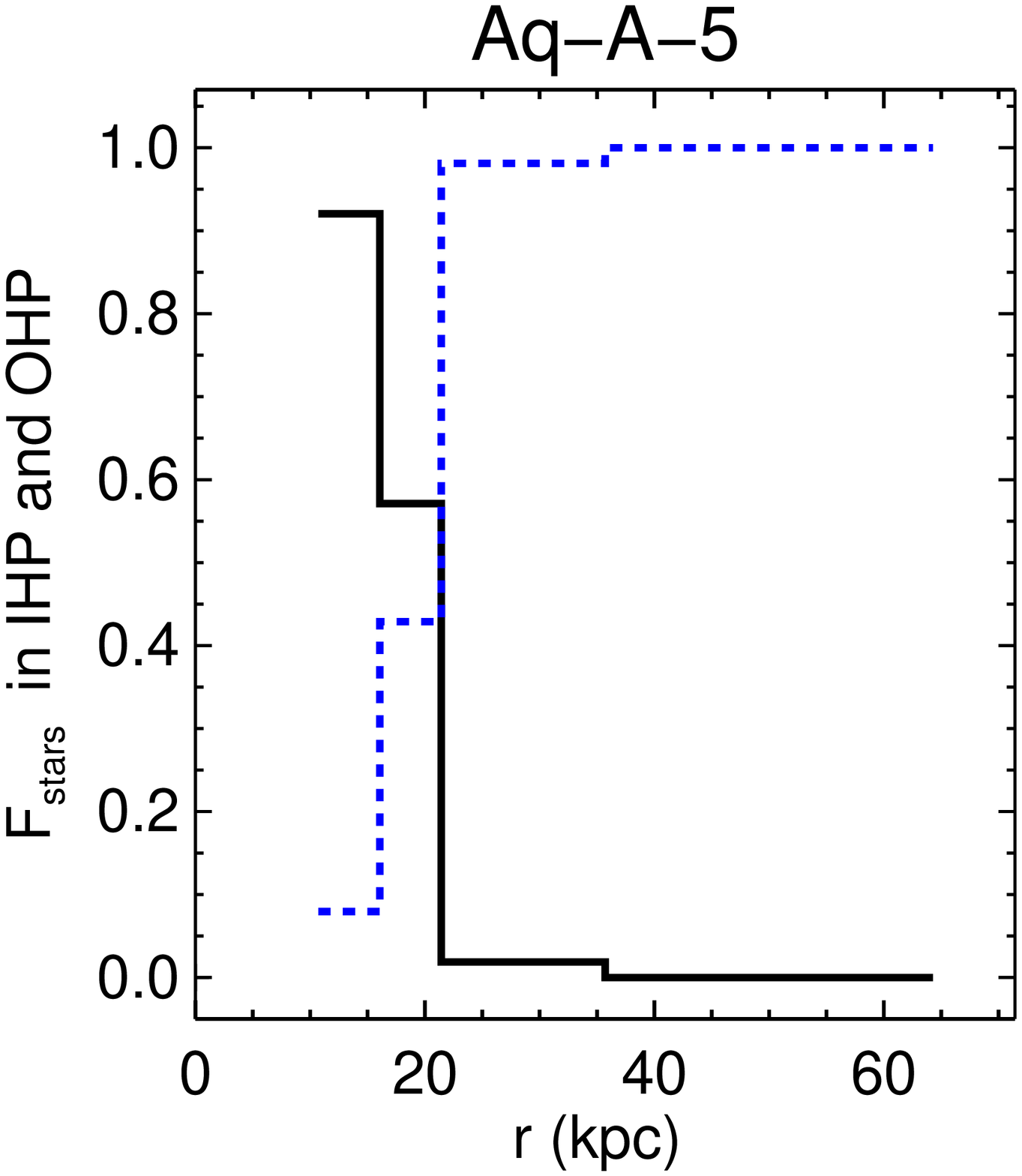}}
\hspace*{-0.02cm}\resizebox{4.0cm}{!}{\includegraphics{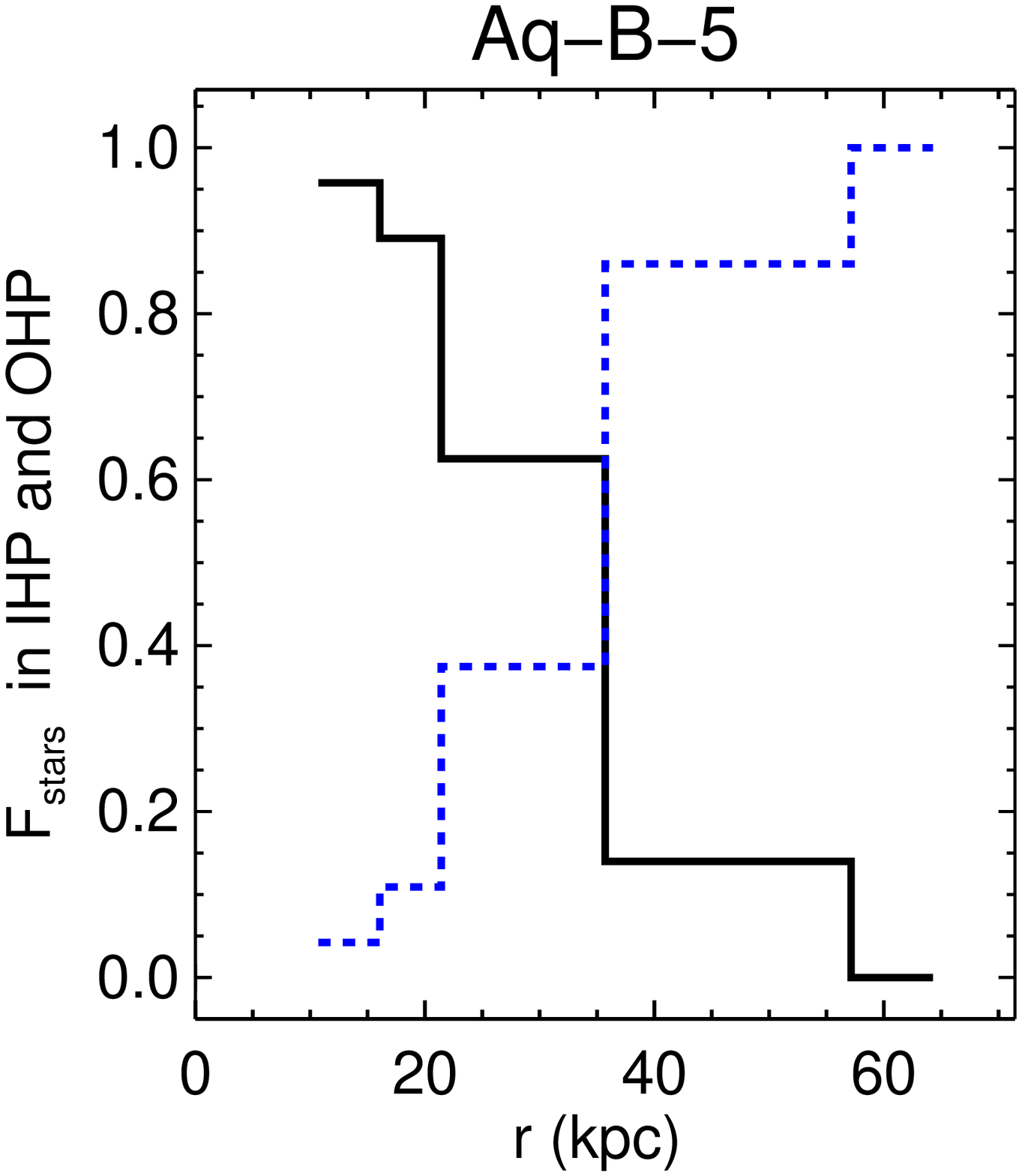}}\\
\hspace*{-0.02cm}\resizebox{4.0cm}{!}{\includegraphics{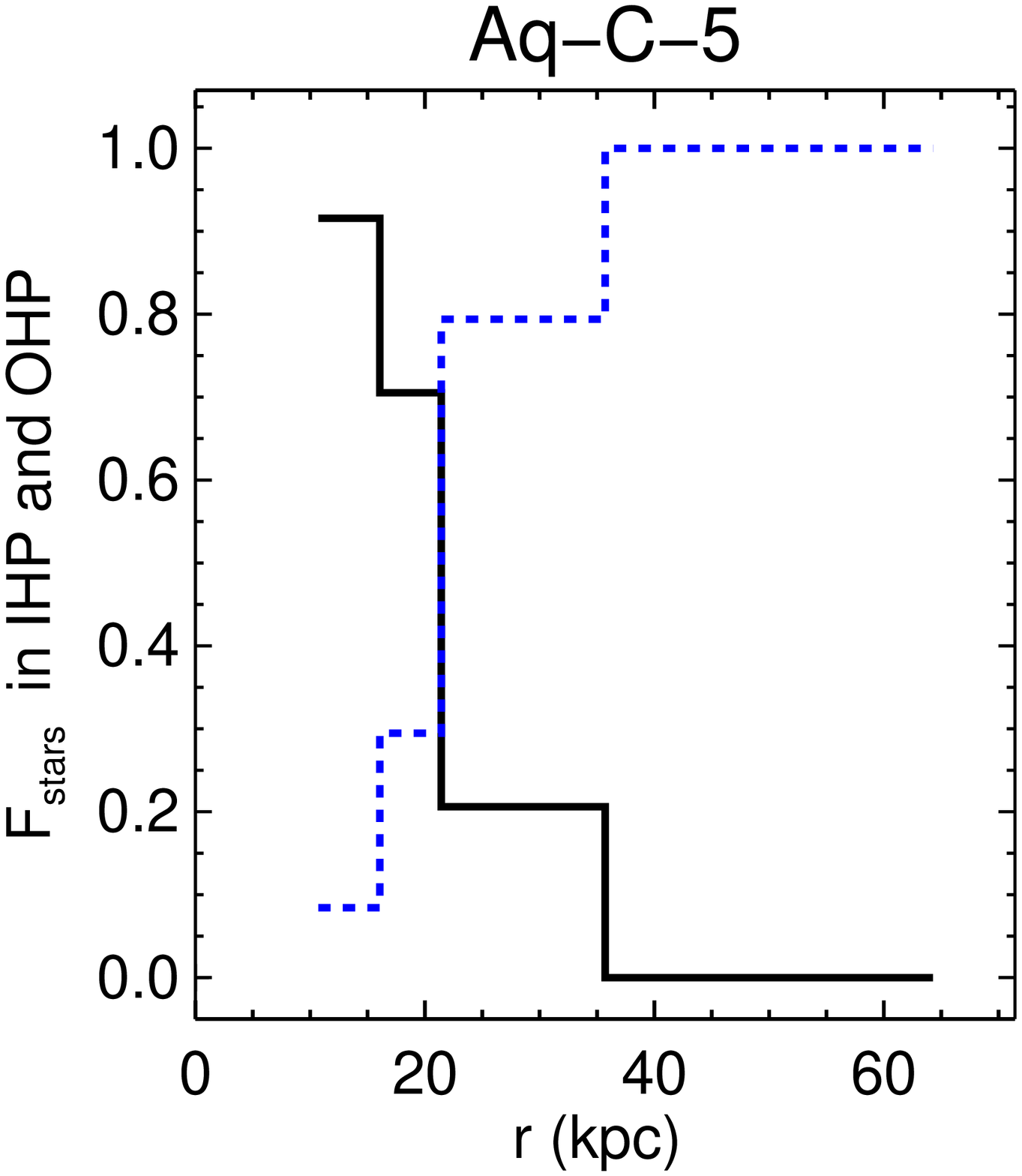}}
\hspace*{-0.02cm}\resizebox{4.0cm}{!}{\includegraphics{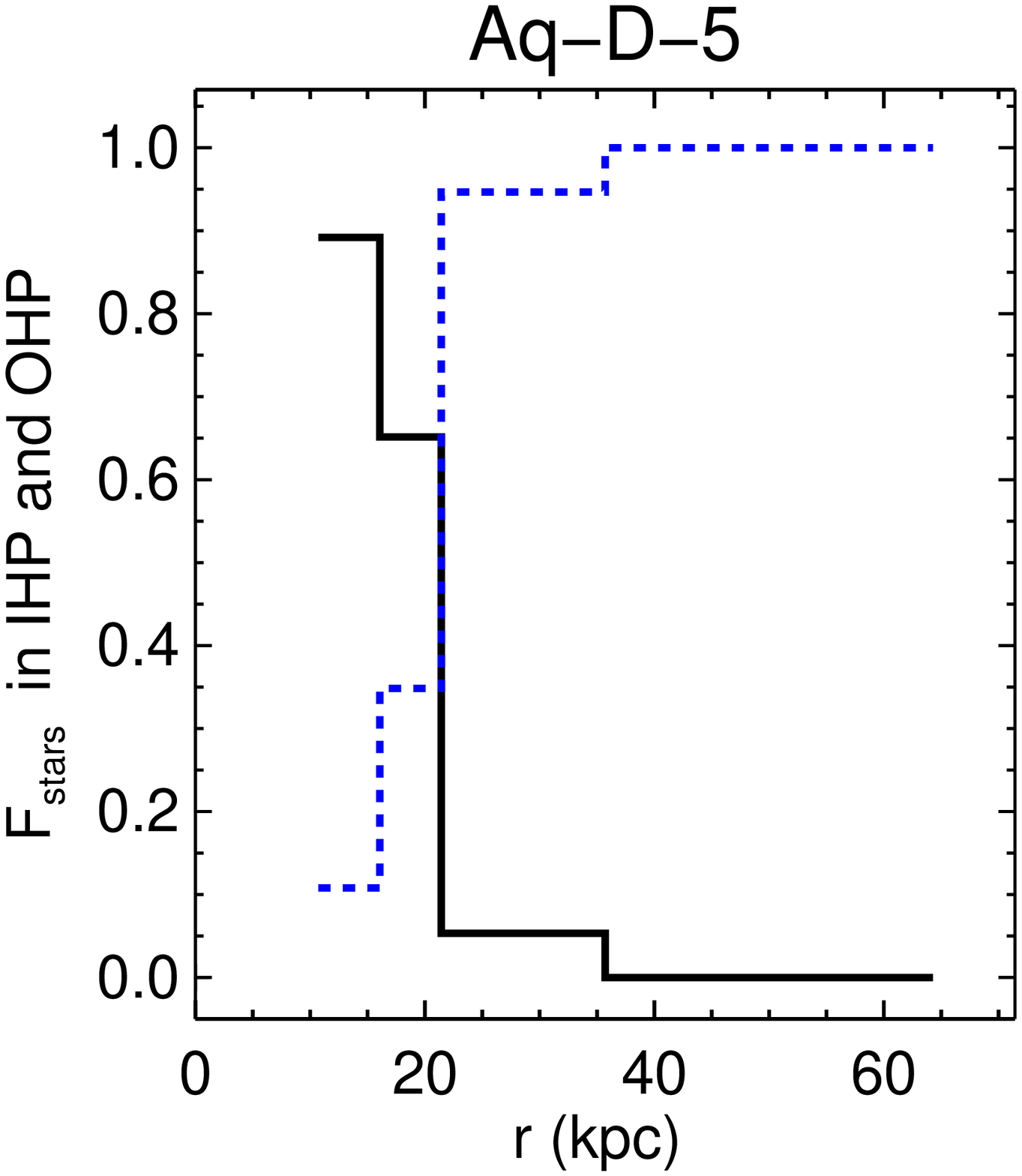}}\\
\hspace*{-0.02cm}\resizebox{4.0cm}{!}{\includegraphics{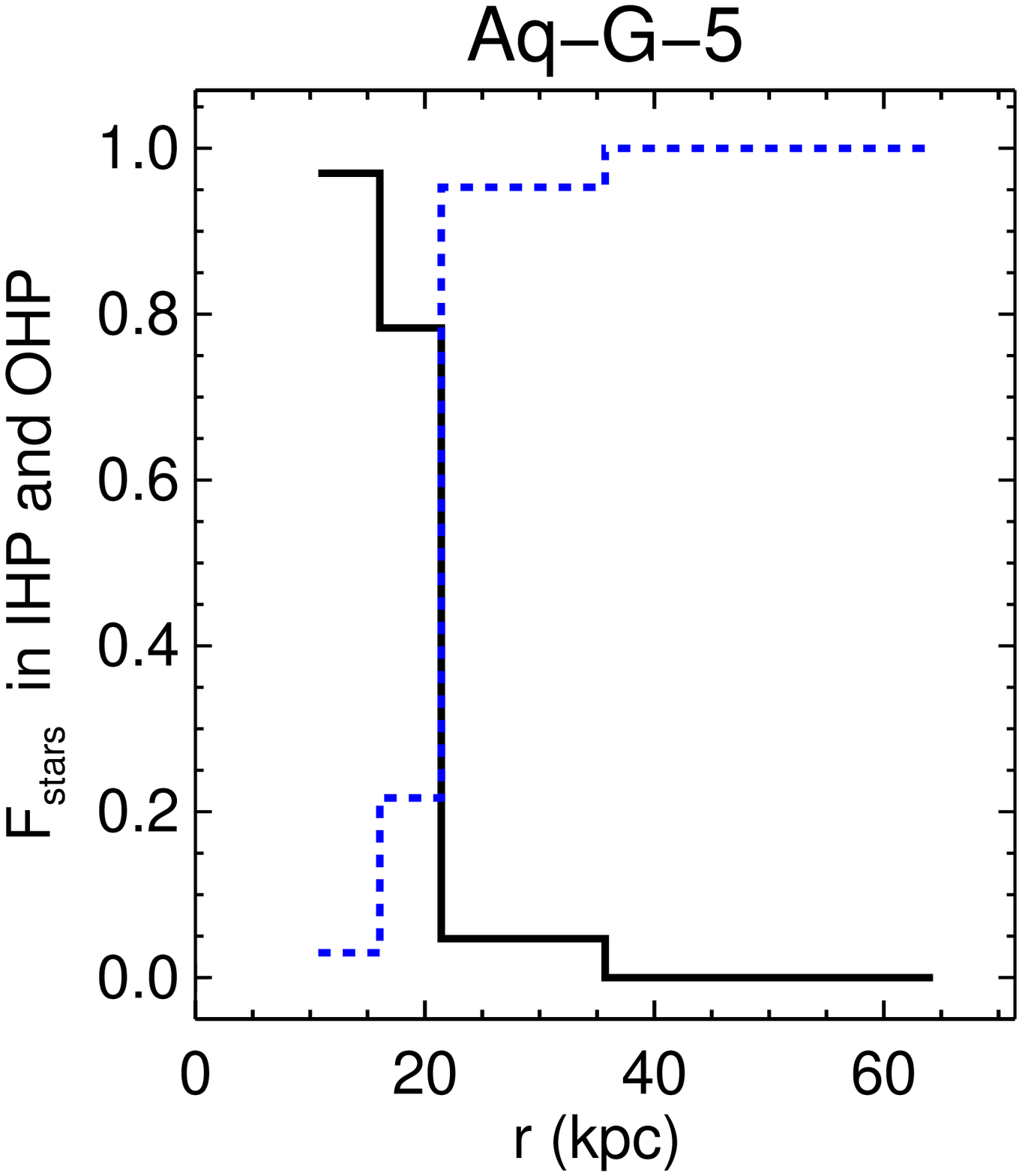}}
\hspace*{-0.02cm}\resizebox{4.0cm}{!}{\includegraphics{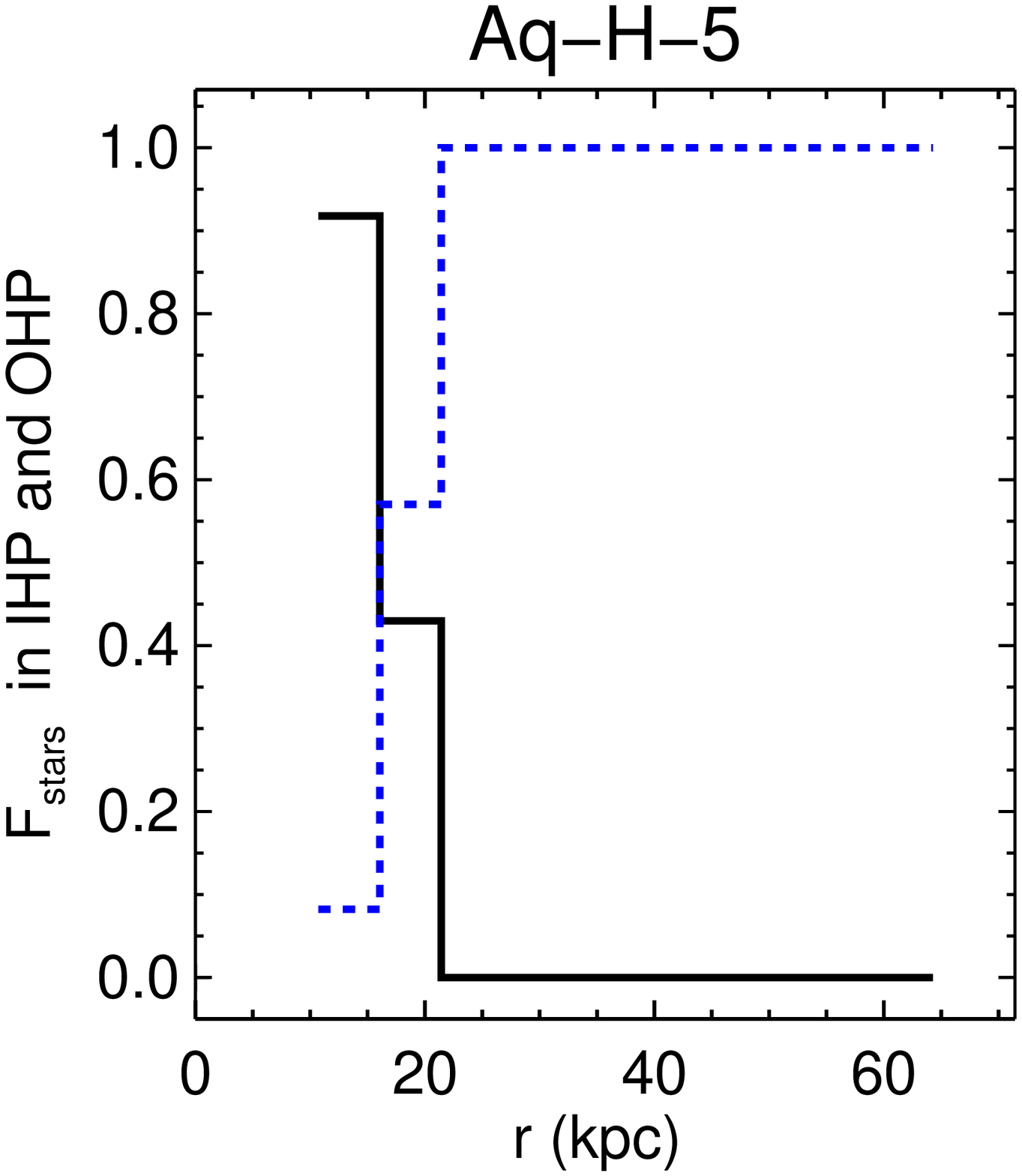}}
\caption{Stellar mass fraction in the IHPs and OHPs, as a function of
radius.  The radial distance at which the two populations change in
relative dominance is taken as the halo transition radius, $r_{\rm
HTR}$. The IHRs are dominated by the IHPs, but there exists a
contribution from the OHPs of $\sim 10-40$ per cent within $r_{\rm HTR}$,
increasing dramatically for larger radii.}
\label{frac}
\end{figure}

We analyse how the IHPs and OHPs are superposed with one another within
the  main-galaxy virial radius. Fig.~\ref{frac} shows the stellar mass
fraction of these populations in equal-radial intervals for each
simulated halo. This immediately allows for the determination of the
spatial limits  of the IHR and OHR for each halo, specified here as
$r_{\rm HTR}$. From inspection of this figure, we can clearly see that
the IHPs dominate within $r_{\rm HTR} $, while for larger radii, the
OHPs are the main contributors. We find $r_{\rm HTR} \sim 15-20 $ kpc
for five out of the six analysed haloes, similar to the inferred
transition region for the MW by \citet{caro2007,caro2010} and
\citet{deJong2010}.
 The only exception is Aq-B-5,  
which exhibits a transition between the
IHR and OHR around $r_{\rm HTR} \sim 36$ kpc, similar to that found for
M31 by \citet{gilbert2013}. In the case of the simulation, this is due to the important contribution of 
accreted stars  to  the IHP in comparison to the \insitu components.
This halo is the only one where the accreted stars dominate over the \insitu stars in the IHP, as can
be seen from fig. 3 in Paper II (note that the \insitu stars tend to be more concentrated and have lower
binding energies). 
The contribution of the OHPs within $\sim
10$ kpc varies between $\sim 10$ per cent and $ \sim 40 $ per cent,
while the IHPs can extend farther into the OHRs, with contributions
smaller than $\sim20$ per cent in our simulated systems. As shown in
Papers~I and II, the OHPs are mainly formed from debris stars, while the
IHPs received significant contributions from \insitu (disc-heated and
endo-debris stars) as well as from debris stars. Nevertheless, the
fraction of debris stars within $5-20$ kpc varies from $\sim 30$ per cent to
$\sim 60$ per cent of the combined stellar haloes (i.e., the IHPs + OHPs
considered together). Within the IHRs, the OHPs and IHPs are superposed,
and present a challenge for interpretation of the individual
subpopulations if a hierarchical clustering scenario applies. The
transition between the IHPs and OHPs is expected to produce a transition
in the observed MDFs as one progresses from the IHRs to the OHRs.

\subsection{The diffuse stellar halo MDF}

The previous analysis suggests that the expected MDFs in the diffuse
IHRs of MW-mass galaxy haloes are not the same as the MDFs in the
diffuse OHRs of the haloes. According to our models, as one moves
outwards, the MDF should change in nature (and shape), decreasing in
average metallicity as the different populations contribute in different
proportion. How these changes are produced, and whether they store
information on the assembly history of a given galaxy, are the questions
on which we focus in this section.

\begin{figure*}
\hspace*{-0.2cm}\resizebox{3.5cm}{!}{\includegraphics{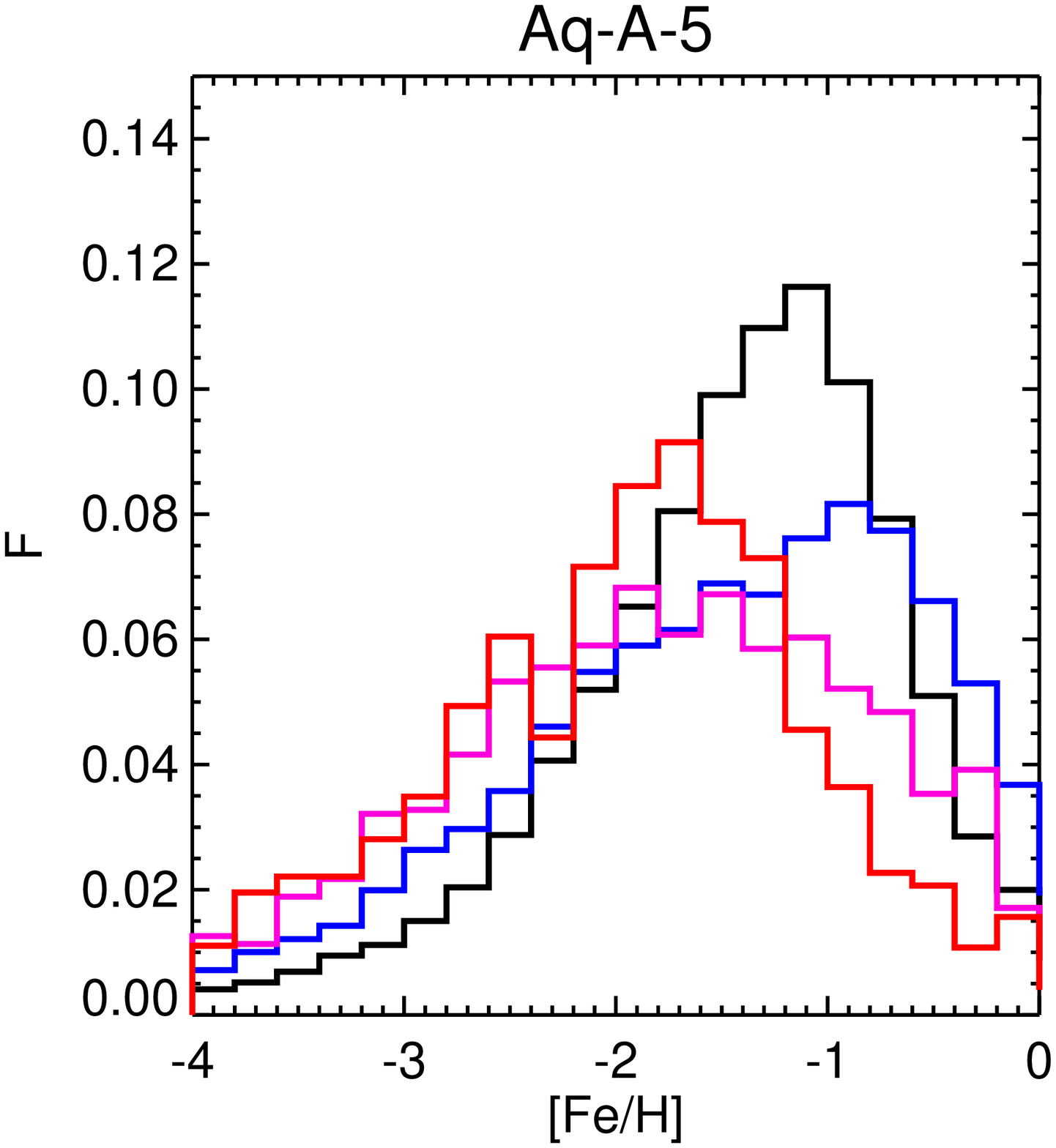}}
\hspace*{-0.2cm}\resizebox{3.5cm}{!}{\includegraphics{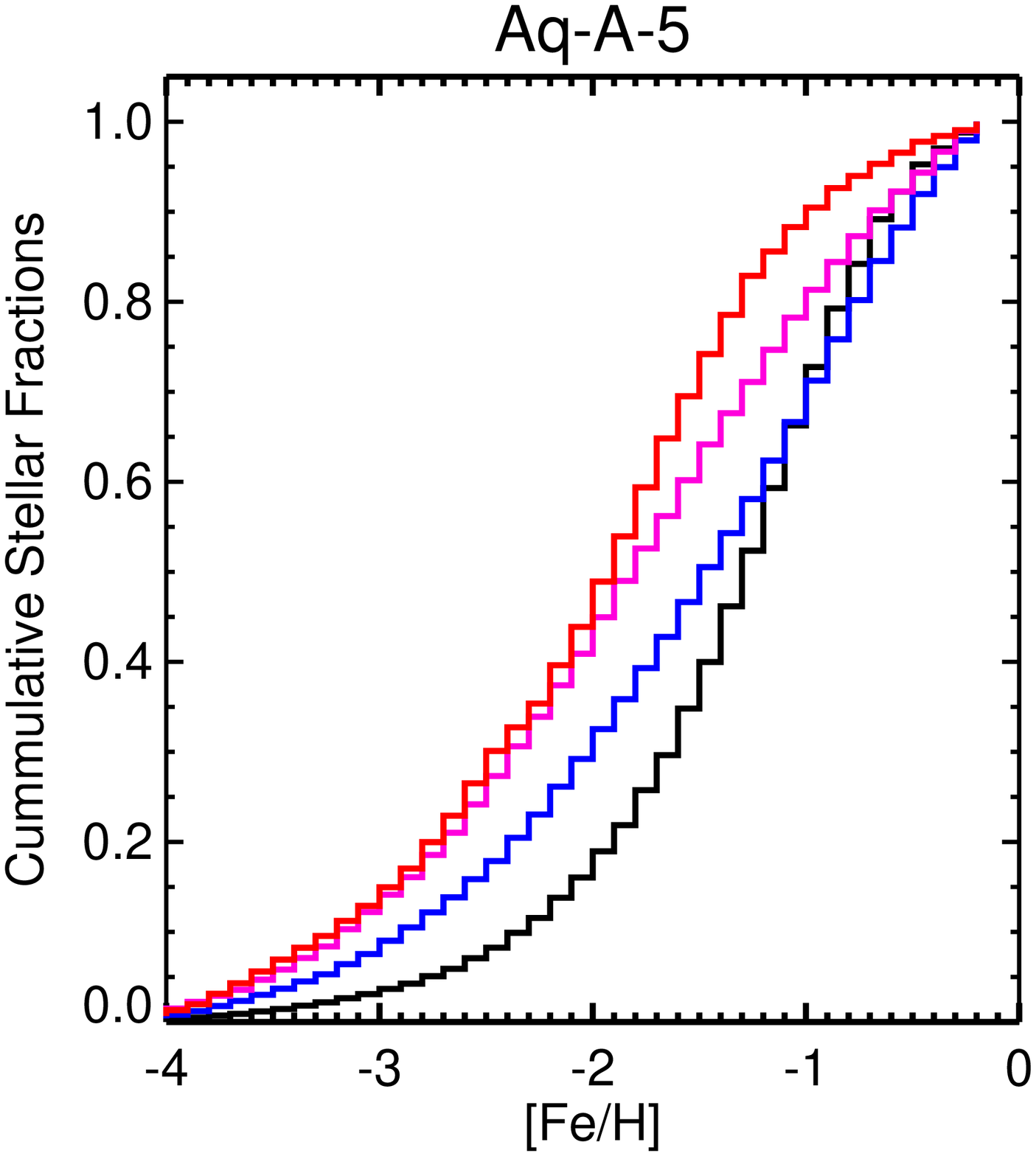}}
\hspace*{-0.2cm}\resizebox{3.5cm}{!}{\includegraphics{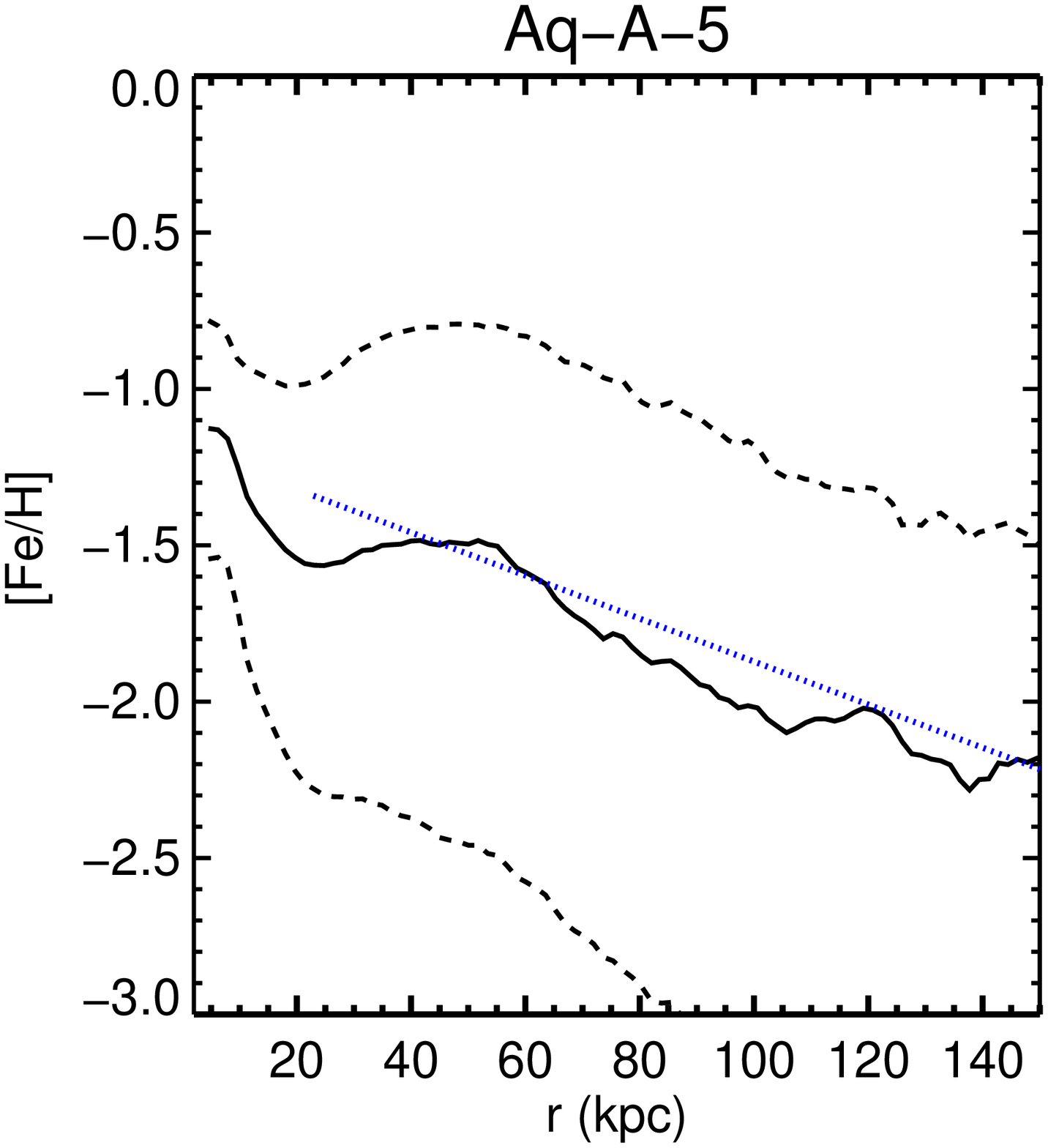}}\\
\hspace*{-0.2cm}\resizebox{3.7cm}{!}{\includegraphics{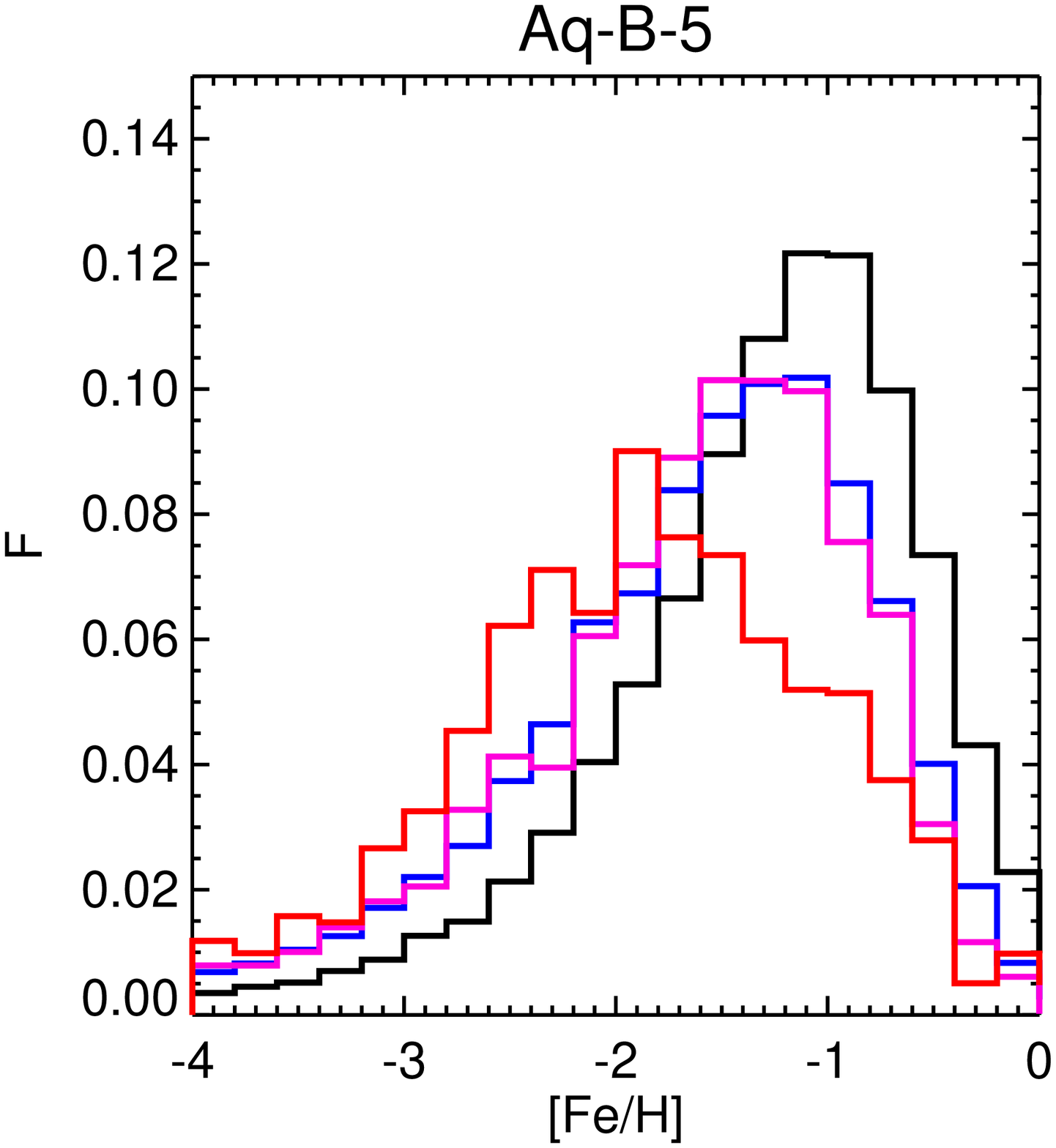}}
\hspace*{-0.2cm}\resizebox{3.5cm}{!}{\includegraphics{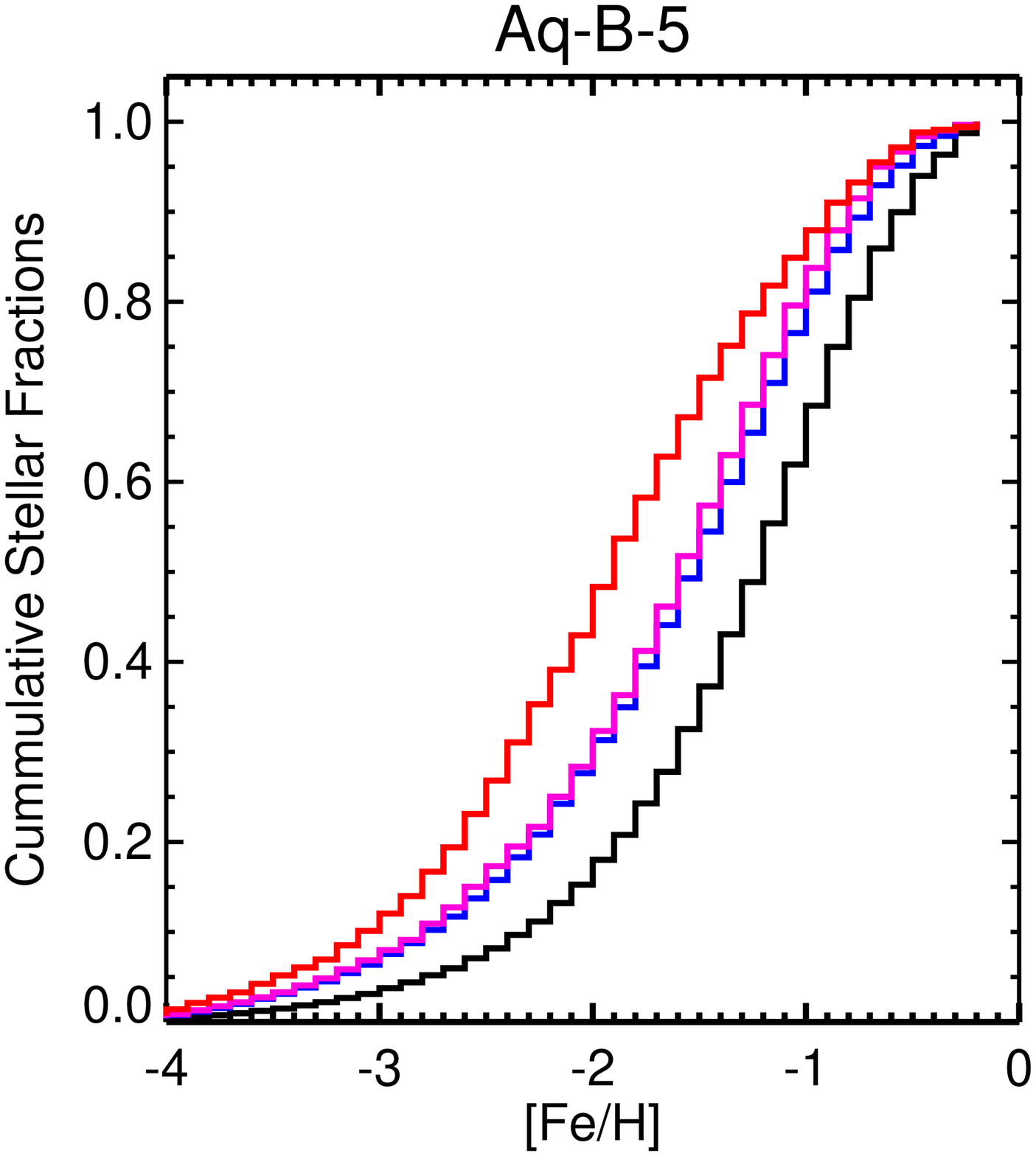}}
\hspace*{-0.2cm}\resizebox{3.5cm}{!}{\includegraphics{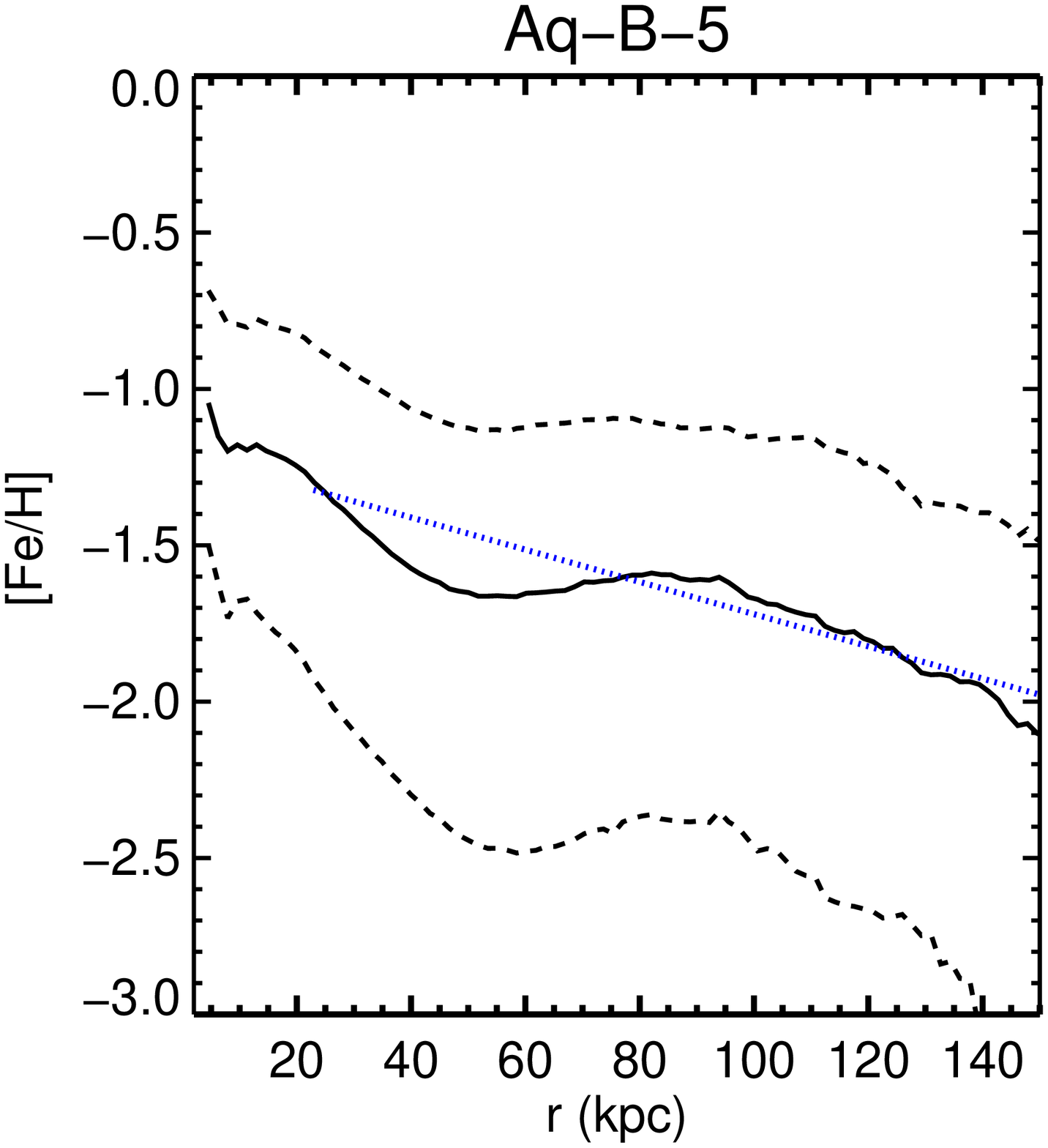}}\\
\hspace*{-0.2cm}\resizebox{3.5cm}{!}{\includegraphics{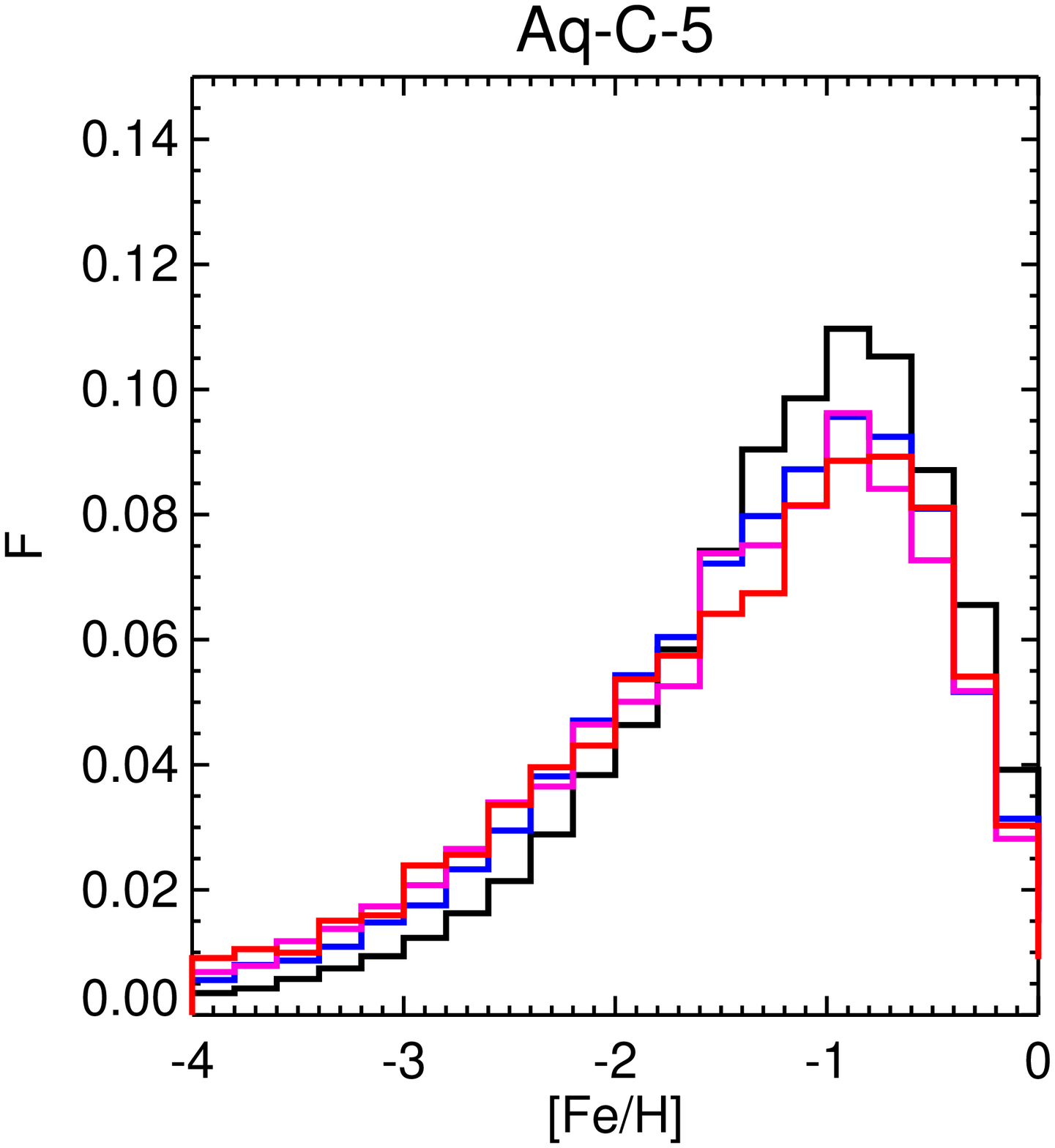}}
\hspace*{-0.2cm}\resizebox{3.5cm}{!}{\includegraphics{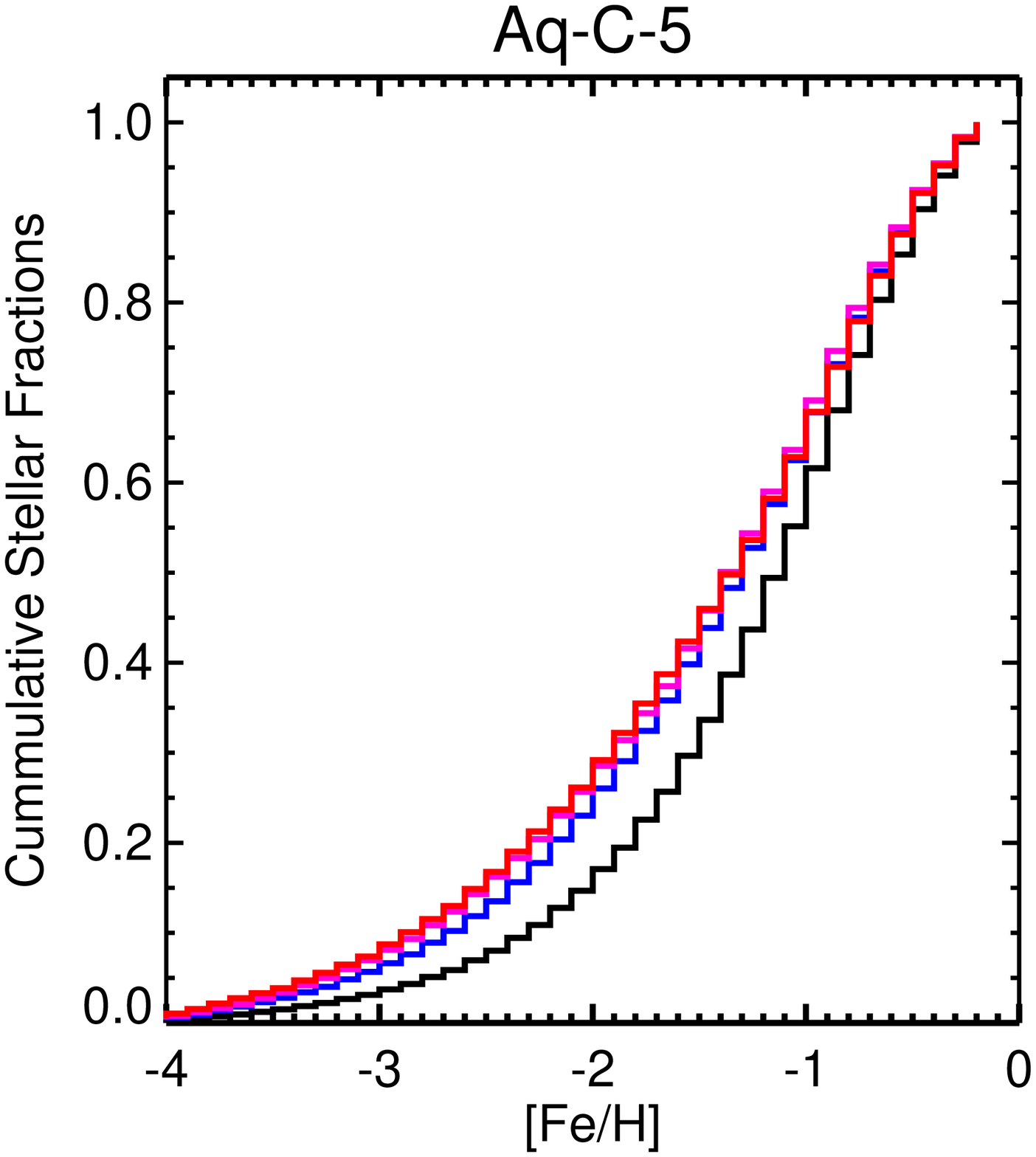}}
\hspace*{-0.2cm}\resizebox{3.5cm}{!}{\includegraphics{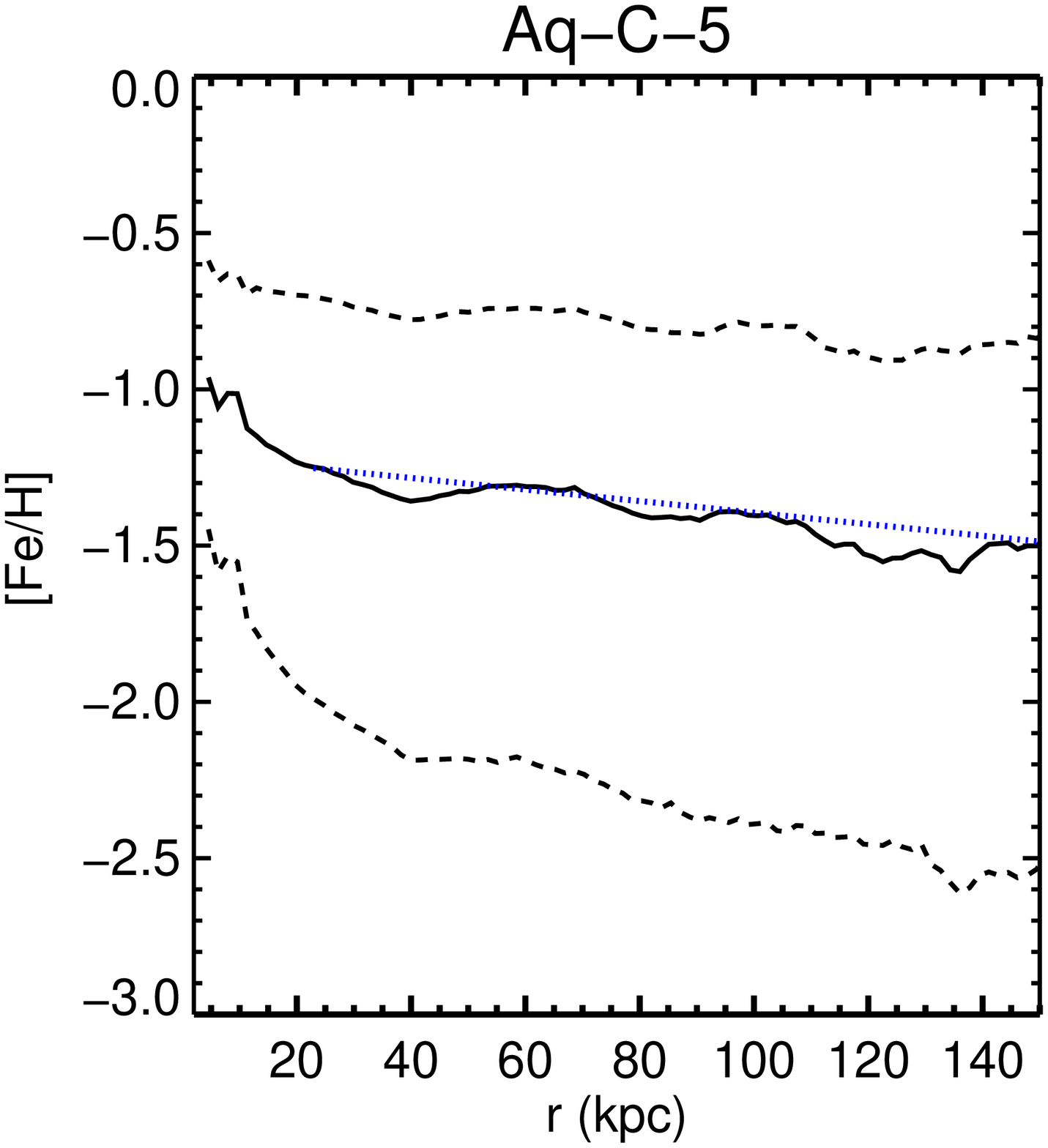}}\\
\hspace*{-0.2cm}\resizebox{3.5cm}{!}{\includegraphics{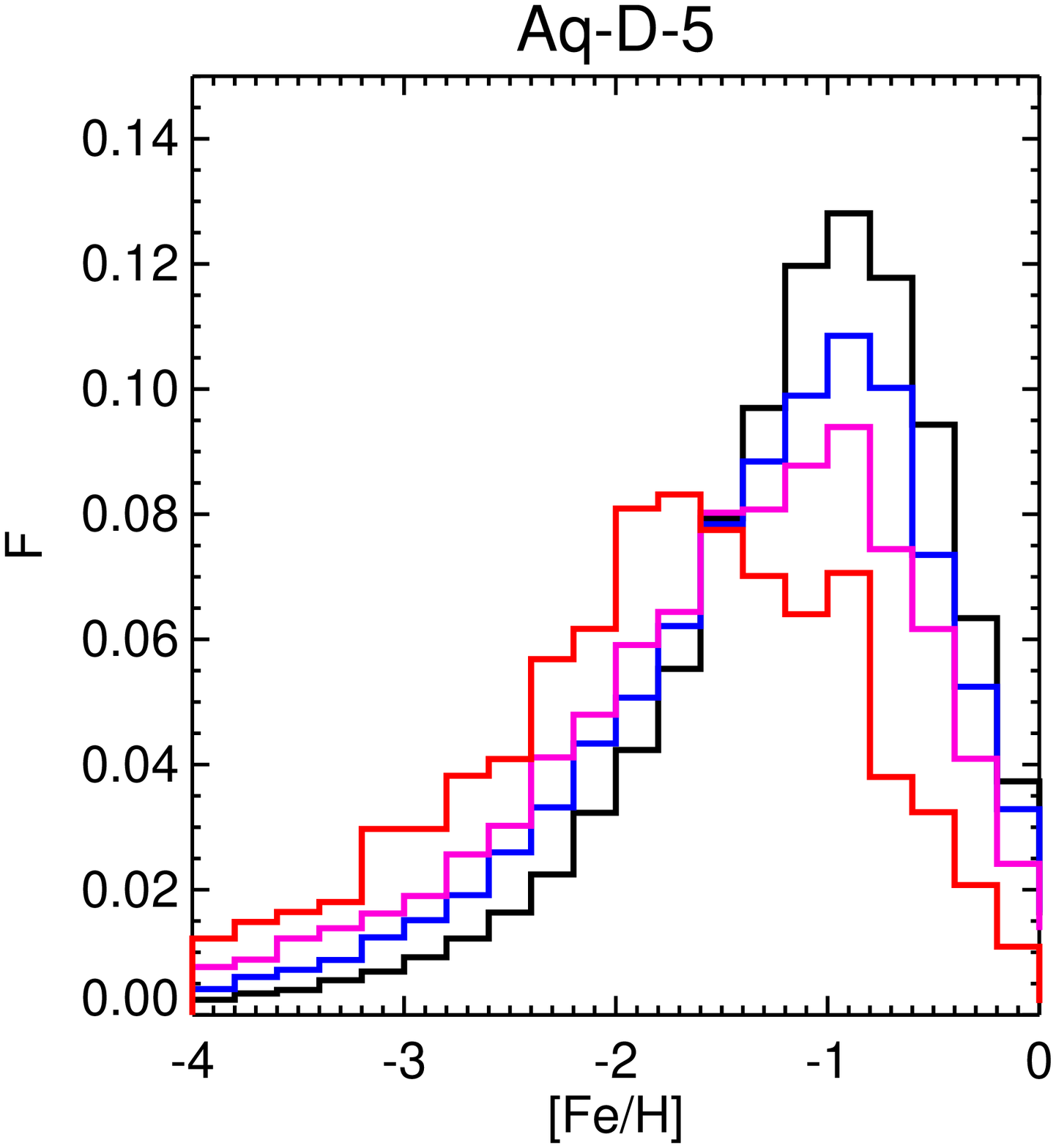}}
\hspace*{-0.2cm}\resizebox{3.5cm}{!}{\includegraphics{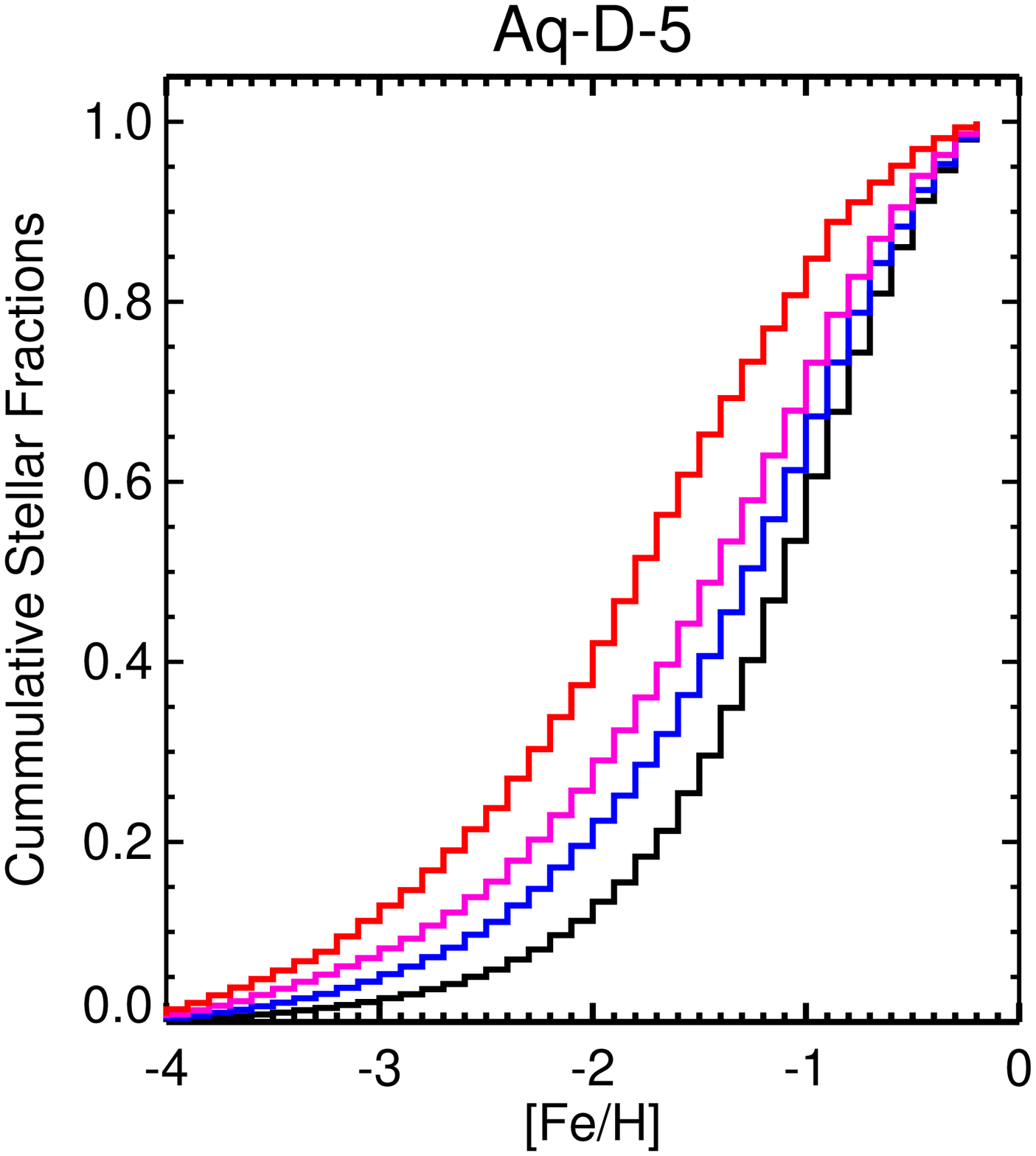}}
\hspace*{-0.2cm}\resizebox{3.5cm}{!}{\includegraphics{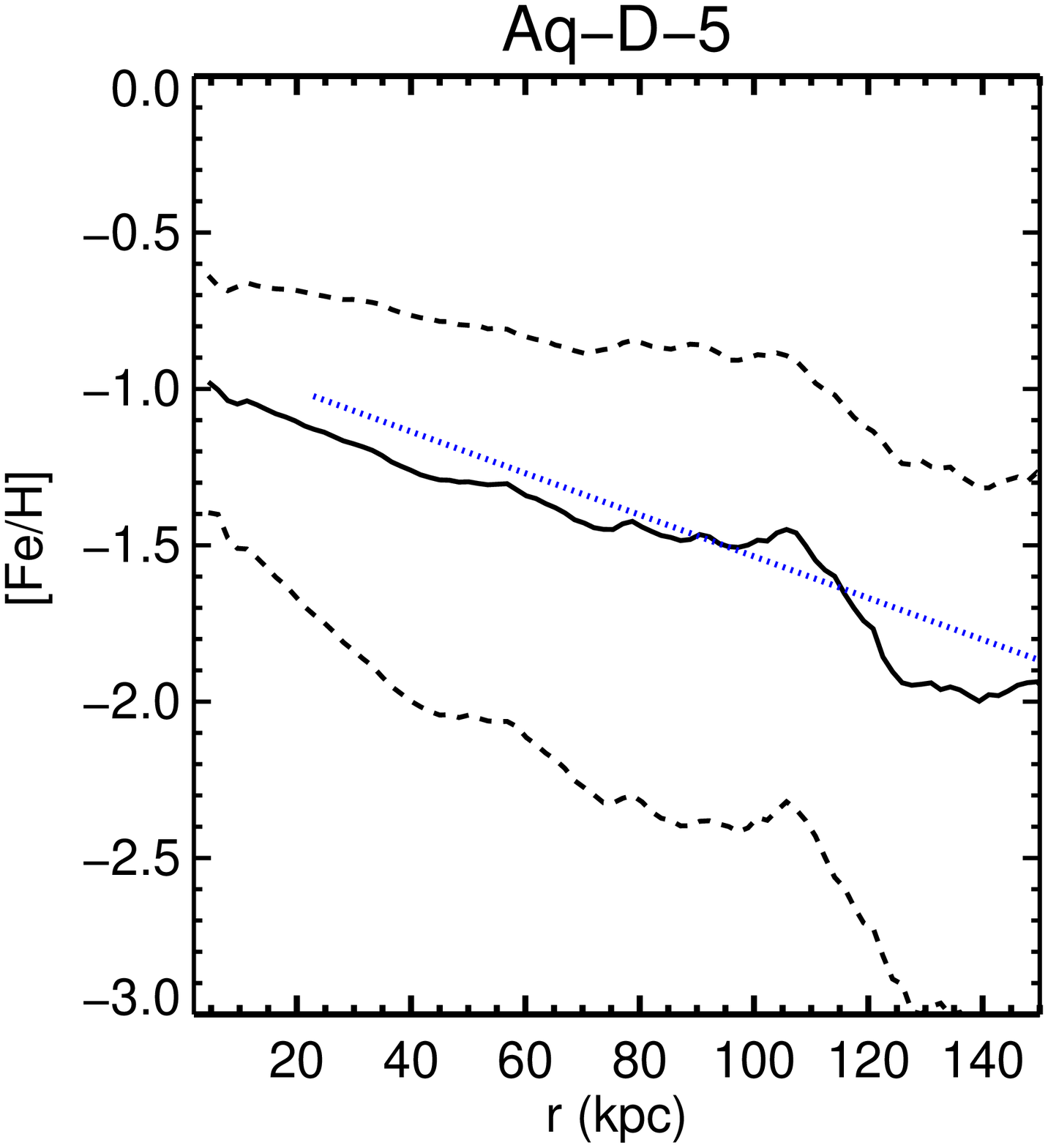}}\\
\hspace*{-0.2cm}\resizebox{3.5cm}{!}{\includegraphics{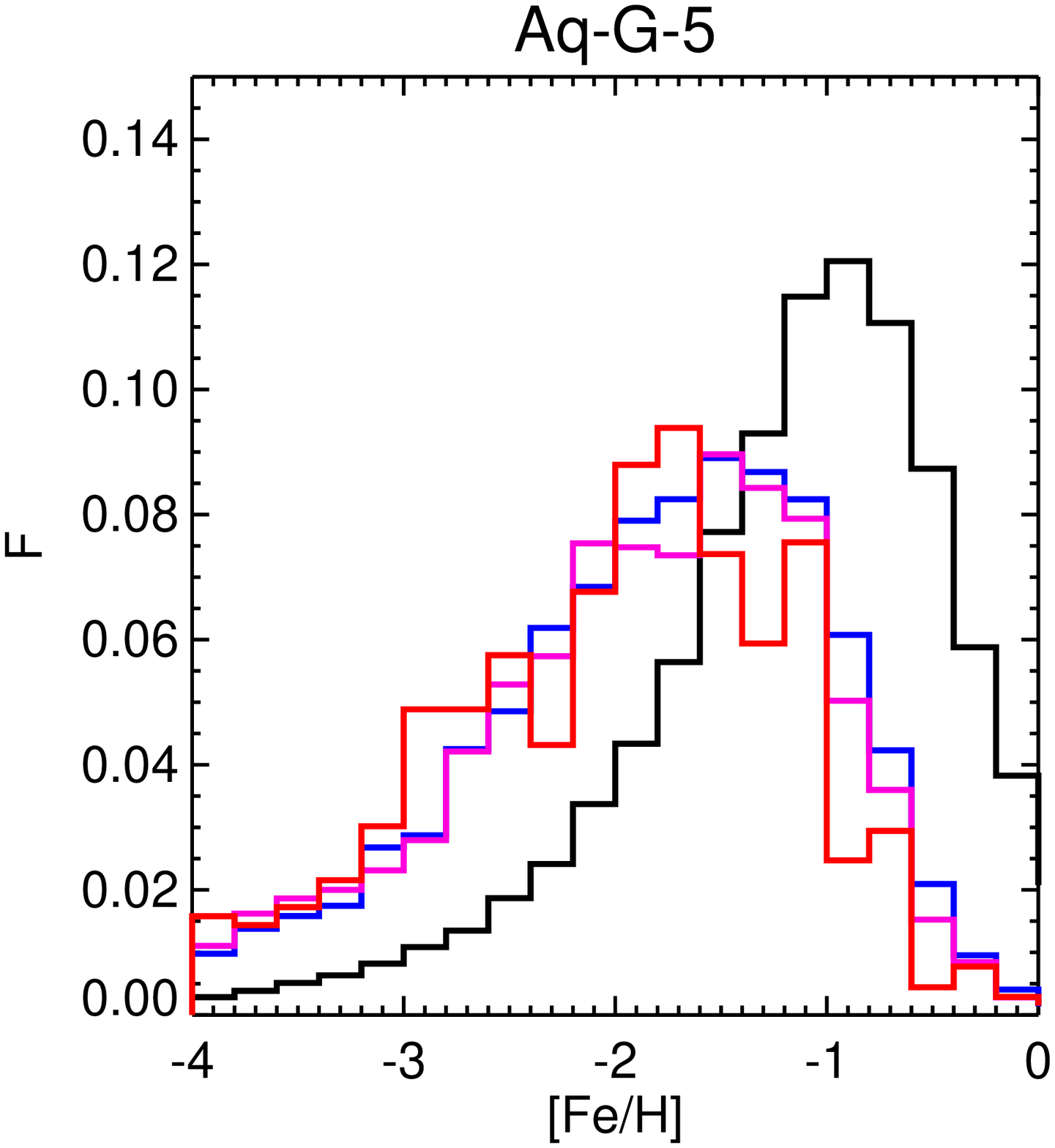}}
\hspace*{-0.2cm}\resizebox{3.5cm}{!}{\includegraphics{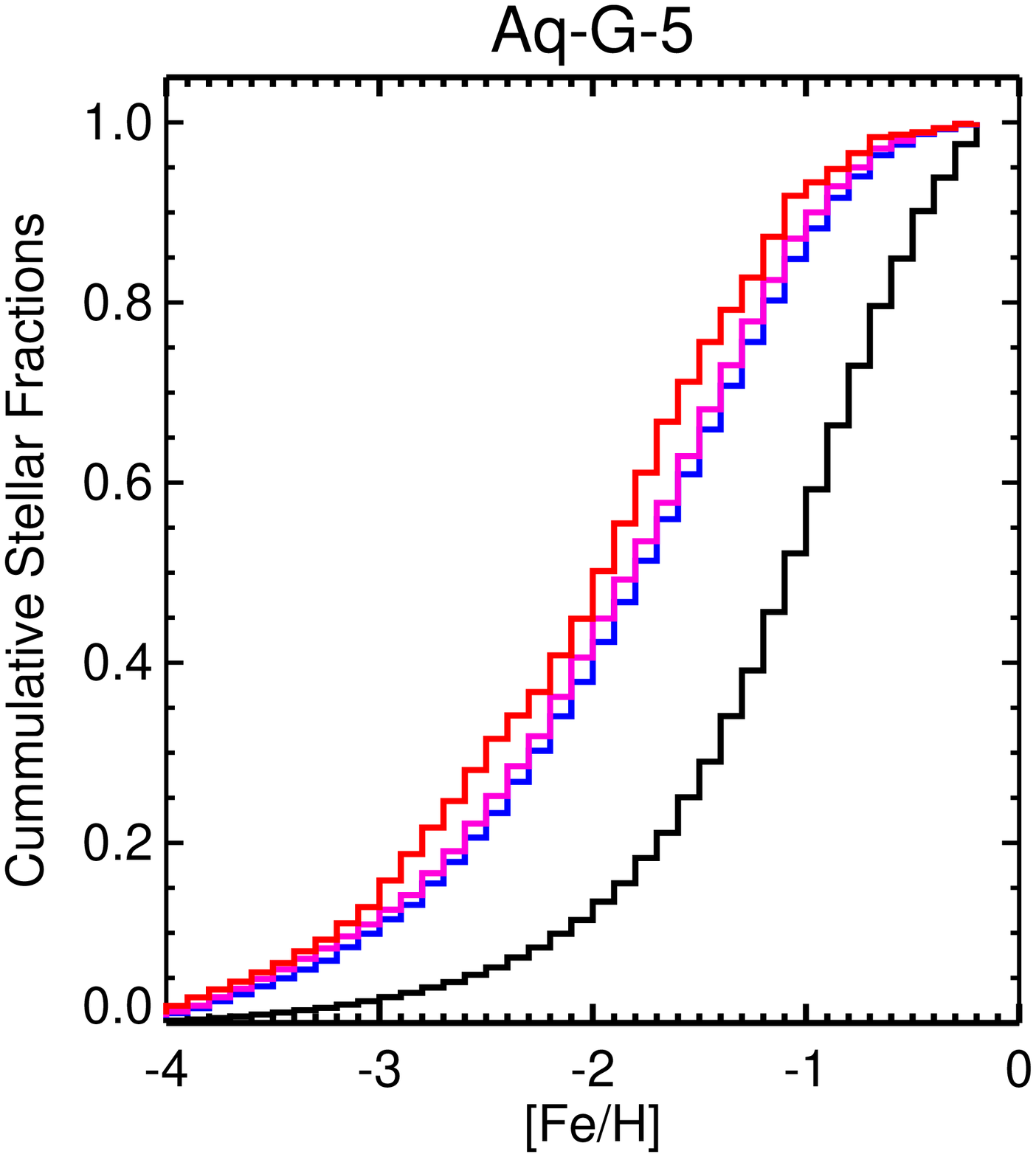}}
\hspace*{-0.2cm}\resizebox{3.5cm}{!}{\includegraphics{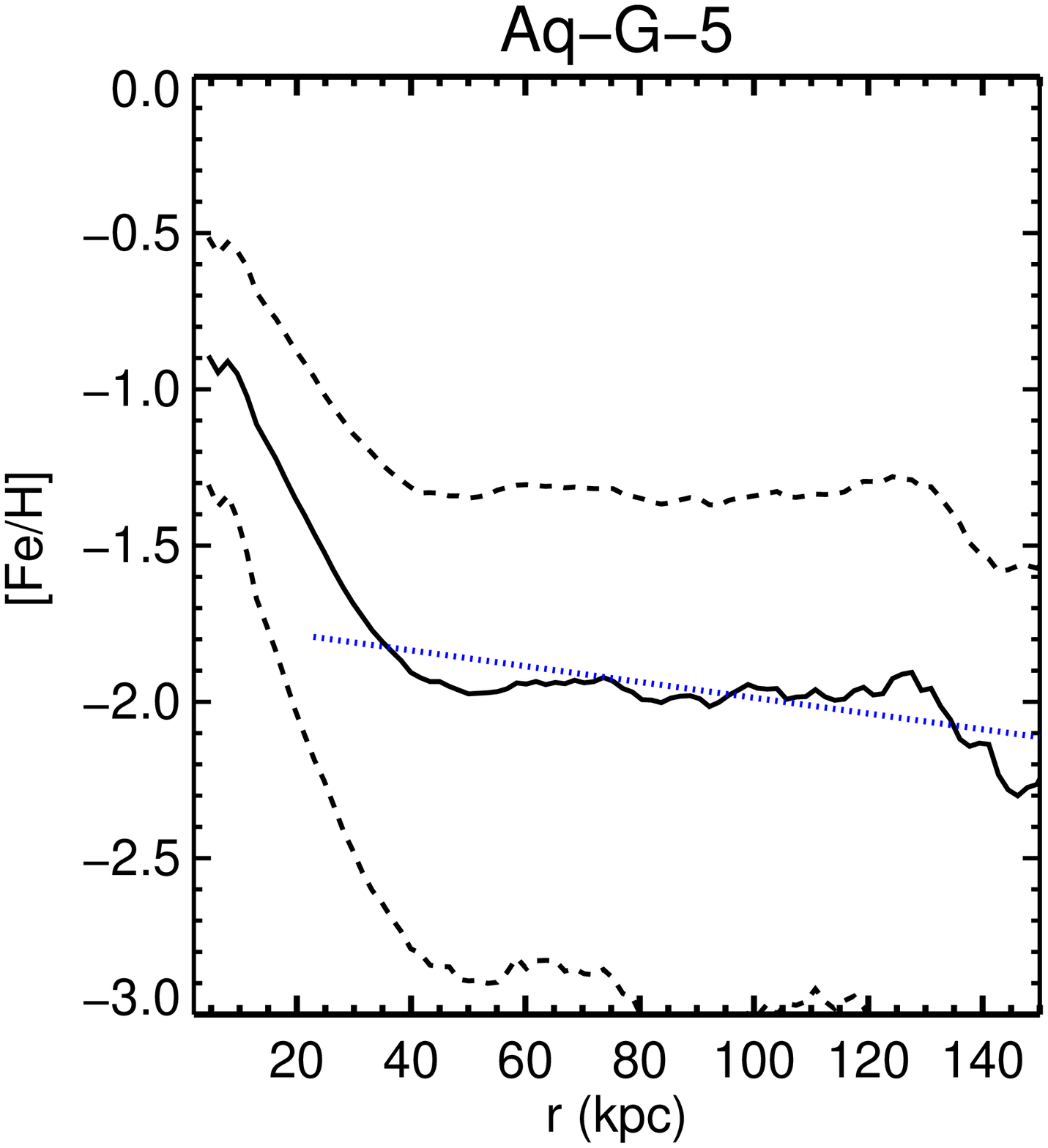}}\\
\hspace*{-0.2cm}\resizebox{3.5cm}{!}{\includegraphics{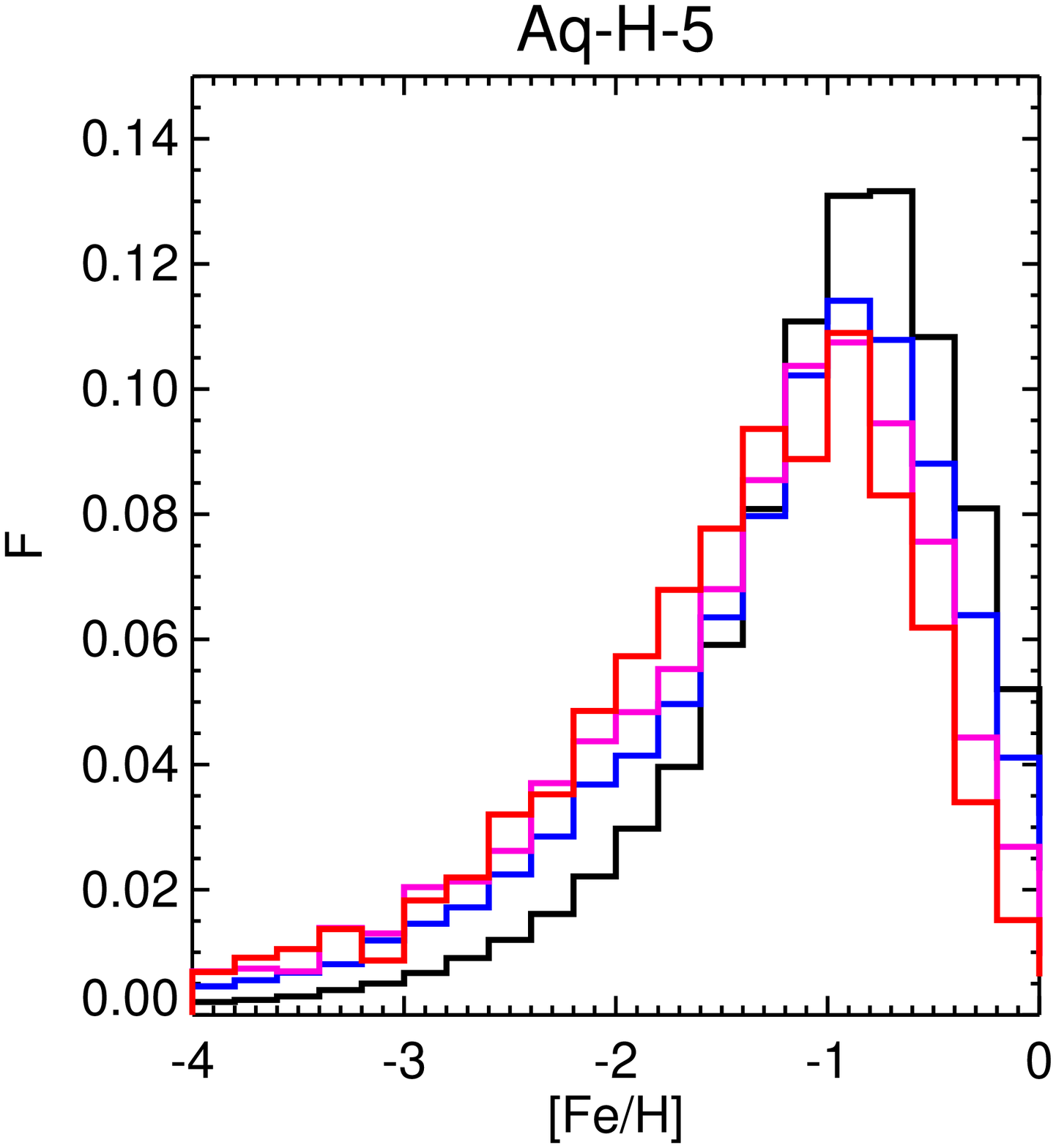}}
\hspace*{-0.2cm}\resizebox{3.5cm}{!}{\includegraphics{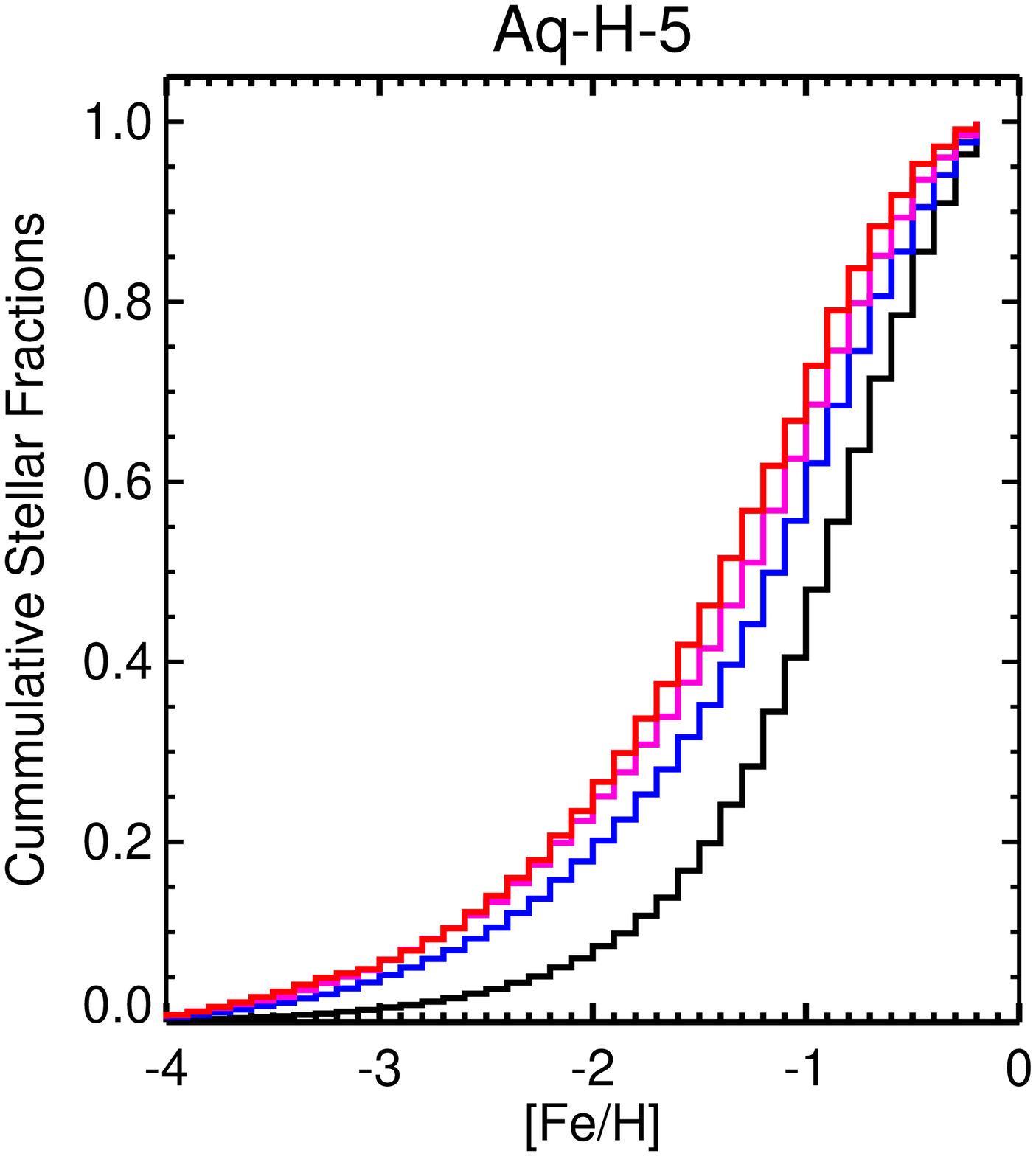}}
\hspace*{-0.2cm}\resizebox{3.5cm}{!}{\includegraphics{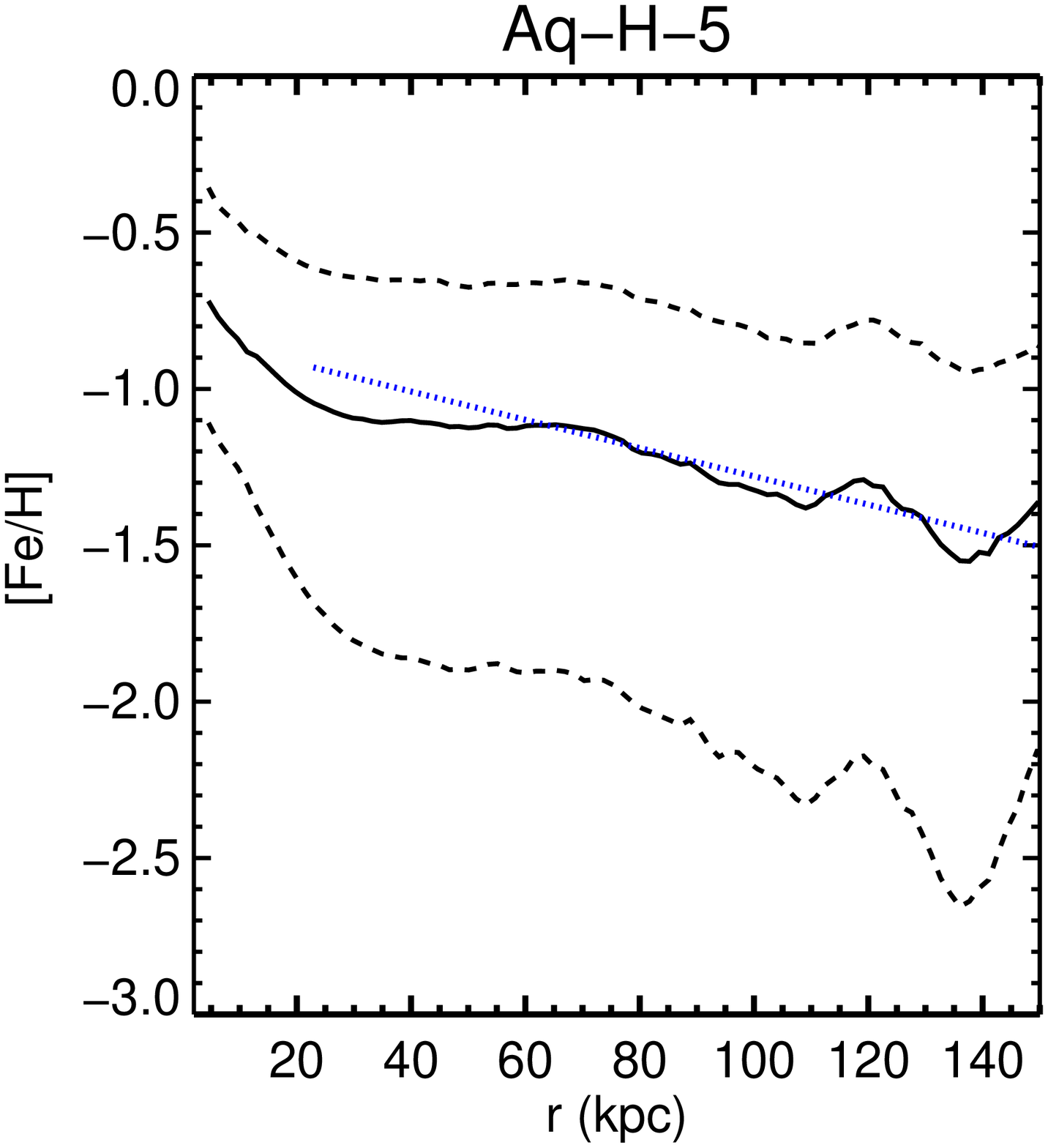}}
\caption{The predicted stellar MDFs (left-hand panels),
the cumulative stellar
mass (middle panels), and the median
metallicity profiles (right-hand panels), as a function of
galactocentric distance, for the entire stellar haloes
(i.e., combined IHPs and OHPs), estimated in four concentric radial bins
at: $<20$ kpc (black lines), $20-40$ kpc (blue lines), $60-80$ kpc
(magenta lines) and $80-120$ kpc (red lines). The blue line in the
right-hand panels show the best fit to the median metallicity profiles
for stars in the OHRs of each simulation; the dashed lines represent
the second and fourth quartiles.}
\label{mdftotal}
\end{figure*}

Fig.~\ref{mdftotal} shows the predicted MDFs  for stars in the
entire stellar haloes, estimated in concentric shells of $ 20$ kpc
width (left-hand panels). As the distance to each shell increases, the
MDFs tend to shift to lower metallicity, although each of simulated
stellar haloes behaves differently in detail. Some haloes exhibit large
displacements toward lower [Fe/H], while others show only small changes.
In all cases, the IHRs exhibit the highest level of chemical enrichment.
In particular, it is interesting to note that the simulated halo of
Aq-C-5 exhibits negligible abundance variations, even though this halo,
just like the others, was formed in a hierarchical clustering scenario.
Clearly, a wide range of behaviour can be expected from halo system to
halo system.

The different behaviours in the simulated MDFs as a function of distance
can be understood from inspection of the cumulative mass fractions of
stars as a function of [Fe/H], for different distances, shown in the
middle panels of Fig. ~\ref{mdftotal}. Although the simulated stellar
haloes share some features in common, they also exhibit a variety of
differences in the MDFs from shell to shell. While some systems exhibit
similar fractions of very metal-poor (VMP; [Fe/H] $< -2$) stars at all
radii (e.g., Aq-C-5), others exhibit a tripling of this fraction from
the inner shells to the outer shells (e.g., Aq-B-5). The change in the
curvature of the cumulative stellar mass fractions reflects the relative
contribution of low- and high-metallicity stars as one moves outwards.
In general, in the IHRs, the stellar mass fraction of stars with
[Fe/H]~$> -1$ are between $\sim 40-60$ per cent, 
while in the OHRs they
represent less than $\sim 20$ per cent of the total stellar mass. VMP stars
can represent up to $\sim 60$ per cent of the total stellar mass in the OHRs
of the simulated haloes. 

Another view of the variation in the predicted MDFs with distance is
provided by the mean [Fe/H] profiles, shown in the right-hand panels of
Fig. ~\ref{mdftotal}. The nature of the IHPs determine the slope within
the IHRs, but have little effect on the characteristics of the MDF in
the OHRs, as we have already seen. Regarding the central regions, where
the IHPs dominate, the profiles are strongly affected by the presence of
disc-heated and endo-debris stars. In our simulated stellar haloes, the
OHPs can contribute up to $\sim 40$ per cent of the total stellar mass within
the IHRs, as can be seen from Fig. ~\ref{frac}.  Disc-heated and
endo-debris stars tend to have lower binding energy, and to be more
concentrated in the central regions. As shown in fig. 1 of
Paper~I, they have a strong influence on the level of enrichment, while
both the relative distribution of accreted stars and \insitu stars
determines the final metallicity profiles within $r_{\rm HTR}$. 

To better quantify the different behaviours seen in the
metallicity distribution profiles, we performed linear regressions, of
the form [Fe/H]$=a * r + b$, from $r_{\rm HTR}$ to 150 kpc, in order to
avoid the noise associated with the low number density of stars at the
outer limit of the haloes. Excluding the IHRs also makes it easier to
estimate a linear regression since, for some haloes, there is a very
steep increase of the metallicity in these regions, which precludes an
adequate fit with a single linear relation. Two stellar haloes (Aq-C-5
and Aq-G-5) exhibit quite flat slopes in their OHRs ($-0.002$ dex
kpc$^{-1}$), while the rest have larger slopes, up to $\sim -0.01$ dex
kpc$^{-1}$ (see Table \ref{tab1}). We note that, outside the $r_{HTR}$,
the dominant populations are the OHPs, formed mainly by debris stars (a
small fraction of endo-debris stars also contribute in this region).

The masses of the accreted subgalactic systems, together with their gas
richness, their baryonic physics (which regulate their star-formation
histories), and how the disruption of the satellites occurred within the
dark matter haloes, are important mechanisms that shape the MDFs in the
OHRs; these are naturally taken into account in our simulations. Indeed,
\citet{cooper2010} and \citet{gomez2012} used analytical models grafted
onto the dark-matter only simulations of the same haloes (albeit with
higher numerical resolution than we explore here), and found very weak
or flat slopes. In the case of \cite{cooper2010}, the authors pointed
out that those haloes which experienced massive accretions exhibited
steeper profiles. This is consistent with our findings, although we
obtain larger (negative) gradients in most of our stellar haloes, which
can be understood when the different assembly histories are considered,
along with a consistent treatment of the microphysics of the baryons. 

The manner in which the subgalactic systems are disrupted and
distributed into the haloes is important as well, since the more-massive
subgalactic systems will be able to survive farther into the main
galaxy, at least in general. Although dynamical friction also depends on
their orbital parameters, and hence can modulate the effects of their
masses, from our analysis it is clear that high-metallicity stars are
more gravitationally bound than low-metallicity stars, as was shown in
Paper~II by consideration of the distribution of binding energy as a
function of metallicity. We thus seek to explore in more detail how
subgalactic systems of different masses contributed to the formation of
each of the simulated stellar haloes, and how to link them to the
characteristics of the MDFs as a function of radius.

\section{Accreted subgalactic systems}

\begin{figure*}
\hspace*{-0.2cm}\resizebox{6.5cm}{!}{\includegraphics{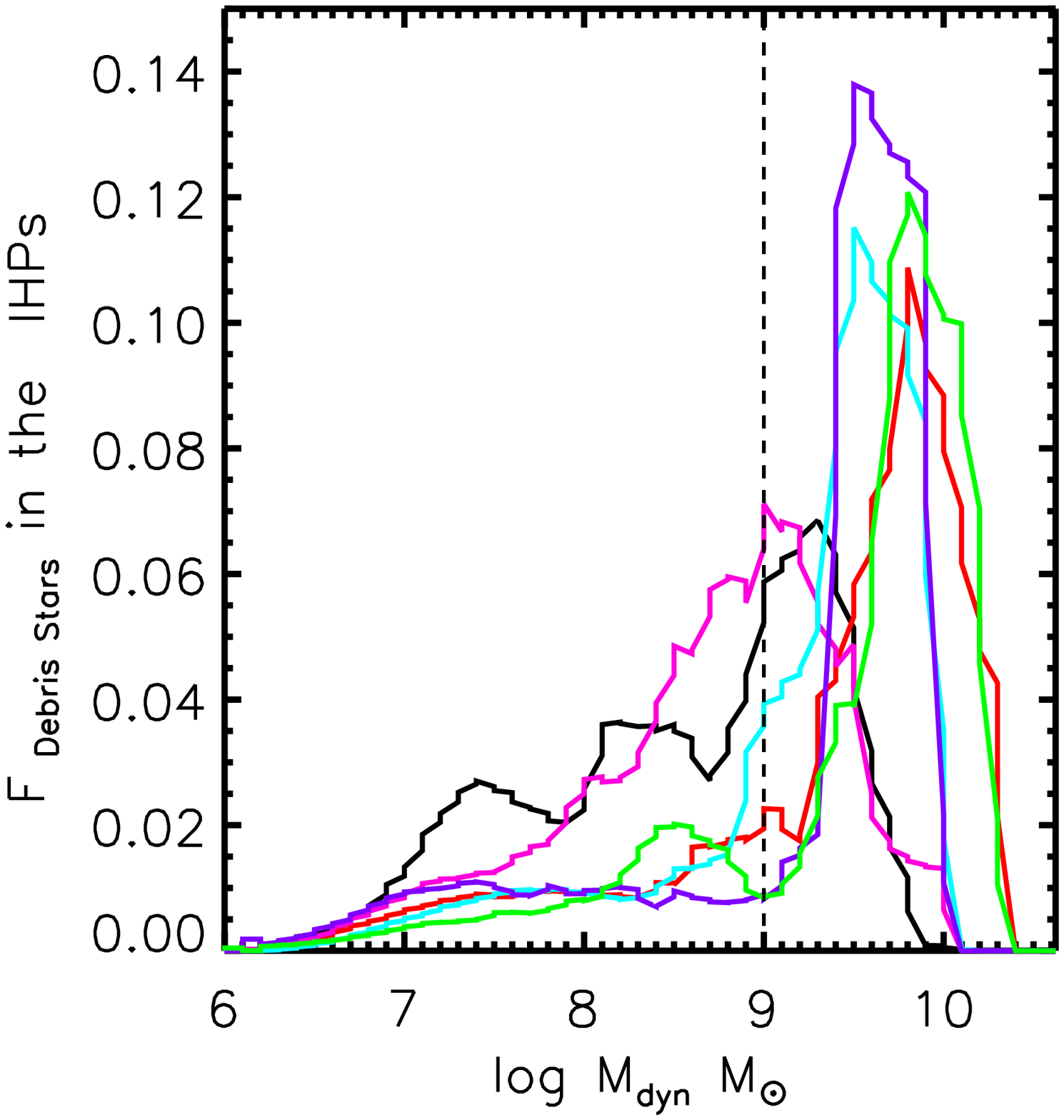}}
\hspace*{-0.2cm}\resizebox{6.5cm}{!}{\includegraphics{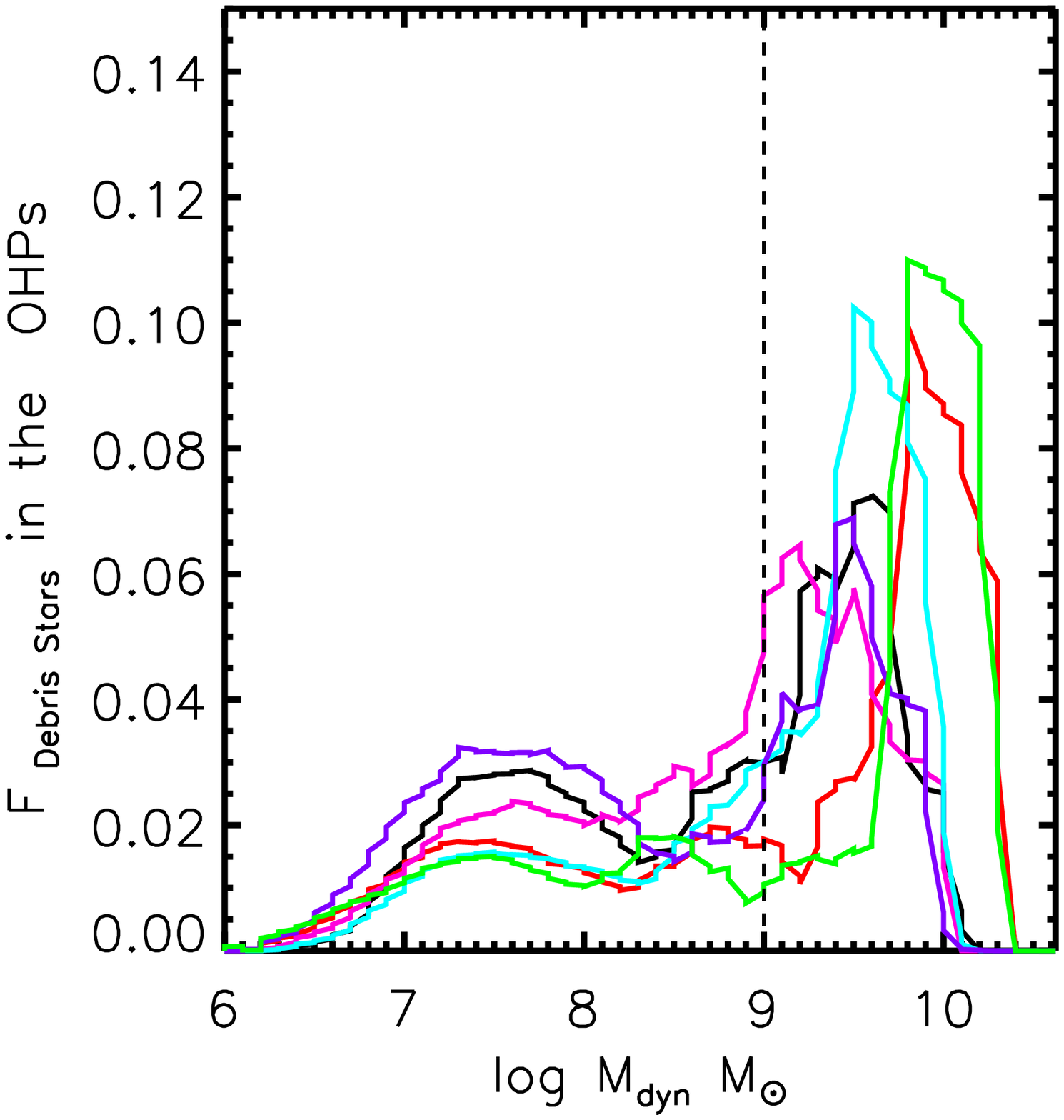}}\\
\hspace*{-0.2cm}\resizebox{6.5cm}{!}{\includegraphics{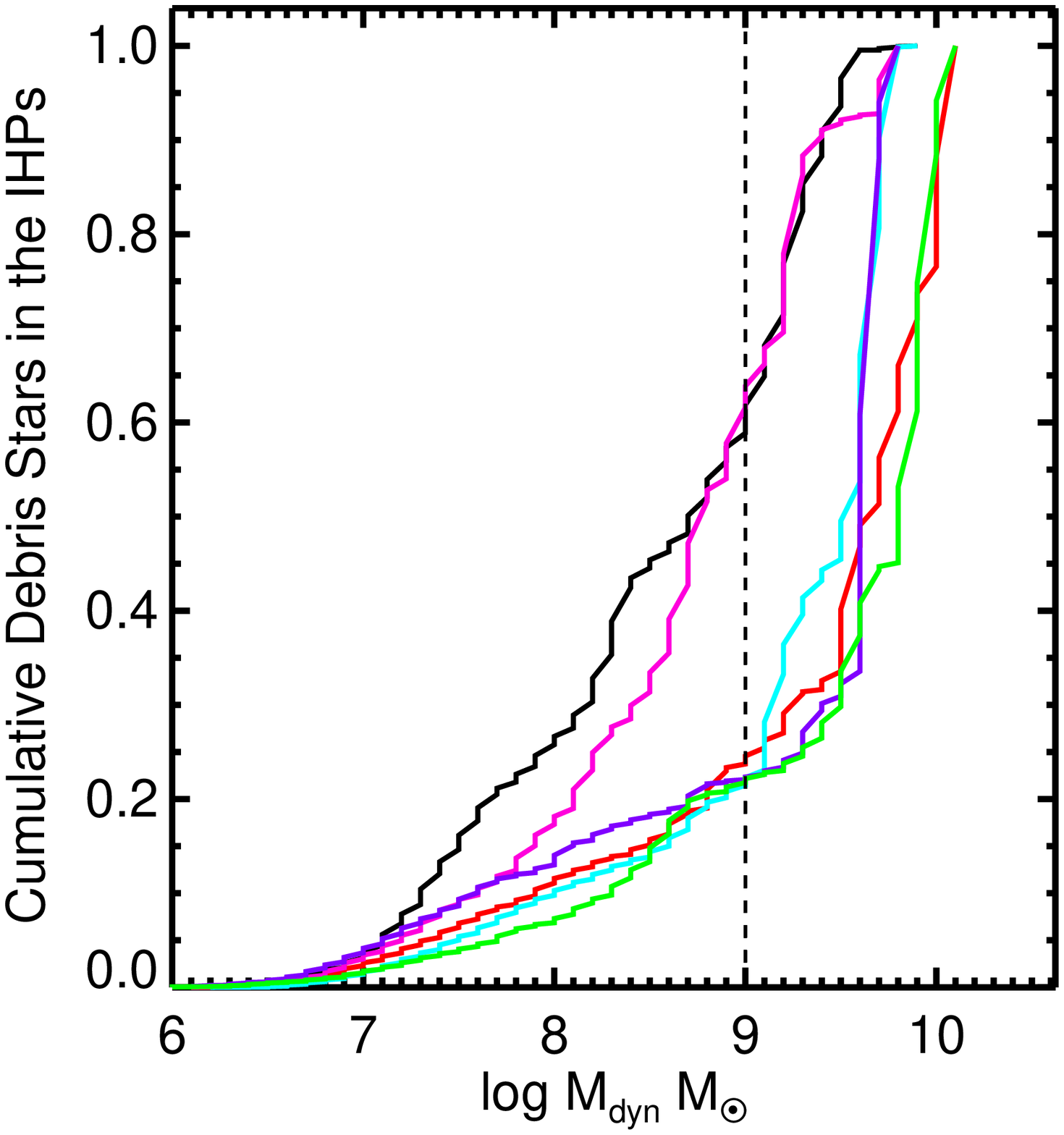}}
\hspace*{-0.2cm}\resizebox{6.5cm}{!}{\includegraphics{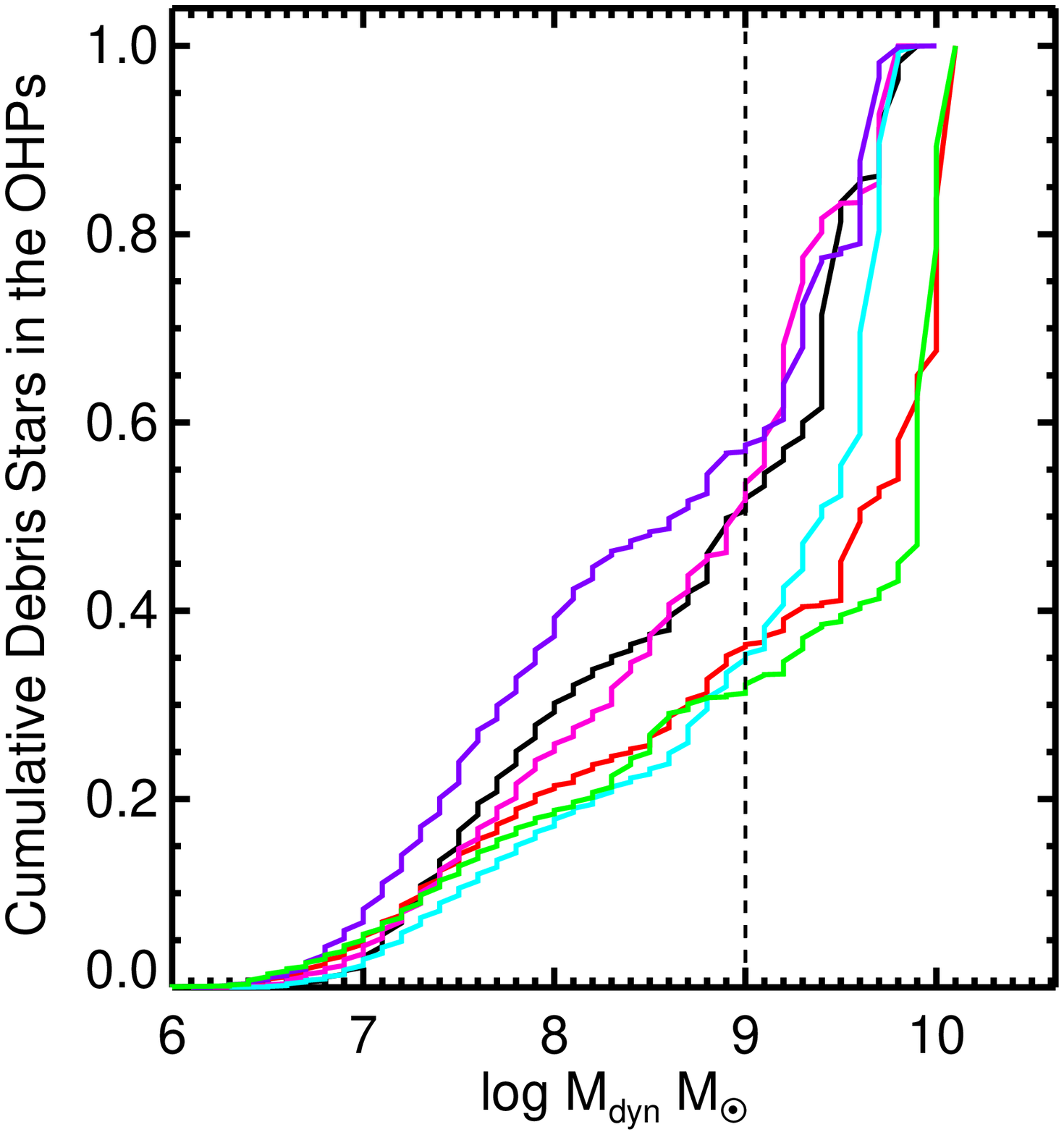}}
\caption{ Mass fraction (upper panels) and cumulative mass (lower panels) of debris stars in the IHPs (left panels) and OHPs
(right panels),  as a function of the dynamical masses of the systems
from which they formed.  The vertical dashed lines denote 
the reference value, M $= 10^{9}$ M$_{\odot}$, used
to separate more-massive from less-massive satellites. See Table 1 for the colour code.}
\label{sate1}
\end{figure*}

The variation of the properties of the diffuse stellar halo MDFs is
determined by how the stellar haloes were assembled. Can the diffuse
stellar halo MDFs tell us anything about the systems that contributed to
their formation? The upper panels of Fig.~\ref{sate1} show the fraction
of stellar-debris mass as a function of the total mass of the
subgalactic systems from which it formed, for the IHPs and OHPs of the
six analysed Aquarius haloes. The lower panels display the same
information, but as cumulative stellar mass. The vertical dashed lines
represent the threshold mass adopted to classify the subgalactic systems
into more-massive and less-massive ones ($M=10^{9}$ M$_{\odot}$).

The contributions of stars from low-mass subgalactic systems are larger
for the OHPs than for the IHPs, as can be clearly seen from both the
accreted-mass and the cumulative-mass fractions. For the IHPs, four out
of the six simulated haloes have less than $\sim 20$ per cent of their
stellar masses formed in low-mass subgalactic systems. The other two,
Aq-A-5 and Aq-C-5, have a larger contribution from low-mass systems, so
that only $\sim 40$ per cent of the stellar masses were formed in
massive subgalactic units.

In the case of the OHPs, there are also clearly different behaviours
from one simulated halo to the next. Three of the OHPs have more than
$\sim 60$ per cent of their stellar masses formed in low- and
intermediate-mass subgalactic systems (Aq-A-5, Aq-C-5 and Aq-G-5), while
the other three (Aq-B-5, Aq-D-5 and Aq-H-5) have more than $\sim 70$ per
cent formed from more-massive systems. Paper~II showed that there was a
correlation between the slope of the metallicity profiles of the OHPs
and the fraction of stars formed in massive systems, due to the fact
that they are found to have larger fractions of more-bound,
higher-metallicity stars. We confirm this finding here, since Aq-B-5,
Aq-D-5 and Aq-H-5 are the haloes exhibiting the steeper metallicity
slopes, as can be seen from Fig.~\ref{mdftotal} and Table \ref{tab1}.

\begin{figure*}
\hspace*{-0.2cm}\resizebox{3.5cm}{!}{\includegraphics{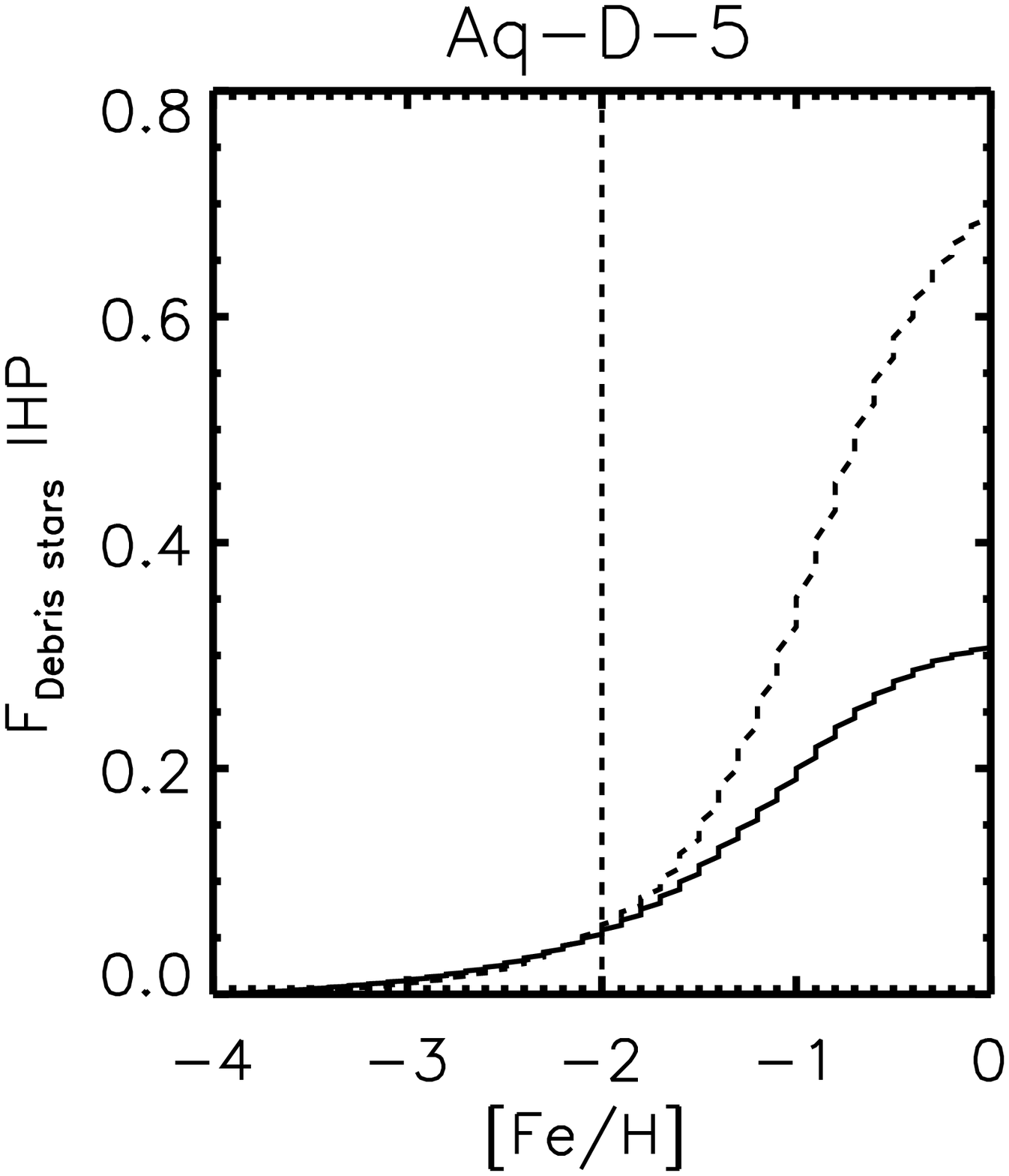}}
\hspace*{-0.2cm}\resizebox{3.5cm}{!}{\includegraphics{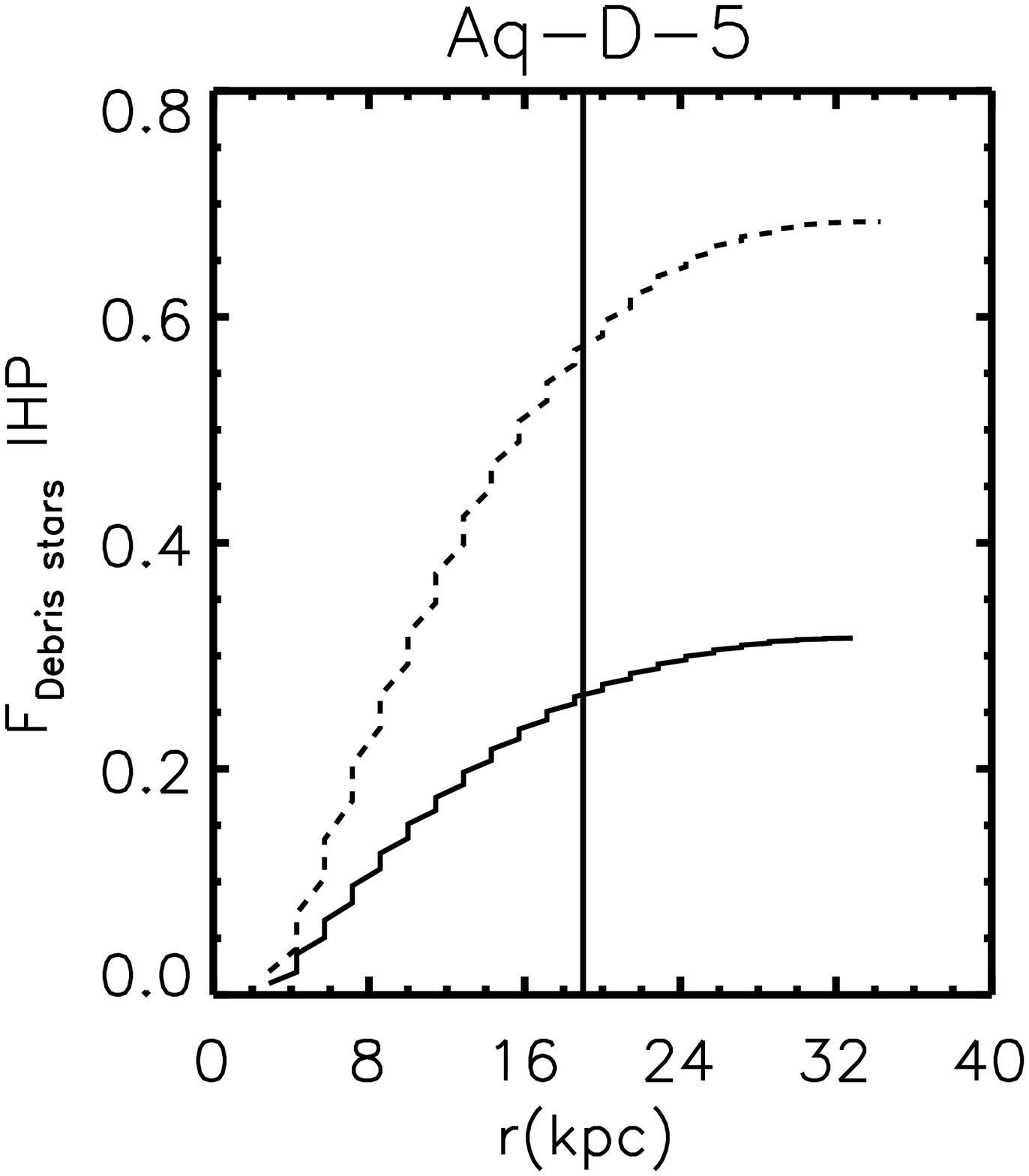}}
\hspace*{-0.2cm}\resizebox{3.5cm}{!}{\includegraphics{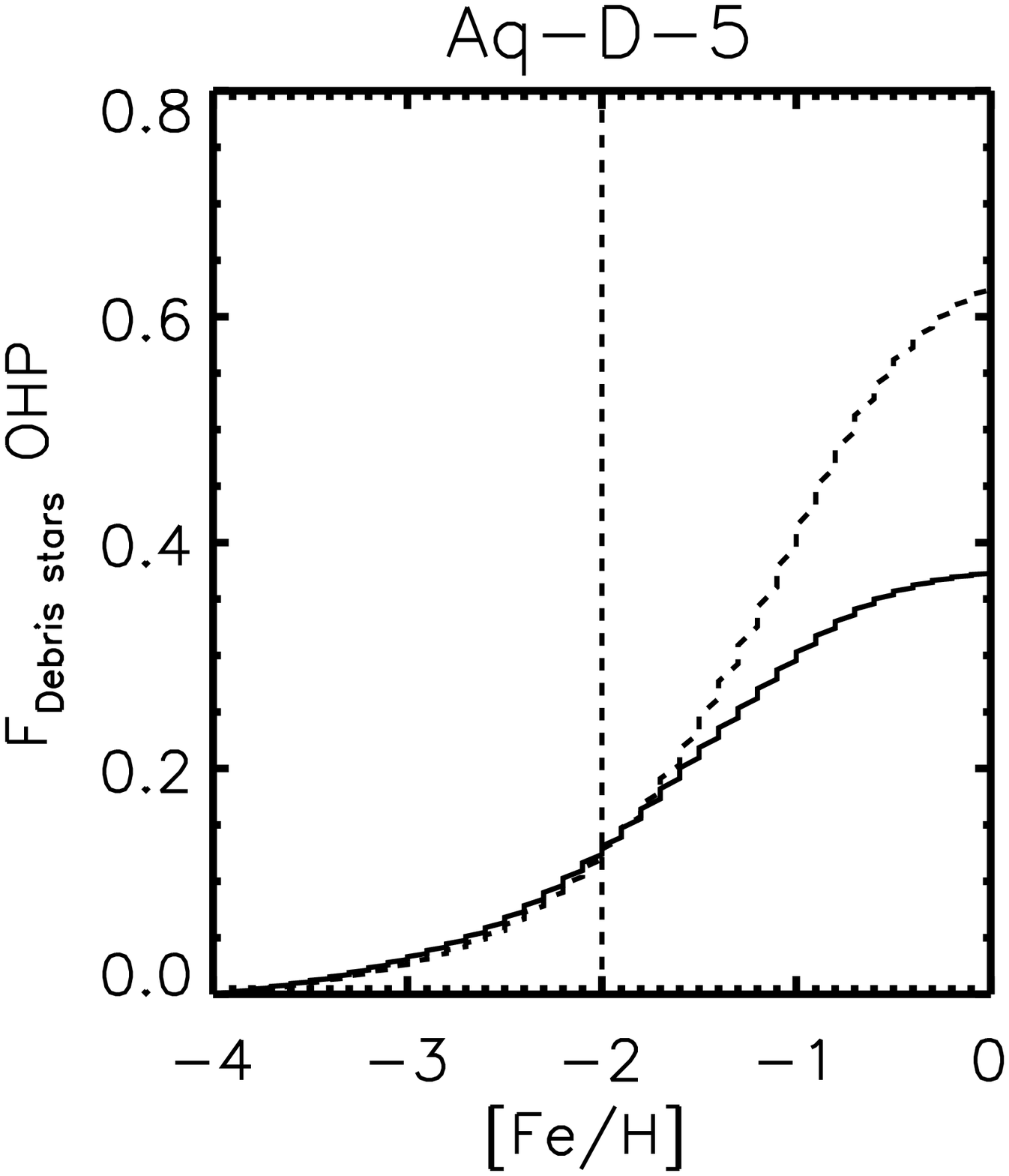}}
\hspace*{-0.2cm}\resizebox{3.5cm}{!}{\includegraphics{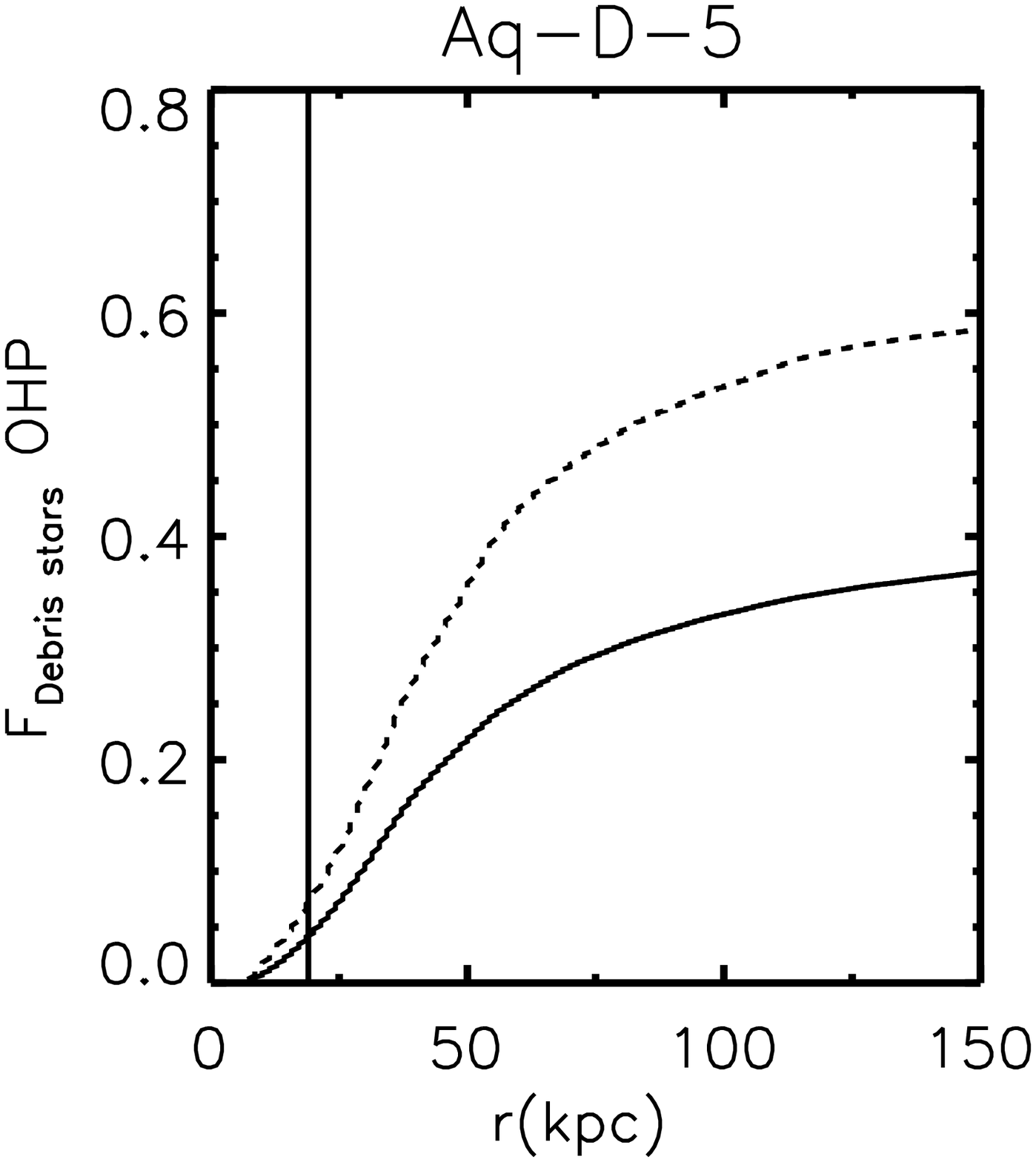}}\\
\hspace*{-0.5cm}\resizebox{3.5cm}{!}{\includegraphics{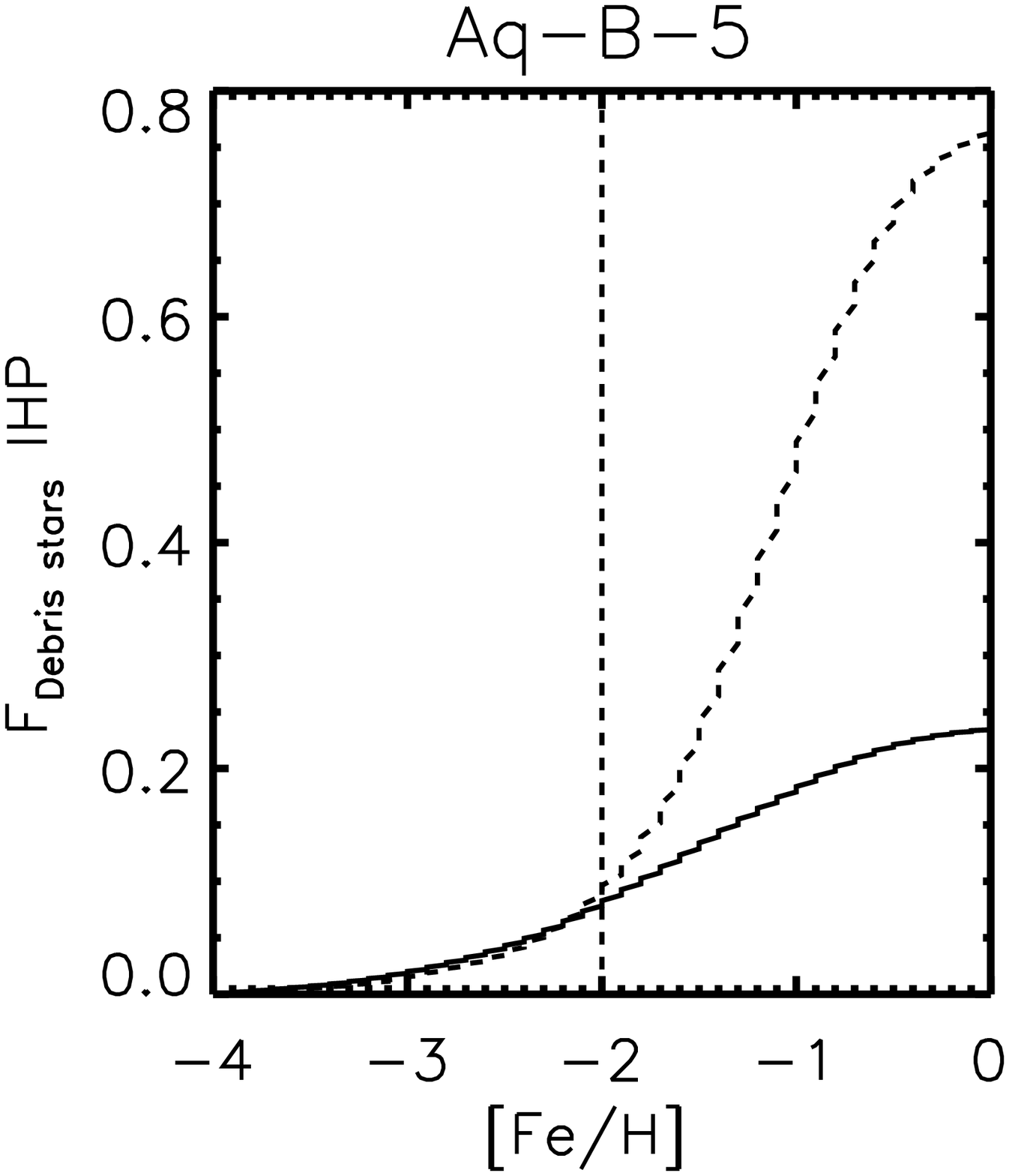}}
\hspace*{-0.2cm}\resizebox{3.5cm}{!}{\includegraphics{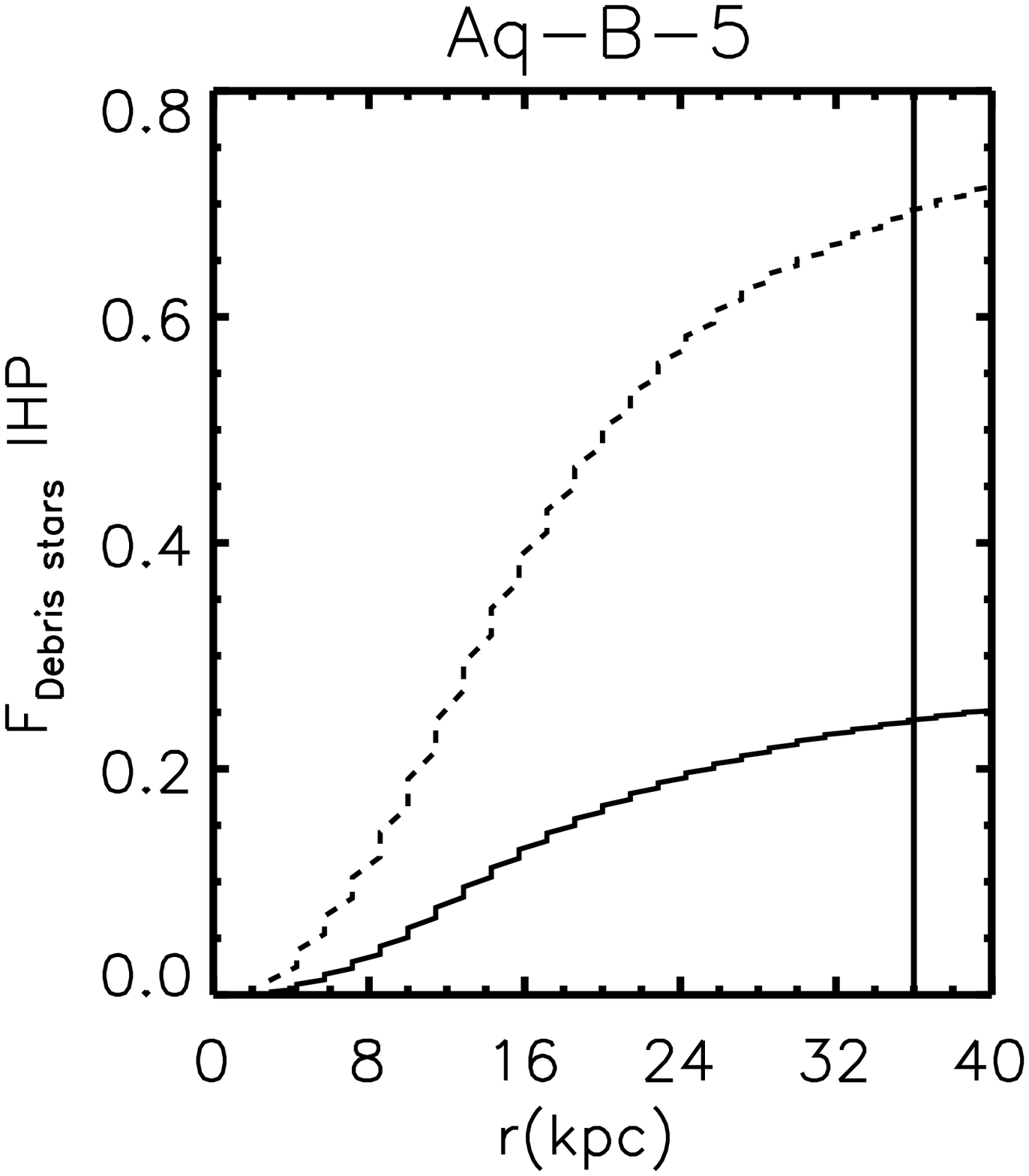}}
\hspace*{-0.2cm}\resizebox{3.5cm}{!}{\includegraphics{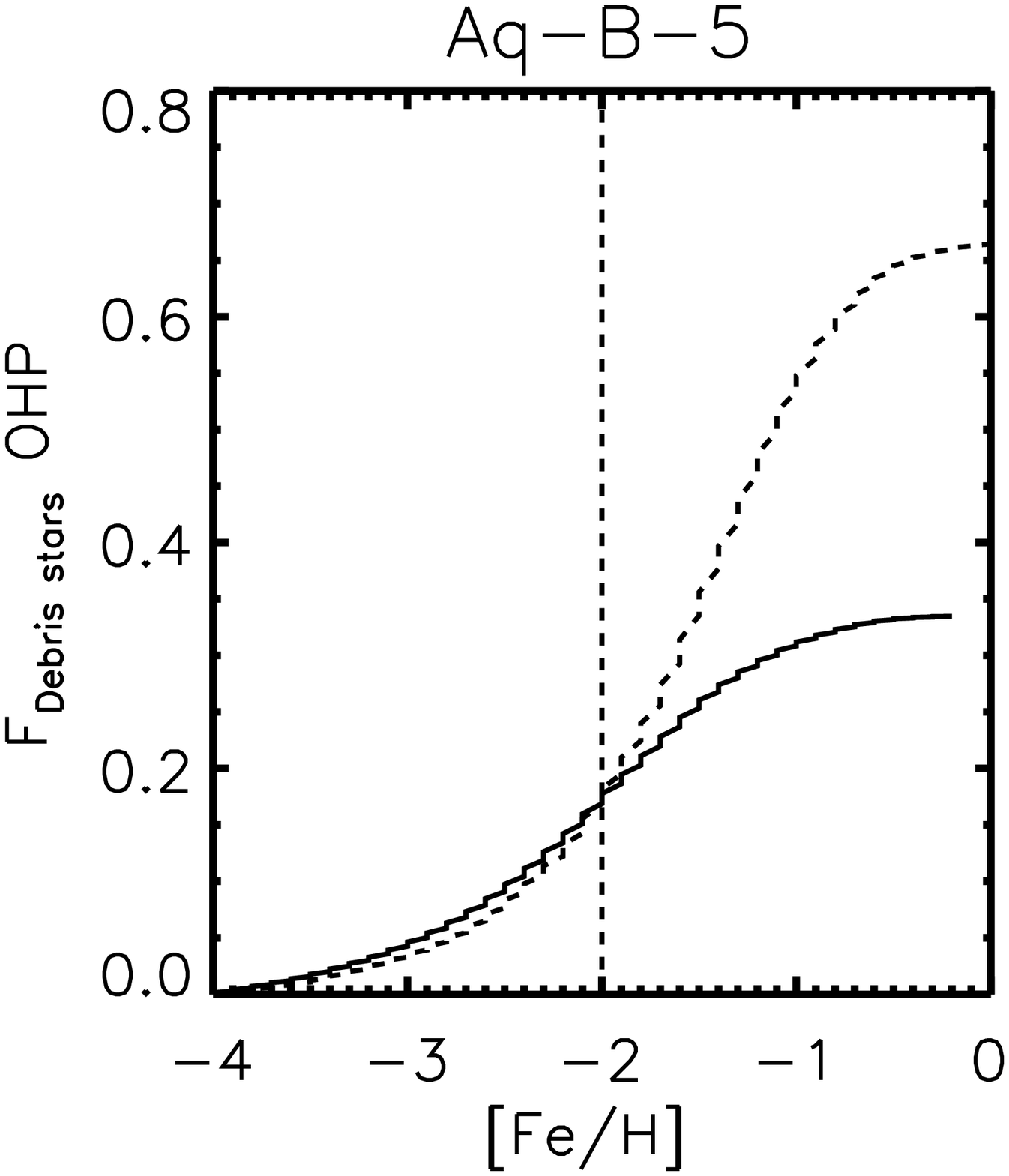}}
\hspace*{-0.2cm}\resizebox{3.5cm}{!}{\includegraphics{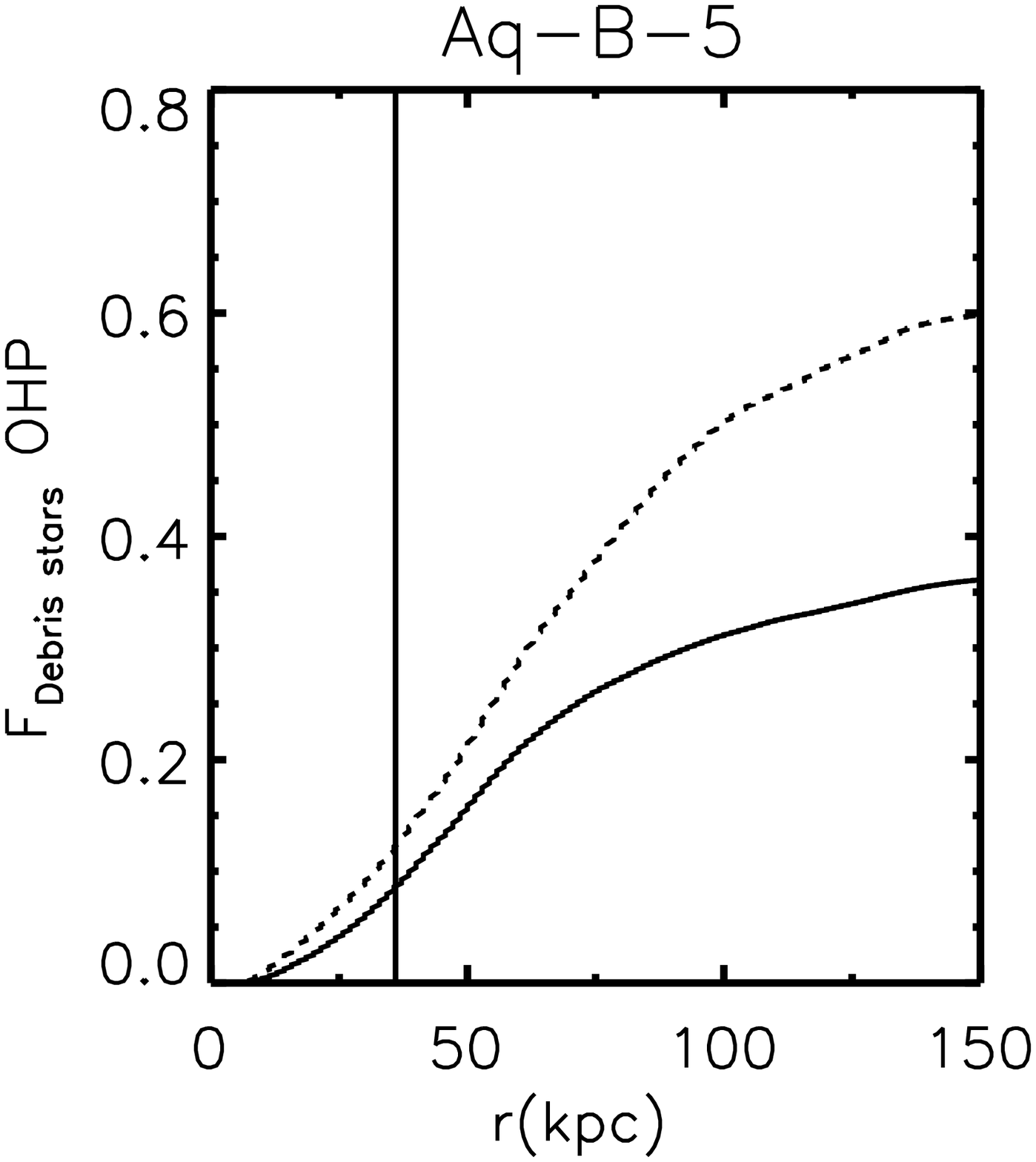}}\\
\hspace*{-0.5cm}\resizebox{3.5cm}{!}{\includegraphics{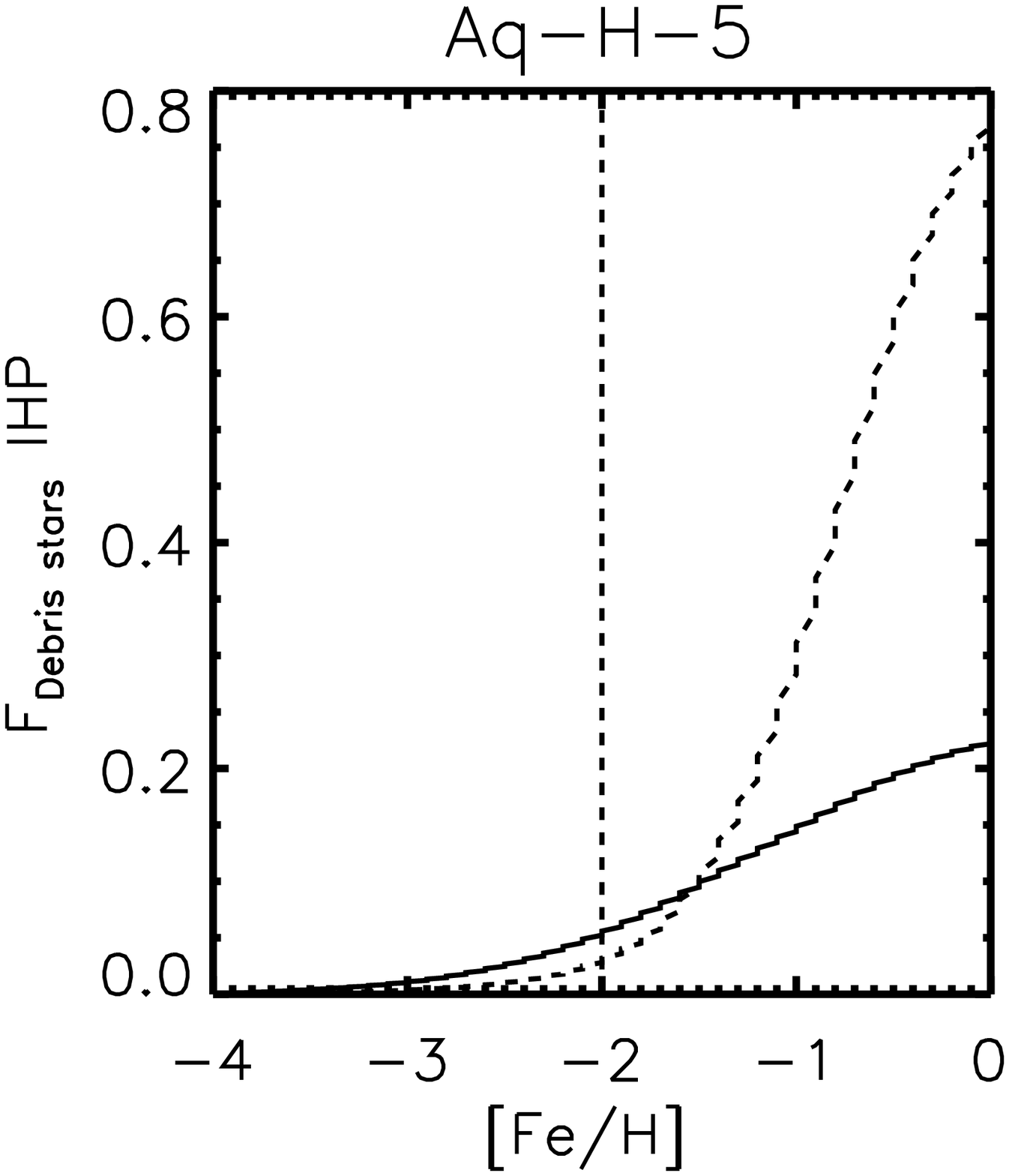}}
\hspace*{-0.2cm}\resizebox{3.5cm}{!}{\includegraphics{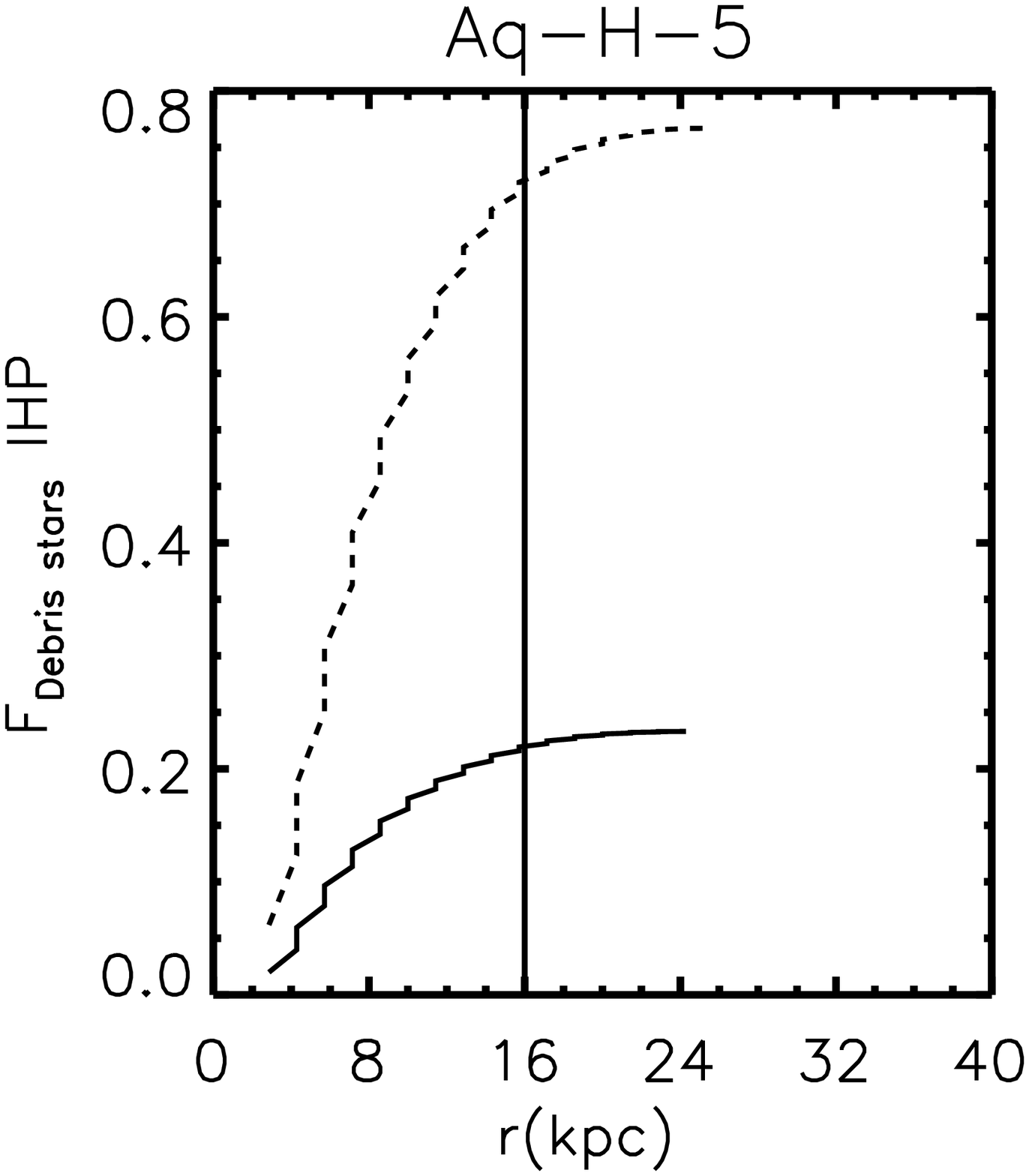}}
\hspace*{-0.2cm}\resizebox{3.5cm}{!}{\includegraphics{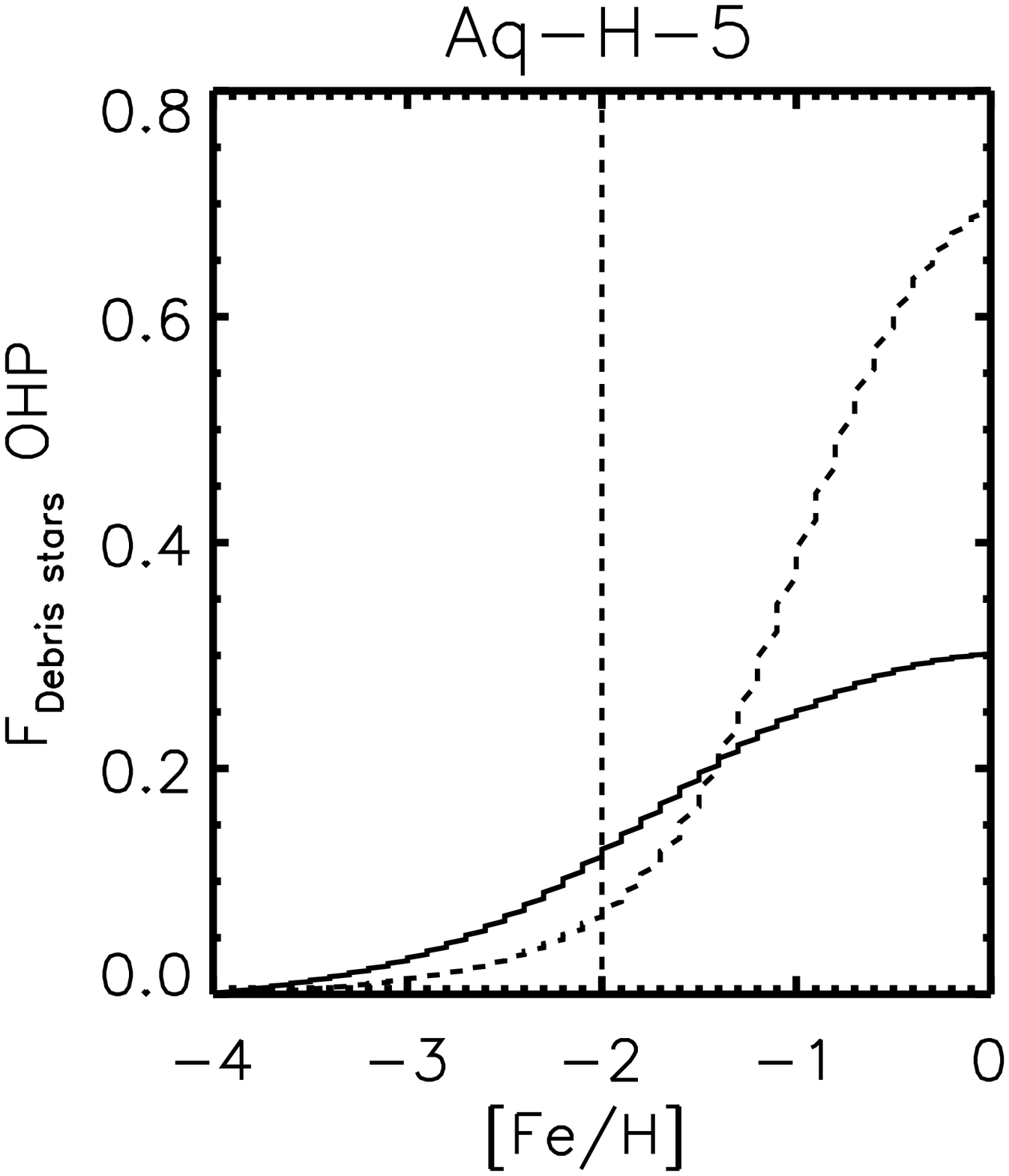}}
\hspace*{-0.2cm}\resizebox{3.5cm}{!}{\includegraphics{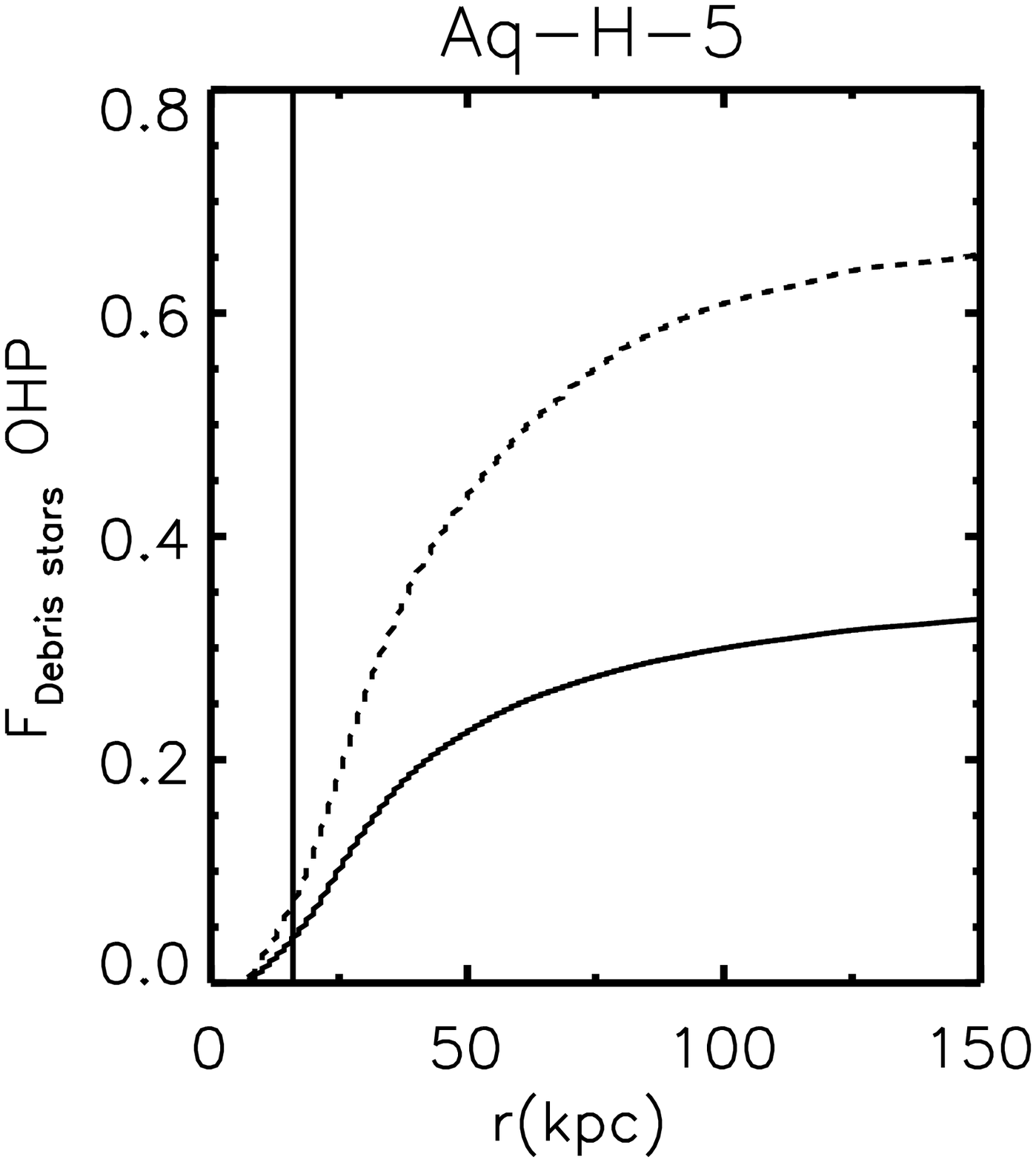}}\\
\hspace*{-0.5cm}\resizebox{3.5cm}{!}{\includegraphics{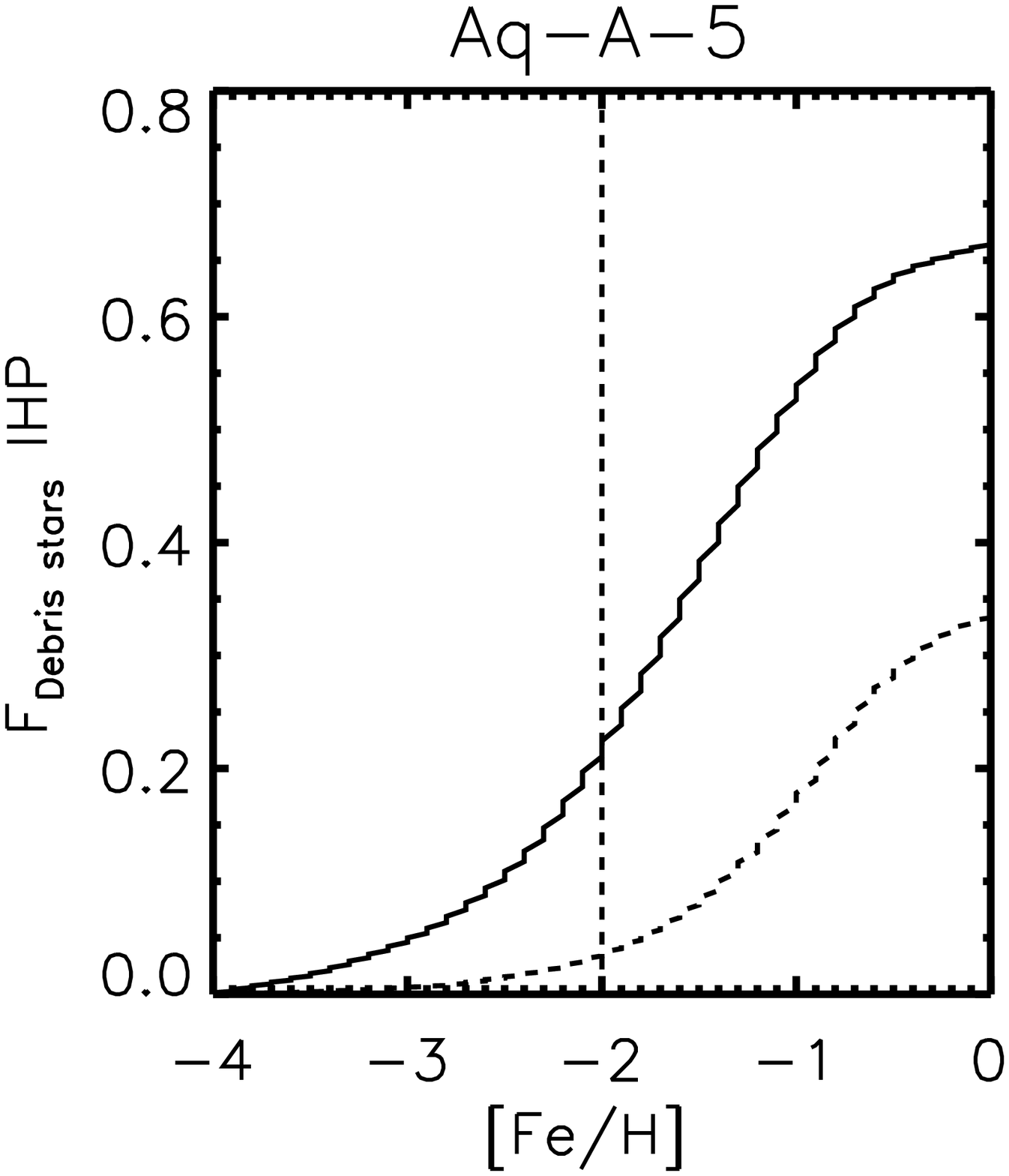}}
\hspace*{-0.2cm}\resizebox{3.5cm}{!}{\includegraphics{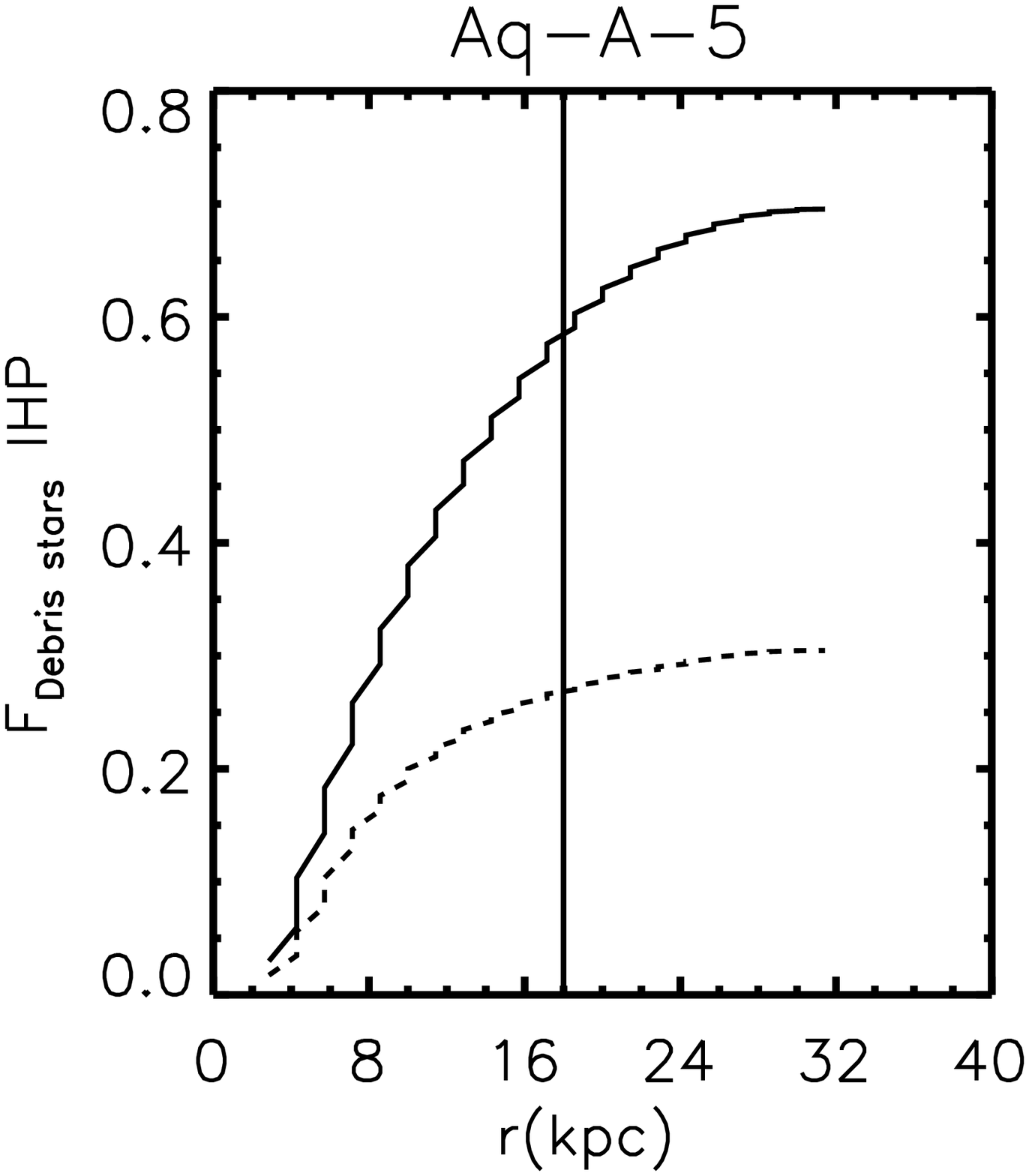}}
\hspace*{-0.2cm}\resizebox{3.5cm}{!}{\includegraphics{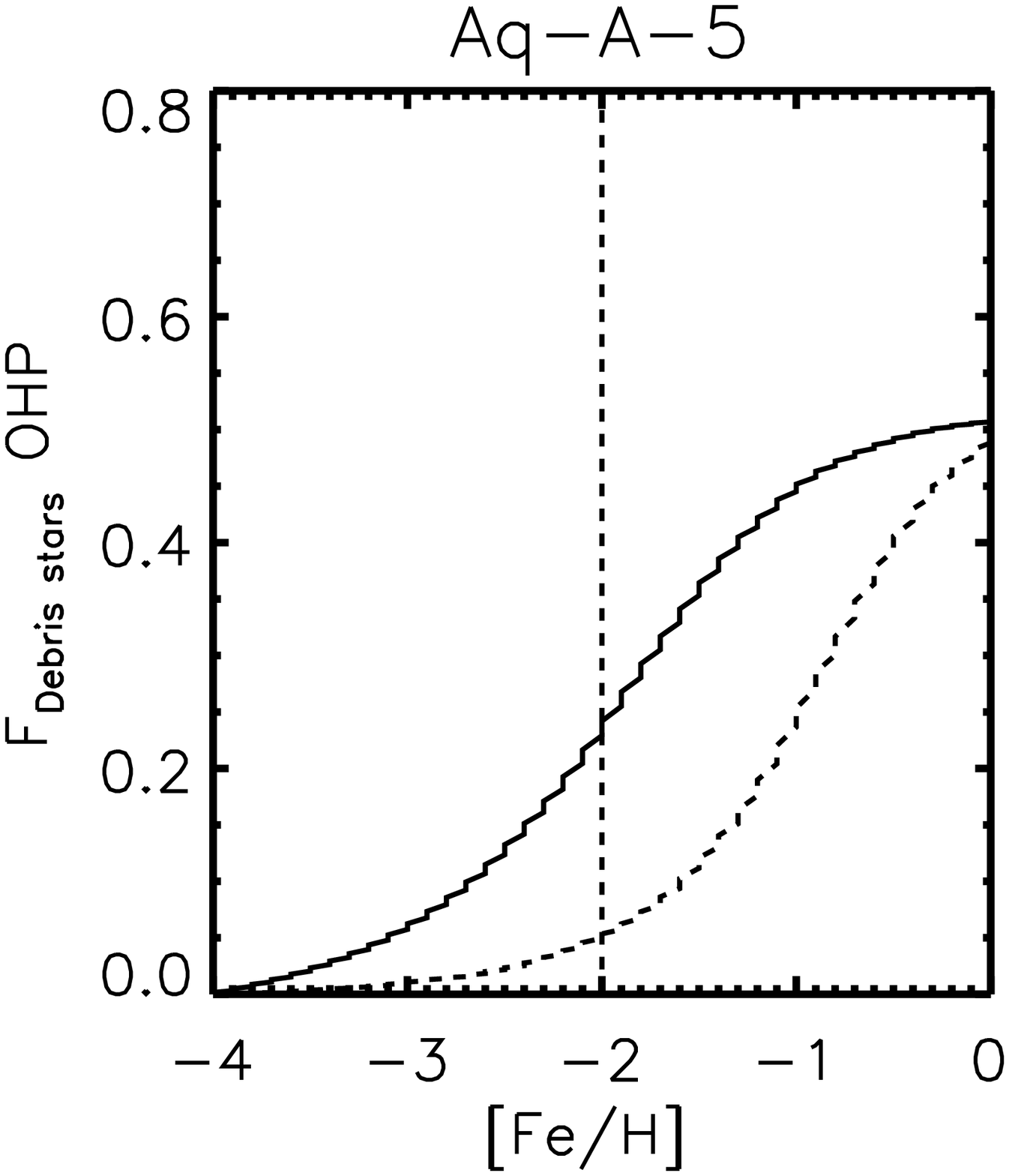}}
\hspace*{-0.2cm}\resizebox{3.5cm}{!}{\includegraphics{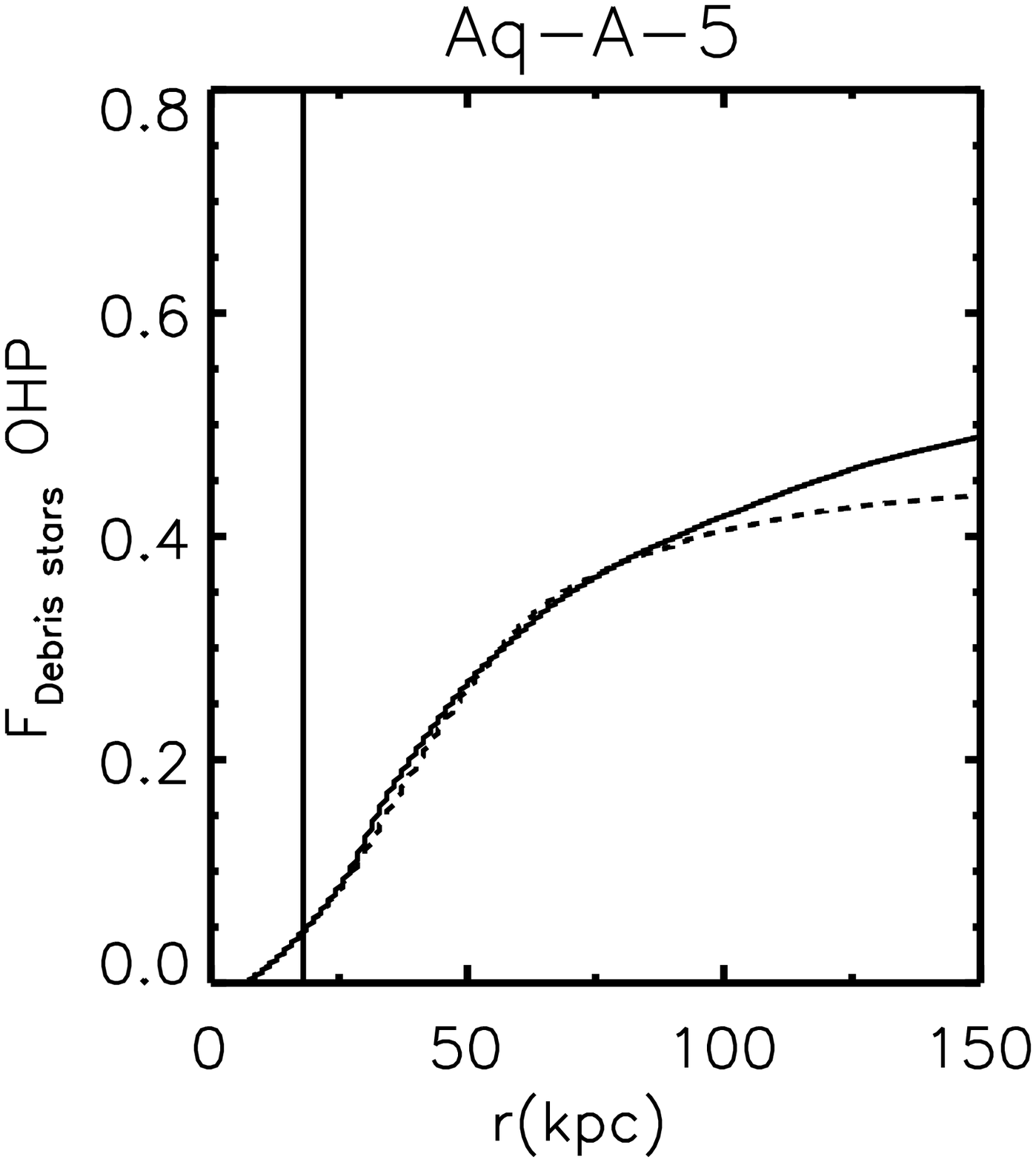}}\\
\hspace*{-0.5cm}\resizebox{3.5cm}{!}{\includegraphics{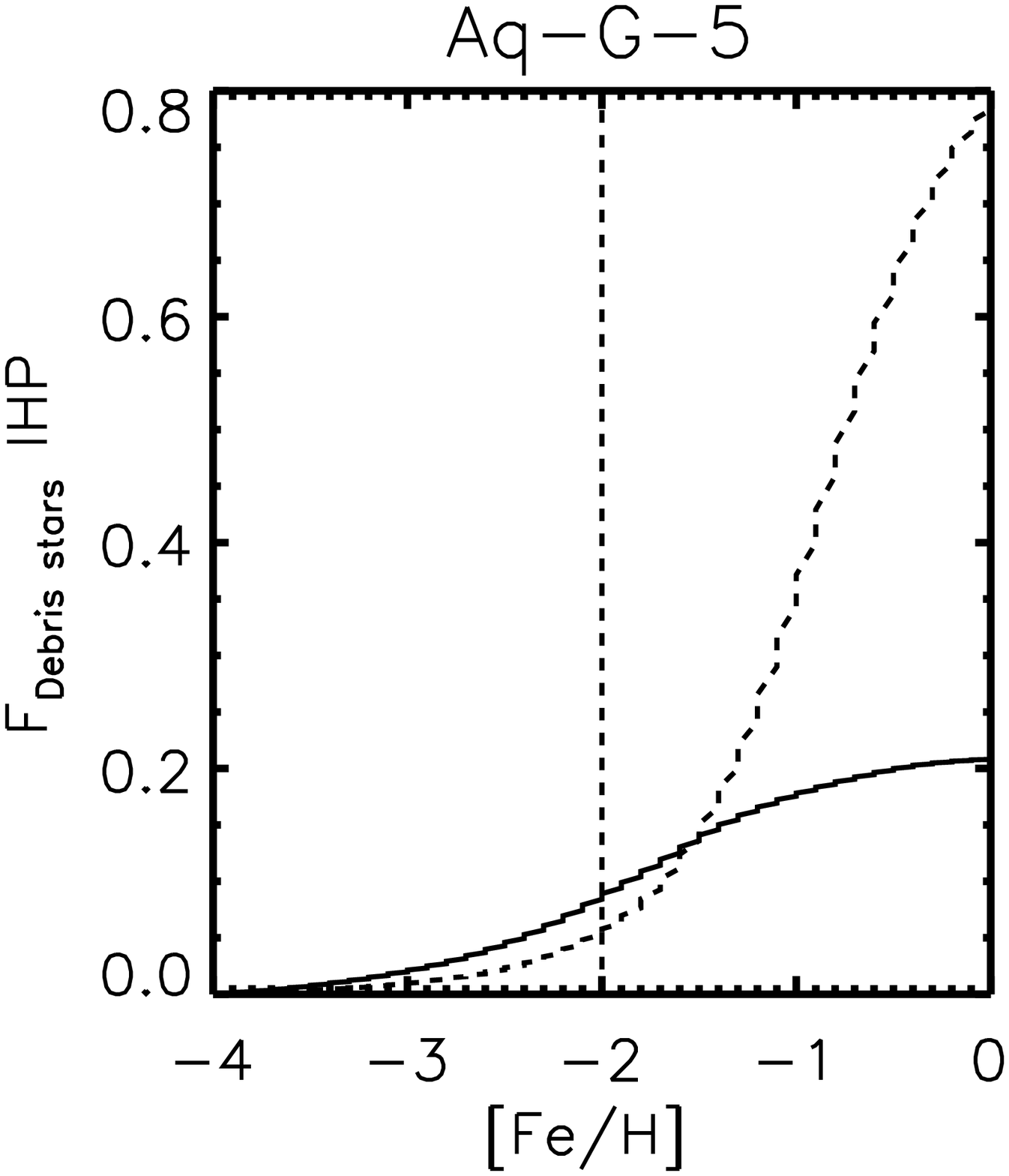}}
\hspace*{-0.2cm}\resizebox{3.5cm}{!}{\includegraphics{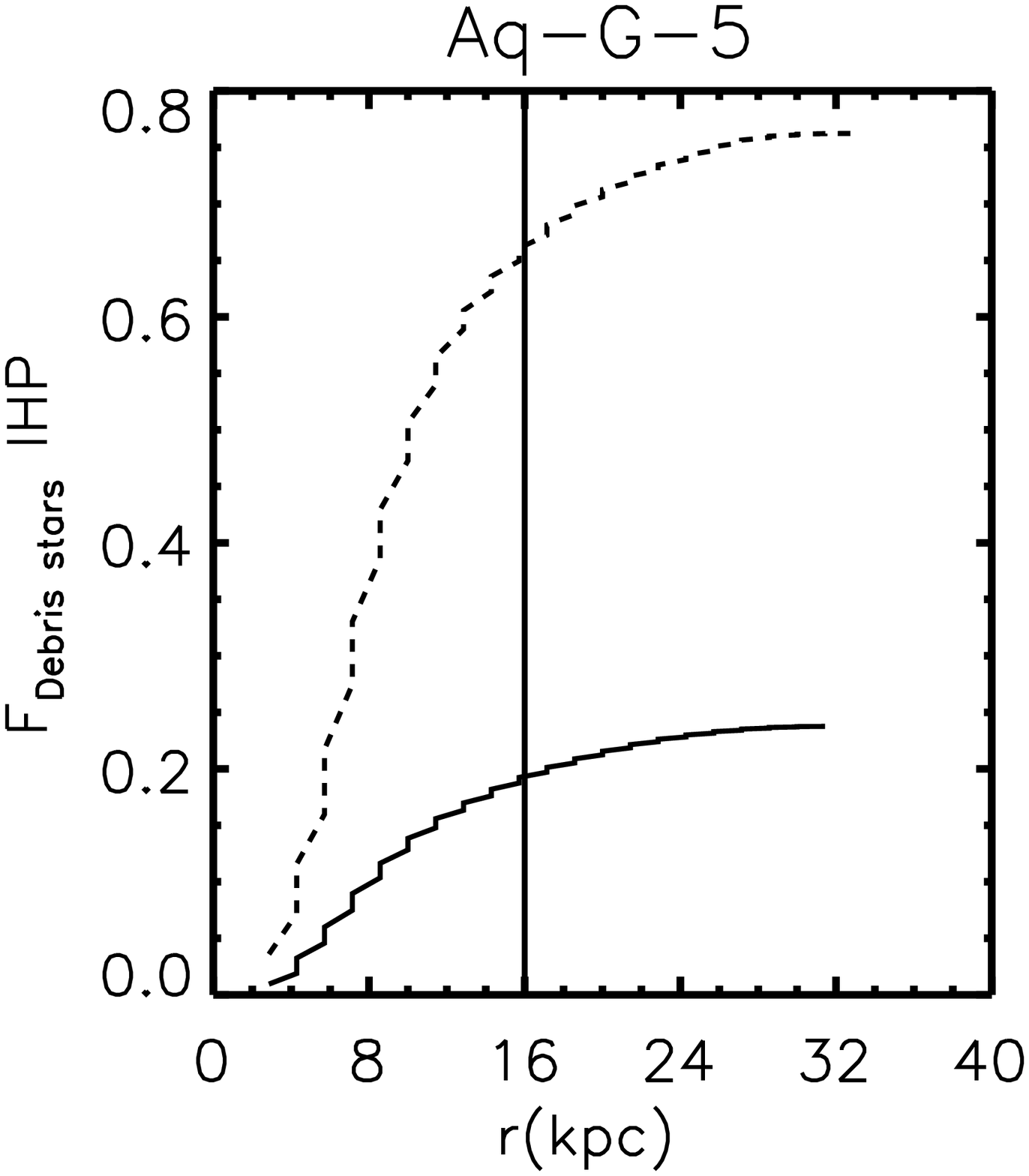}}
\hspace*{-0.2cm}\resizebox{3.5cm}{!}{\includegraphics{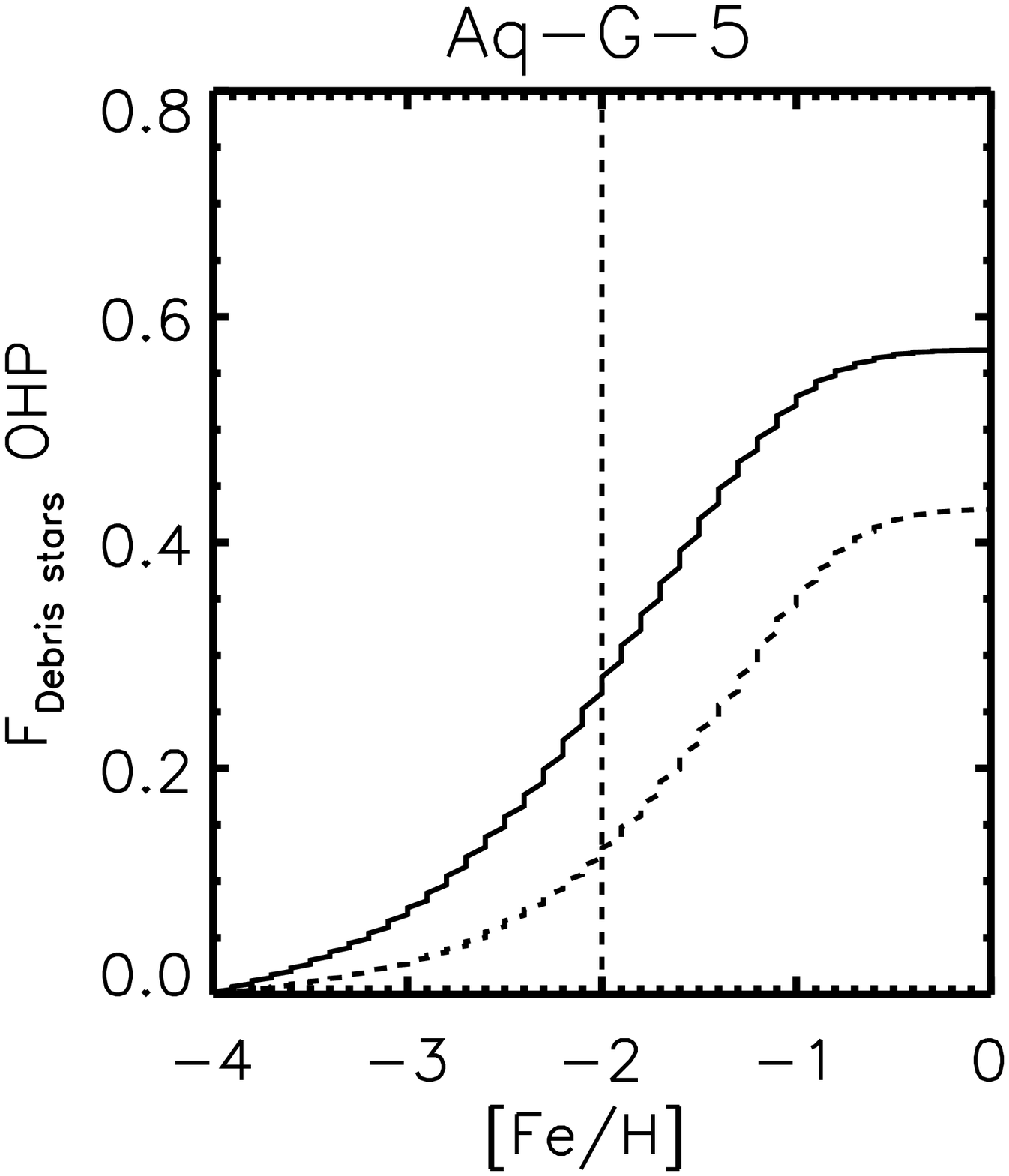}}
\hspace*{-0.2cm}\resizebox{3.5cm}{!}{\includegraphics{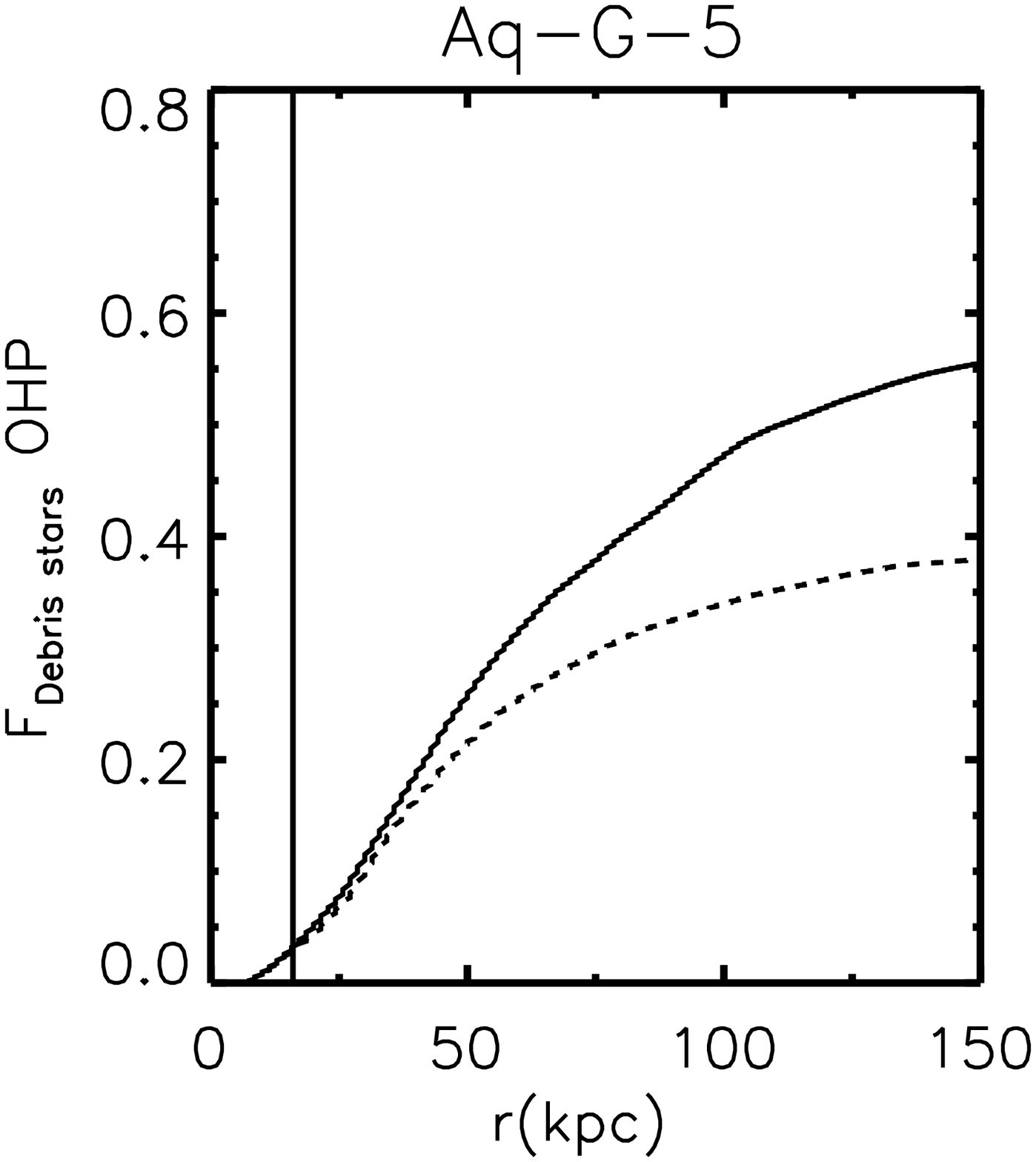}}\\
\hspace*{-0.5cm}\resizebox{3.5cm}{!}{\includegraphics{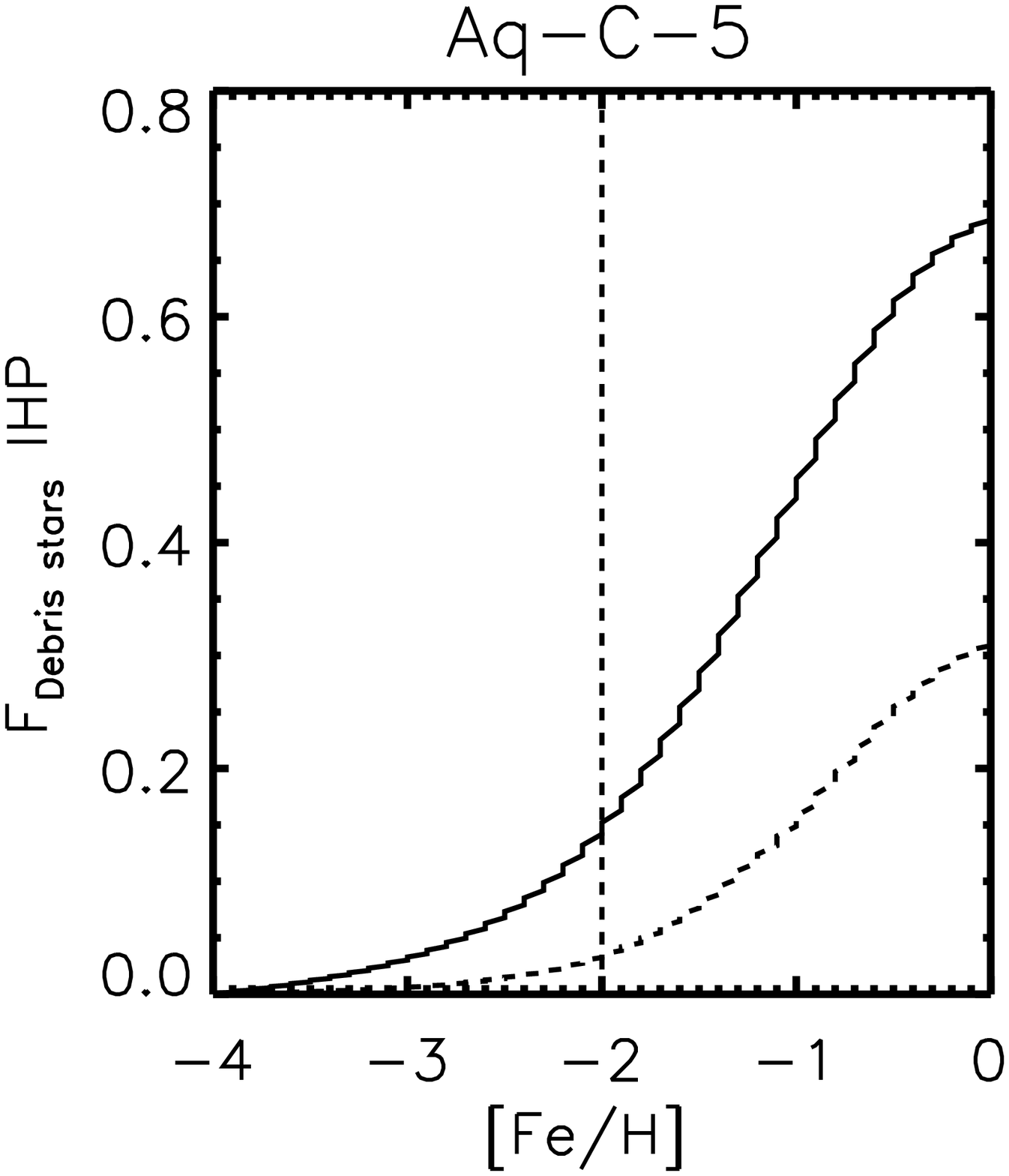}}
\hspace*{-0.2cm}\resizebox{3.5cm}{!}{\includegraphics{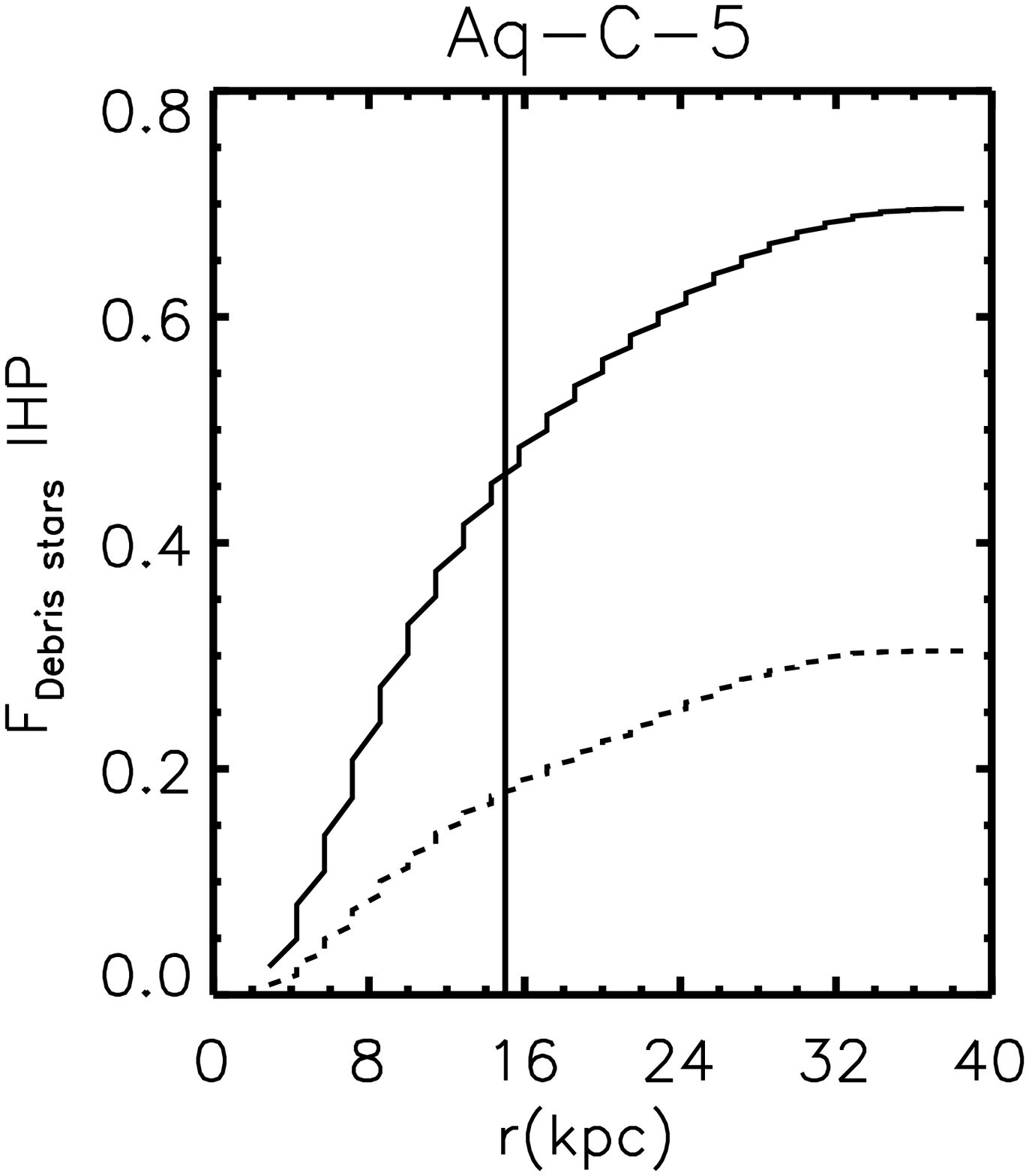}}
\hspace*{-0.2cm}\resizebox{3.5cm}{!}{\includegraphics{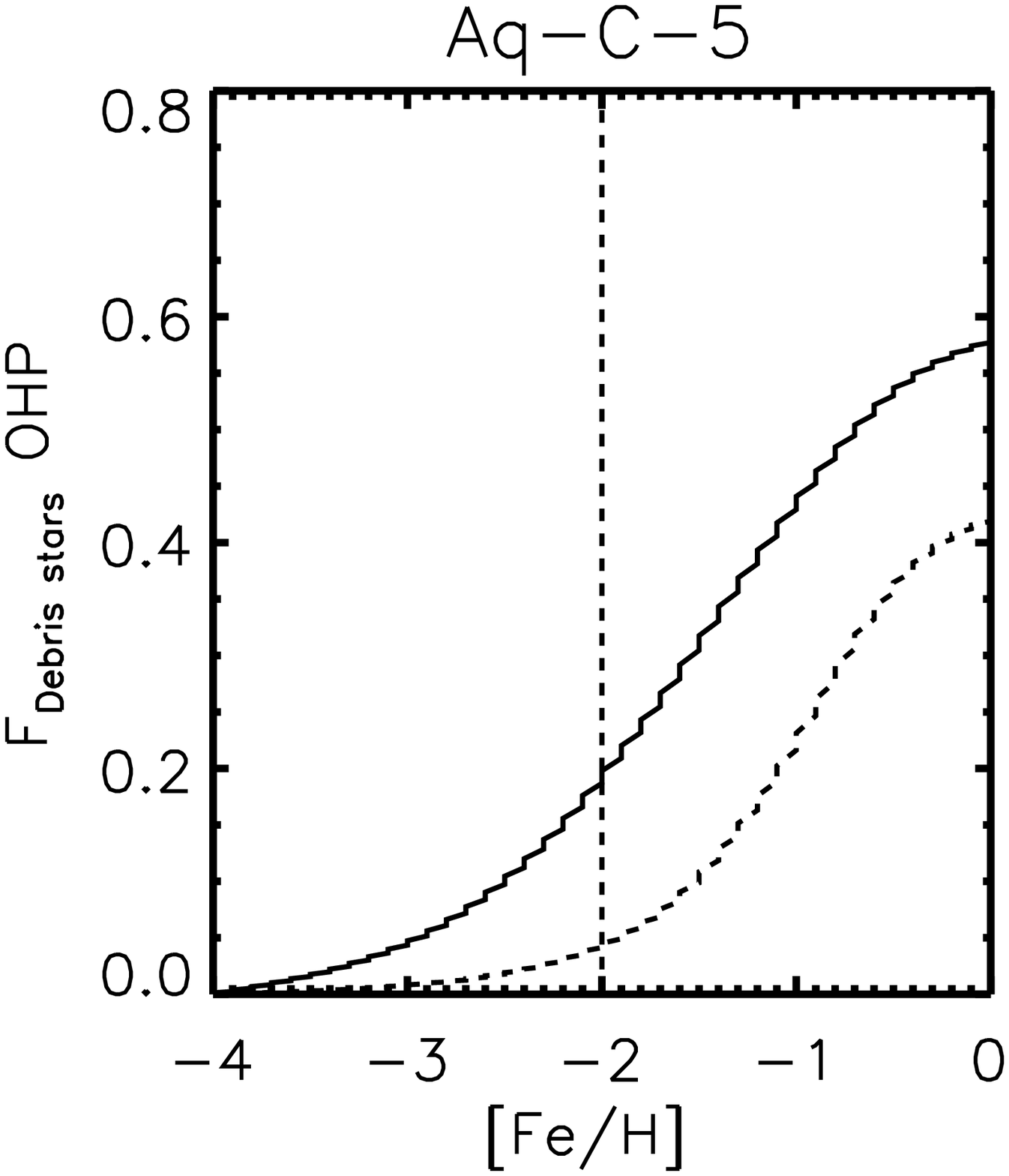}}
\hspace*{-0.2cm}\resizebox{3.5cm}{!}{\includegraphics{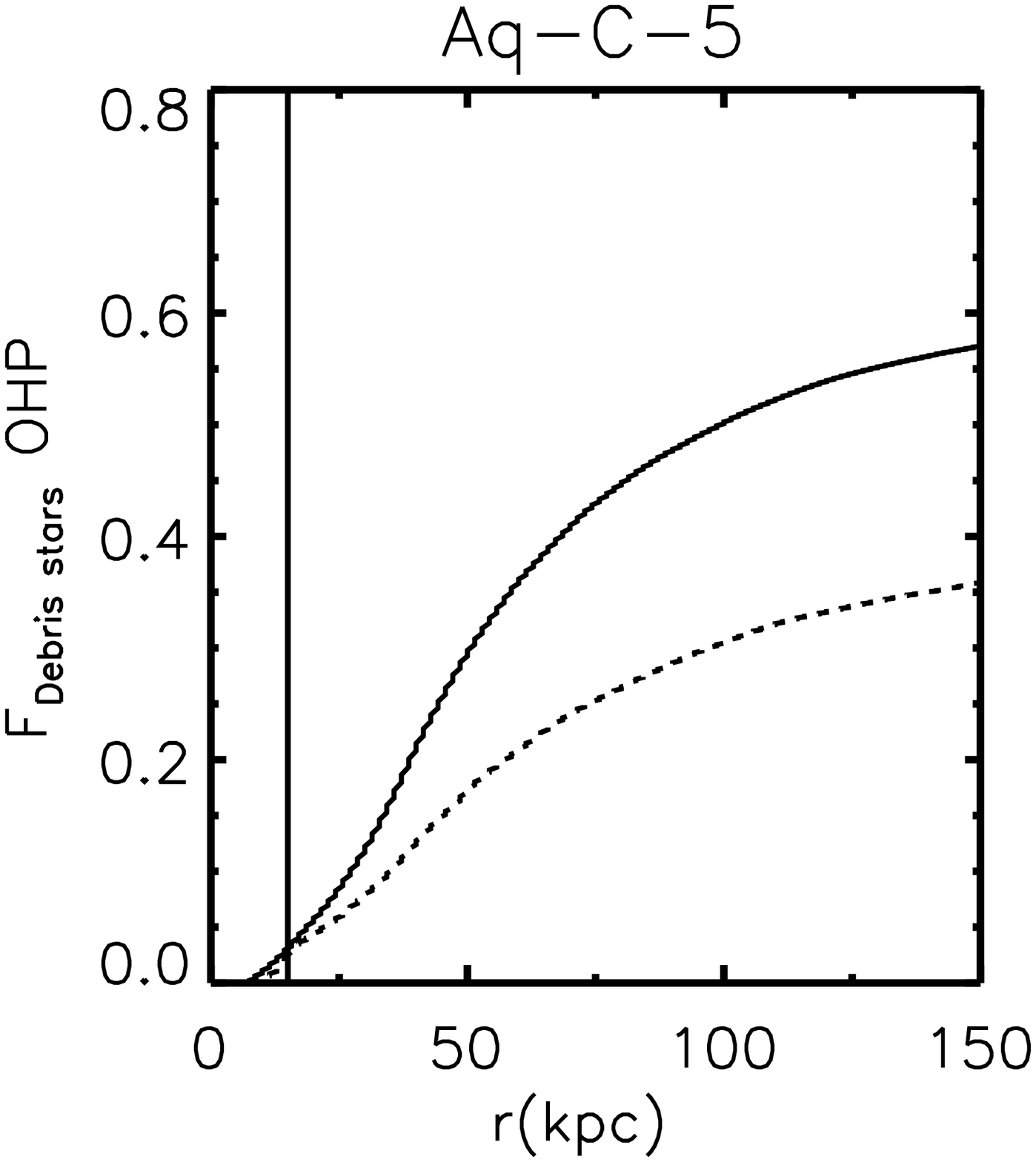}}\\
\caption{Cumulative stellar mass, as a function of the metallicity of
debris stars (first and third columns), and as a function of
galactocentric distance (second and fourth columns), formed in
less-massive (M $< 10^{9}$ M$_{\odot}$, solid lines), and more-massive
(M $> 10^{9}$ M$_{\odot}$, dashed lines) subgalactic systems, which were
accreted to form the IHPs and OHPs of our simulated haloes. The vertical
dashed lines  denote [Fe/H], while the vertical solid lines show the
$r_{\rm HTR}$. The stellar haloes have been ordered from those
exhibiting overall steeper to flatter metallicity-profile slopes, from
the upper left to the lower right. Flatter overall metallicity slopes
can be associated with haloes that have been assembled with a more
significant contribution  from low-mass satellites. There is a clear
transition in the relative contribution of  low- and high-mass
satellites, which is reflected in the MDF characteristics. There are
systems that exhibit quite different behaviours, as expected in
hierarchical clustering scenarios.
}
\label{MDFsatelites}
\end{figure*}

 The relative contribution of low-mass and high-mass subgalactic systems to
the IHPs and OHPs can be better illustrated from
Fig.~\ref{MDFsatelites} (second and fourth columns, respectively).
From this figure, we can see that the stellar haloes with the 
steeper  [Fe/H] profiles have more important contributions from stars formed in
massive satellites at all radii. As one moves to haloes with flatter
metallicity profiles, the relative contribution of  low-mass satellites
increases, so that the stellar halo with the flatter slope is
dominated by the contribution of stars formed in low-mass satellites at all
radii. In the case of Aq-G-5, we can clearly see the difference in the
debris assembly between the IHP and OHP, which is responsible for the
sharp change in the MDF.

Massive subgalactic systems are expected to contribute a significant
fraction of high-metallicity stars to both the IHPs and OHPs. They will
also survive farther into the haloes, contributing to setting the nature
of the metallicity profiles. In order to clearly show how subgalactic
systems contribute stars of different metallicities,
Fig.~\ref{MDFsatelites} shows the cumulative mass fraction as a function
of [Fe/H], for debris stars formed in more-massive and less-massive
accreted satellites, for both the IHPs and OHPs of our simulated haloes
(first and third columns, respectively). They have been ordered, from
upper left to lower right, according to the slope of the total [Fe/H]
profiles, from the steepest to the flattest ones. Regarding the IHPs,
more than $\sim80$ per cent of the debris stars come from satellites 
more massive than $10^{9}$ M$_{\odot}$, which also contribute most of
the high-metallicity stars. Aq-A-5 and Aq-C-5 are the only simulations
that exhibit significant contributions from lower-mass subgalactic
systems for the IHPs (recall that the IHPs also received important
contributions from \insitu stars, which can modify their metallicity
profiles).

From Fig. ~\ref{MDFsatelites} it is clear that, while massive
subgalactic systems contribute most of the high-metallicity stars, the
low-metallicity stars can come from low-mass, intermediate-mass and
massive systems. However, stellar haloes with the flatter slopes tend to
have important contributions from low-mass subgalactic systems.
This figure allows us to understand the shape of the cumulative stellar
mass shown in Fig.\ref{mdftotal} for Aq-C-5 and A-G-5. The convex shape
of the latter is produced by more significant contributions from
low-metallicity stars formed in lower-mass subgalactic systems, as
compared to Aq-C-5. 

The fact that the MDF of the diffuse stellar halo for a given galaxy
does not change strongly as one moves from the IHR to the OHR does 
not refute its hierarchical assembly, but rather, suggests that the
assembly history of such a galaxy might involve similar contributions
from satellites of different masses, which are better mixed within the
potential well of the main system, with a clear trend to have the
flatest slope in those systems with larger contribution from low-mass
satellites.

\subsection{Confrontation with observations}

Numerous observational efforts are beginning to provide information, not
only on the existence of extended, diffuse stellar haloes in the MW and
other nearby galaxies, but on their chemical and dynamical properties as
well. It is still premature to carry out a detailed statistical
comparison between these observations and the simulations,
since both lack sufficient numbers of objects (or realizations) to fully
represent the range of phenomena that exist. However, even with the
limited information available at present, we can compare with the
observed variation of the MDFs of the detected stellar haloes in spiral
galaxies as one moves from the IHRs to the OHRs, in order to see if our
simulated haloes are indeed providing useful insight.

\begin{figure}
\hspace*{-0.2cm}\resizebox{7.5cm}{!}{\includegraphics{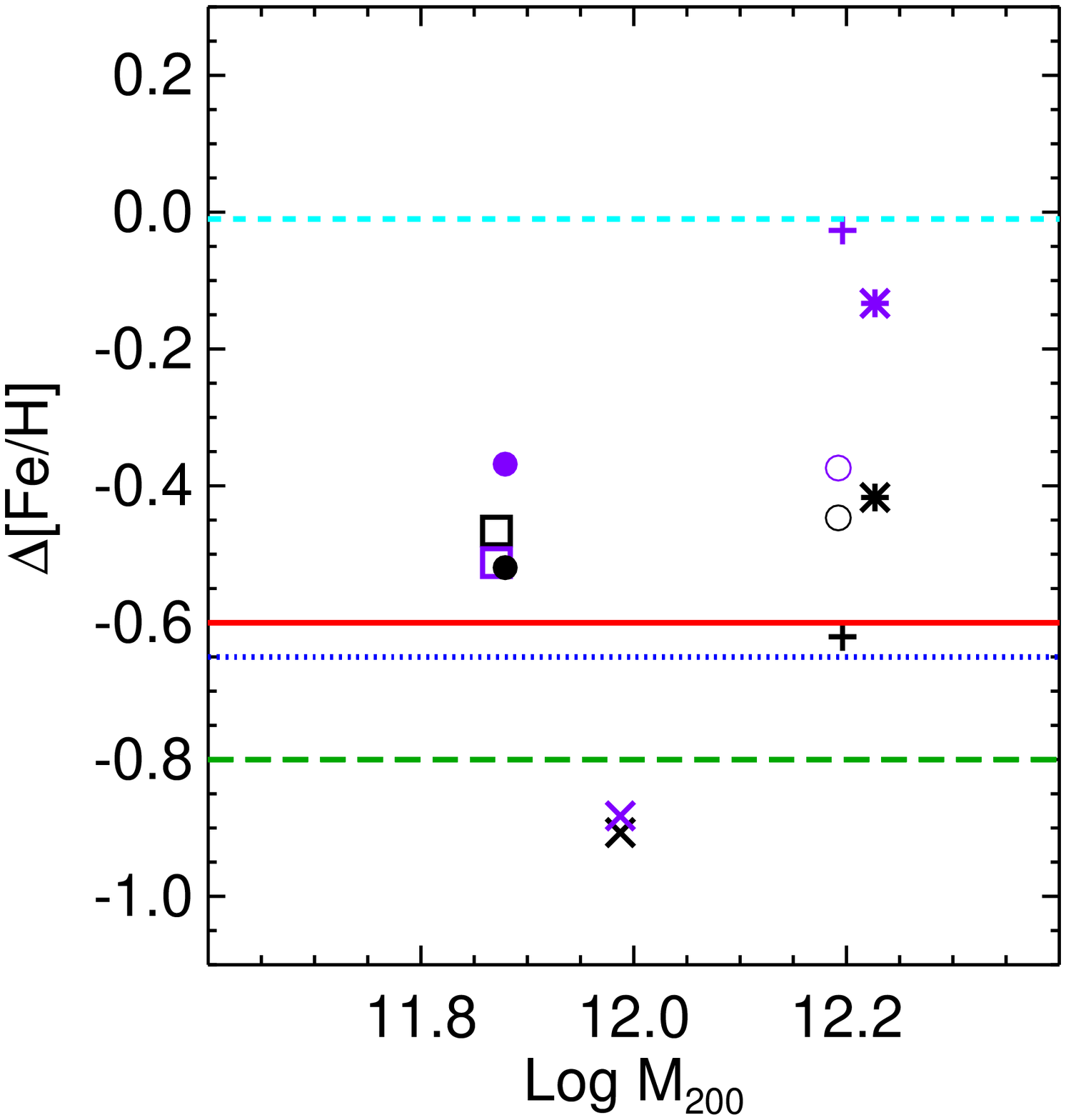}}
\caption{Difference between the median [Fe/H] of the IHPs and OHPs,
$\Delta$[Fe/H], as a function virial mass,
using the entire stellar populations (black symbols) and the debris
subpopulations only (violet symbols). The horizontal lines denote the
observed values for the MW  \citep[red solid line]{caro2010}, M32 \citep[blue
dotted line]{saraje2012} and M31
\citep[green dashed lines]{gilbert2013,ibata2013}. We also include a CMD-deduced
value for the [Fe/H] profile for M81 \citep[cyan dashed
line]{monachesi2013}.
}
\label{diffFeH}
\end{figure}

For the purpose of this exercise, we adopt a simple difference between
the medians of the [Fe/H] distributions for the IHPs and OHPs,
$\Delta$[Fe/H]. In the case of the observations, $\Delta$[Fe/H] is
estimated using the peak values of the MDFs reported by observers to be
associated with the likely IHPs and OHPs of the galaxy under
consideration, or else is inferred from the CMD-based metallicity
profiles. The simulated haloes exhibit a mean $\Delta$[Fe/H]$\sim -0.5$
dex (see also Paper~I), well within the range in this parameter covered
by current observations (see Fig. ~\ref{diffFeH}). 

We have also estimated $\Delta$[Fe/H] from the simulations by only
considering the portion of the stellar IHPs and OHPs that was
contributed by the debris stars. This produced a decrease in the
metallicity differences, with a mean $\Delta$[Fe/H]$\sim -0.35 $ (Fig.
~\ref{diffFeH}, violet points). The inclusion of the \insitu
subpopulations generally increases the value of $\Delta $[Fe/H], since
the latter contribute more chemically-enriched stars (in particular to
the IHPs), although there are outliers. The simulated halo Aq-G-5, for
example, exhibits a very large metallicity difference even in the case
when only debris stars are considered, and is only weakly affected by
addition of the \insitu component. Such behaviour results from the
particular assembly history of this halo, which produced a very steep
MDF profile in the IHP as a result of the contribution from massive
satellites, as can be seen in Fig.~\ref{MDFsatelites}. The other
interesting system is Aq-B-5, which has an IHP and OHP dominated by
debris stars, but only negligible contributions from the disc-heated
 subpopulation.

Fig.~\ref{diffFeH} also displays the observed values reported for the
MW, M31 and M32 \citep{brown2007,caro2010,saraje2012,gilbert2013},
indicated by the different-coloured horizontal lines. We also include an
estimate from the CMD metallicity profiles reported by
\citet{monachesi2013} for M81, which implies a flat metallicity
gradient, and consequently we assume a $\Delta$[Fe/H]$\sim 0 $. As noted
above, the observed ranges are in good agreement with our results.
According to our simulations, in order to obtain the metallicity
difference between the IHPs and OHPs observed for the MW, M32 or M31, a
contribution of \insitu stars would be required, which suggests the
accretion of massive satellites or/and the presence of disc-heated
stars. Conversely, for M81, our simulations suggest an assembly history
that is dominated by debris stars formed in a satellites of various
masses that contributed similar fractions to the stellar halo, with
almost no contribution from \insitu stars. 
However, we note that this comparison is fairly rough, as it does not
provide clear information on the metallicity profiles.

\section{Very Metal-Poor Stars}

The VMP stars, defined as those with [Fe/H] $< -2$, can provide information
about the nature of the interstellar medium from which they were born,
at early stages of the star-formation process. Hence, they have been
claimed to provide local windows on the formation of the first
subgalactic systems. The commonly adopted hypothesis is that the
lower-metallicity stars would come primarily from debris stars, while
the higher-metallicity stars would have a larger probability to have
been formed in situ. In general terms, VMP stars might better trace the
formation of the OHPs. Our simulations agree with these assumptions, at
least in the OHRs, but within the IHRs, a more careful analysis is
needed.

\begin{figure}
\hspace*{-0.2cm}\resizebox{7.5cm}{!}{\includegraphics{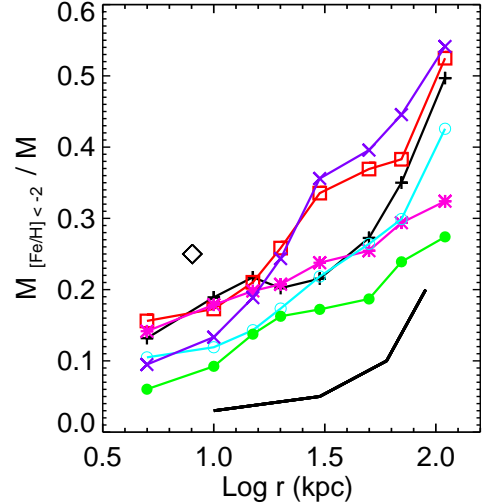}}
\caption{Fraction of the total stellar mass associated with stars with [Fe/H]$ <
-2$, the VMP stars (solid lines, see colour and symbol coding from Table 1). For
comparison we include observational estimates for the MW
\citep[large diamond]{an2013} and M31 \citep[black thick line]{gilbert2013}. }
\label{LowMet2}
\end{figure}

First, we calculated the frequency of VMP stars in our simulated haloes
as a function of galactocentric distance. As shown in
Fig.~\ref{LowMet2}, VMP stars represent less than $\sim 20$ per cent of the
total stellar mass within the IHRs, while their percentage increases
rapidly in the OHRs. For comparison, we include the observed estimates
for the Milky Way reported by \citet{an2013}, based on an analysis of
the photometric MDF of main-sequence stars in SDSS Stripe 82 located
between 5 and 8 kpc from the Sun\footnote{Note that we have plotted the
fraction that \citet{an2013}, associate with the low-metallicity (OHP)
component in their analysis, $\sim 25$ per cent, although they indicate that
the relative contribution of the OHP component increases with declining metallicity.
For [Fe/H] $< -2$, this fraction may rise to as high as $\sim 50$ per cent.},
and for M31 by \citet{gilbert2013}. The M31 metallicities are
photometric estimates for stars that are known members of the M31 giant
branch, based on spectroscopic observations. The line shown for this
sample represents the fraction of stars bluer than a 10 Gyr, [Fe/H]$=-2.0$, solar [$\alpha$/Fe] isochrone \citep{vandenberg2006}. Our
simulated values are within the limits established by these two
galaxies. In the case of M31, we can appreciate a similar increase of
the VMP stars as a function of radius, as expected from our simulations,
although it is clear that the halo system of M31 is generally more
metal-rich than our simulations. In our stellar haloes, VMP stars are
primarily associated with debris subpopulations. The \insitu stars
represent a small fraction of the VMP stars, with only a very weak
dependence on radius. Between $\sim 50$ per cent and $\sim 90$ per cent
of the VMP debris stars are found to be born in subgalactic systems with
$M<10^{9}$ M$_{\odot}$.
 
Within the IHRs, which includes the fiducial solar neighbourhood, the fraction of
VMP stars coming from \insitu subpopulations is significant, although it
is always less than $\sim 40$ per cent . Hence, a selection of VMP stars within 
$r_{\rm HTR}$ would provide, in general, information on the properties
of debris stars, but it also could be (in some cases more than others)
confounded by the presence of \insitu subpopulation stars. 

The situation changes, however, when additional
information, such as that provided by the $\alpha$-elements, is
considered. Within the IHRs, between $\sim 55$ per cent and $\sim 90$
per cent of VMP stars with [O/Fe]$ > 0$ come from accreted subgalactic
systems. As expected, these fractions increase for larger radii. Again,
the rate of increase depends on the assembly history of each simulated
halo. We estimate that the fraction of VMP debris stars originating in
low-mass subgalactic systems ranges between $\sim 60-90$ per cent. An
interesting point is that the contribution from lower-mass subgalactic
systems is quite uniform as a function of radius. This is in agreement
with the results reported in Paper~II, where it was shown that VMP stars
cover the entire range of binding energies, while higher-metallicity
stars were concentrated towards lower binding energies.

In these simulated haloes, within a fiducial solar neighbourhood, debris
stars represent $\sim 50$ per cent of the total stellar mass of VMP stars. This
fraction increases to $\sim 60-90$ per cent if the condition of [O/Fe]$> 0$
is imposed. More than $\sim 70$ per cent of VMP stars were formed in
subgalactic systems with $M < 10^{9} M_{\odot}$. These fractions should
be taken as indicative, not definitive, since they would likely change
from halo to halo, and also with the adopted chemical-evolution model.
 There are  results which suggest that metal diffusion might affect the
frequency of very  low-metallicity stars \citep[e.g.][]{pilkington2012},
but, in this case, since it affects all simulated systems equally, the
trends reported in this work are not expected to change.  Moreover, the
treatment of metal diffusion should maintain such a trend, since it seems
to be already in agreement with observations.

\section{Conclusions}

This paper focusses on the study of how the MDF changes in the stellar
haloes from the IHRs to the OHRs, and how the characteristics of these
changes can be linked to the assembly history of the haloes in
hierarchical formation scenarios. Observations have now begun to yield
important information on the dynamics and chemical properties of the
stellar populations in the MW and nearby galaxies, which can be
understood within the current cosmological context. 

Our six simulated stellar haloes have total masses between
$\sim 10^{10} - 3 \times 10^{10}$M$_{\odot}$, values which  agree
with recent  estimates of the total stellar halo of M31 by
\citet{ibata2013}.  In addition, the stellar mass function of the
surviving satellites are well within the observed values for
MW and M31 reported recently by \citet{mcconnachie2012} and \citet{watkins2013}. The mean
metallicity of the stellar haloes correlate with  the fraction of
stars accreted from massive systems, as discussed in Paper~I.

 From our study of six MW-mass galaxies simulated within a
$\Lambda$CDM cosmology, we report the following results:

\begin{itemize}

\item The halo transition radius for a given halo, $r_{\rm HTR} $, defined as 
the distance where the relative dominance of the diffuse stellar
population changes from the IHP to the OHP, is typically $\sim 15-20$ kpc
(except for one halo which has $r_{\rm HTR} \sim 36$~kpc). This
characteristic radius defines an inner-halo region (IHR) and an
outer-halo region (OHR); the IHR is dominated by the IHP, and the OHR is
dominated by the OHP. There is a $\sim 20-40$ per cent 
contribution of the OHPs to the IHRs in our simulations, while the IHP
number densities decrease very quickly, and have an almost negligible
presence in the OHRs. 

\item The MDF and the cumulative stellar mass of the total diffuse 
stellar haloes vary as a function of galactocentric radius. The level of
change is different for each simulated system, reflecting their
different formation histories. Within the IHR, the \insitu component can
play a more important role, increasing the mean metallicity. From
$r_{\rm HTR} $ out to the virial radius, the stellar haloes are formed
primarily by the accretion of satellites with different masses.
Nevertheless, some stellar haloes show very small changes from the IHR
to the OHR, while others exhibit large metallicity differences.

\item The metallicity profiles exhibit behaviours that are consistent with
the changes in the MDFs. Our results indicate that mild metallicity
gradients are present in the OHRs of those stellar haloes which received
important contributions from debris stars formed in massive satellites
($M > 10^{9} M_{\odot}$). Flat gradients in the OHRs are found in those
stellar haloes assembled by similar contributions from more-massive and
less-massive accreted satellites; the relative contribution of
lower-mass subgalactic system is larger in this case than for those with
mild gradients.  

\item Indeed, we found that debris from lower-mass subgalactic systems
tends to contribute some $\sim 60-90$ per cent of the VMP ([Fe/H] $< -2$)
stars. These VMP stars cover the entire range of binding energy, hence
they are equally distributed within the virial radius.
Higher-metallicity stars tend to have lower binding energies, and hence,
on average, are more centrally concentrated. VMP stars arise primarily
from the debris subpopulations, and the frequency of VMP stars is an
increasing function of galactocentric distance. VMP stars belonging to
the \insitu subpopulation exhibit no trend with radius in our
simulations, and account for less than 10 per cent of the total numbers.
The increase in the fraction of VMP stars with radius appears to be a
characteristic signature of the hierarchical assembly of the haloes via
subgalactic accretion.

\end{itemize}

The analysis presented in this work shows clearly how the contributions
from subgalactic systems of different masses form stellar haloes with a
range of characteristics, and fix the nature of the chemical properties
at different galactocentric distances. Our results show how hierarchical
clustering scenarios can naturally reproduce both the presence and the
absence of metallicity gradients in the diffuse stellar haloes of large
spirals such as the Milky Way. Even so, that there are different
numerical caveats related to the subgrid physics modeling and numerical
resolution which need to be improved in future revisions of simulations
such as we have explored here. For this purpose, confrontation of the
detailed model predictions with the observed chemodynamical properties
of galaxies remains a powerful and illuminating tool.

\section*{Acknowledgements}
We thank the referee for his/her comments, which helped to improve the
paper. PBT acknowledges  the L'Oreal-Unesco-Conicet Award for
Women in Science, PICT Raices (2011) of the Ministry of Science
and Technology (Argentina), and the Cosmocomp and Lacegal Networks of
FP7 Programme of the European Community. TCB acknowledges partial
funding from grant PHY 08-22648: Physics Frontier Center / Joint
Institute for Nuclear Astrophysics (JINA), awarded by the U.S. National
Science Foundation.

\bibliography{tisserarev2}

\end{document}